\documentclass{aa}
\newcommand{\sftw}[1]{\texttt{#1}}

\usepackage{color}
\usepackage[normalem]{ulem}
\usepackage{amsmath,amssymb}	
\usepackage{textcomp}
\usepackage{booktabs}
\usepackage{xcolor}
\usepackage{nicefrac}
\usepackage{bm}
\usepackage{graphicx}
\usepackage{url}
\usepackage{savesym}
\usepackage[T1]{fontenc}
\usepackage{adjustbox}
\usepackage{orcidlink}

\makeatletter
\renewcommand*\aa@pageof{, page \thepage{} of \pageref*{LastPage}}
\makeatother

\usepackage{natbib,twoopt}
\bibpunct{(}{)}{;}{a}{}{,} 

\usepackage[caption=false]{subfig}

\definecolor{green1}{RGB}{0, 128, 0}

\begin{document} 

   \title{Vertical shear instability in two-moment radiation-hydrodynamical simulations of irradiated protoplanetary disks}

 \titlerunning{VSI in irradiated PPDs I. Angular momentum transport and turbulent heating}
 \authorrunning{Melon Fuksman, Flock \& Klahr}

   \subtitle{I. Angular momentum transport and turbulent heating}

     \author{Julio David Melon Fuksman
          \inst{1}\orcidlink{0000-0002-1697-6433}
          \and
          Mario Flock\inst{1}\orcidlink{0000-0002-9298-3029}
          \and
          Hubert Klahr\inst{1}\orcidlink{0000-0002-8227-5467}
          }

   \institute{Max Planck Institute for Astronomy, K\"onigstuhl 17, 69117 Heidelberg, Germany\\\email{fuksman@mpia.de}}

   \date{Received 31 March 2023 /
         Accepted 9 November 2023}

\abstract
{Hydrodynamical instabilities are likely the main source of turbulence in weakly ionized regions of protoplanetary disks. Among these, the vertical shear instability (VSI) stands out as a rather robust mechanism due to its few requirements to operate, namely a baroclinic stratification, which is enforced by the balance of stellar heating and radiative cooling, and short thermal relaxation timescales.
}
{Our goal is to characterize the transport of angular momentum and the turbulent heating produced by the nonlinear evolution of the VSI in axisymmetric models of disks around T Tauri stars, considering varying degrees of depletion of small dust grains resulting from dust coagulation. We also explore the local applicability of both local and global VSI-stability criteria.
}
{We modeled the gas-dust mixture in our disk models by means of high-resolution axisymmetric radiation-hydrodynamical simulations including stellar irradiation with frequency-dependent opacities. This is the first study of this instability to rely on two-moment radiative transfer methods. Not only are these able to handle transport in both the optically thin and thick limits, but also they can be integrated via implicit-explicit methods, thus avoiding the employment of expensive global matrix solvers. This is done at the cost of artificially reducing the speed of light, which, as we verified for this work, does not introduce unphysical phenomena.
}
{Given sufficient depletion of small grains (with a dust-to-gas mass ratio of $10\%$ of our nominal value of $10^{-3}$ for $<0.25$ $\mu$m grains), the VSI can operate in surface disk layers while being inactive close to the midplane, resulting in a suppression of the VSI body modes. The VSI reduces the initial vertical shear in bands of approximately uniform specific angular momentum, whose formation is likely favored by the enforced axisymmetry. Similarities with Reynolds stresses and angular momentum distributions in 3D simulations suggest that the VSI-induced angular momentum mixing in the radial direction may be predominantly axisymmetric. The stability regions in our models are well explained by local stability criteria, while the employment of global criteria is still justifiable up to a few scale heights above the midplane, at least as long as VSI modes are radially optically thin. Turbulent heating produces only marginal temperature increases of at most $0.1\%$ and $0.01\%$ in the nominal and dust-depleted models, respectively, peaking at a few (approximately three) scale heights above the midplane. We conclude that it is unlikely that the VSI can, in general, lead to any significant temperature increase since that would either require it to efficiently operate in largely optically thick disk regions or to produce larger levels of turbulence than predicted by models of passive irradiated disks.
}
{}


\maketitle

\section{Introduction}\label{S:Introduction}

Understanding the diverse mechanisms that produce turbulence in protoplanetary disks is a necessary step toward improving our understanding of planet formation.
Turbulence directly impacts the transport, settling, coagulation, and fragmentation of dust grains by controlling their diffusivity and collisional velocities \citep{Ormel2007,Johansen2014,Klahr2020}, thus playing an important role in planetesimal formation.
It can also contribute to stellar accretion by producing outward transport of angular momentum, although it is unclear whether the low level of transport produced by different hydrodynamical (HD) and magnetohydrodynamical (MHD) instabilities \citep[e.g.,][]{Lesur2022PPVII} can explain current observations of accretion rates \citep[]{Hartmann1998,Manara2016}, or whether instead accretion is mainly driven by magnetized winds \citep[e.g.,][]{BaiStone2013,Bethune2017,Gressel2020}.
On the other hand, large-scale structures produced by different instabilities, such as zonal flows and vortices, are able to concentrate dust grains, possibly triggering the streaming instability and leading to the formation of planetesimals  \citep{Johansen2007,Gerbig2020,schaefer2020}. In poorly ionized "dead" zones of protoplanetary disks, where the magnetorotational instability \citep[MRI, e.g.,][]{BalbusHawley1991} is expected to be suppressed by nonideal phenomena such as Ohmic and ambipolar diffusion and, depending on the orientation of large-scale magnetic fields, the Hall effect \citep[][]{Bai2015,Simon2018}, 
purely HD instabilities have been predicted to take over as the main mechanisms driving turbulence \citep[e.g.,][]{LyraUmurhan2019,Pfeil2019}. Among these, the vertical shear instability (VSI) stands out as a rather robust phenomenon dominating the production of turbulence in numerous HD \citep[][]{Nelson2013,Manger2018,Flock2017RadHydro} and even nonideal MHD simulations \citep[e.g.,][]{CuiBai2022}.


Local and vertically global stability analyses \citep[e.g.,][]{UrpinBrandenburg1998,Urpin2003,LinYoudin2015} predict that the VSI can be triggered as long as two conditions are met. On the one hand, the disk's pressure and density stratification must be baroclinic ($\nabla p \times \nabla \rho \neq 0$, where $p$ and $\rho$ are the gas pressure and density, respectively), which in vertical and radial hydrostatic equilibrium naturally leads to an azimuthal velocity stratification varying with distance from the midplane, that is, with vertical shear.
Baroclinic stratifications naturally result from temperature gradients either in the radial or in the vertical direction, which are typically enforced by the equilibrium between stellar or turbulent heating and radiative cooling. On the other hand, buoyancy forces, which can restore gas perturbations into equilibrium, must be suppressed.
Under these conditions, a gas parcel moving in a direction between its corresponding constant specific angular momentum surface and the disk rotation axis is accelerated away from is original position due to its angular momentum excess \citep[e.g.,][]{Knobloch1982}.
Globally, this process can destabilize radially traveling inertial waves \citep{Nelson2013,BarkerLatter2015}, leading to the production of vertically elongated vertical flows reaching significant fractions of the sound speed \citep[e.g.,][]{Flock2017RadHydro}.

If the additional buoyancy produced by magnetic tension \citep[e.g.,][]{LatterPapaloizou2018} and dust back-reaction \citep{Lin2019} are negligible, the stabilizing effect of buoyancy can be suppressed for short enough thermal relaxation timescales. A local constraint on the thermal relaxation timescale leading to instability was heuristically derived in \cite{Urpin2003}, following the work by \cite{Townsend1958}, in the spirit of a Richardson number \citep[e.g.,][]{Chandrasekhar1961}, quantifying the ratio between the work of restoring forces and the energy that can be extracted from the vertical shear. Local stability analyses are however insufficient to describe both the linear and nonlinear global evolution of the instability, which produces flows spanning different local conditions over several scale heights in the vertical direction. To overcome this limitation, a global stability criterion was derived in \cite{LinYoudin2015} for disks with vertically uniform temperatures and cooling rates.
Even though global HD simulations support this criterion in both the linear and nonlinear regimes \citep[e.g.,][]{Nelson2013,Manger2021II}, this analysis cannot account for deviations from the cited assumptions.
In such cases, local stability criteria may still be useful to determine the stability of specific regions, but little progress has been made in this direction \citep[see, however,][]{Stoll2016,Pfeil2021,Fukuhara2023}.


If dust and gas species are kept in thermal equilibrium by collisions, the thermal relaxation timescale is determined by the radiative cooling rate. Computing this quantity is in general not an easy task, as it depends on the optical depth along the typical wavelengths of the perturbations produced by the VSI \citep[see, e.g.,][]{LinYoudin2015}. These are currently unknown due to a lack of theoretical predictions, together with the fact that in HD computations the minimum wavelengths are limited by grid diffusion. 
Even if characteristic VSI modes are radially optically thin, in which case the cooling rate becomes scale-independent,
the temperature perturbations produced by the VSI flows may span several characteristic sizes with different optical depths exceeding the typical VSI wavelengths. On top of this, if larger-scale structures such as vortices are produced by secondary instabilities or independent mechanisms \citep[][]{Marcus2015,Barranco2018,Manger2018,Pfeil2021}, their optical depths may largely exceed that of the VSI modes, and hence they may cool at a different rate. Thus, the issue of computing accurate cooling timescales in HD simulations is only avoided if these naturally result from an effective solution of the radiative transfer equation, as it is done in radiation hydrodynamics (Rad-HD) codes. Moreover, unlike for isothermal and $\beta$-cooling prescriptions, such codes can be used to quantify the turbulent heating produced in VSI-unstable disk regions, which is particularly relevant for planet formation models relying on $\alpha$-viscous heating prescriptions \cite[e.g.,][]{Burn2022,Voelkel2022}.

In this work we investigated the nonlinear evolution of the VSI in axisymmetric Rad-HD simulations of protoplanetary disks including heating by stellar irradiation with realistic tabulated opacities, focusing on the VSI growth and saturation phases, the produced transport of angular momentum and heating, and the applicability of local stability criteria.
Both the baroclinic stratification and the thermal relaxation timescales required to trigger the VSI in our simulations arise
from the balance of irradiation heating and radiative cooling. We studied a small disk region between $4$ and $7$ au around a T Tauri star and consider different degrees of depletion of small dust grains due to coagulation and settling, making the disk either vertically thick or thin and resulting in the formation different stability regions.

We neglected departures from thermal equilibrium between dust and gas species occurring in low-density regions, where the thermal relaxation timescale depends on a combination of the radiative cooling timescale and the collisional timescales of dust and gas species \citep[see, e.g.,][]{Malygin2017,Barranco2018}. This effect is considered in an extension of our stability analysis presented in a second part of this work \citep[][Paper II henceforth]{MelonFuksman2023}.

We employed resolutions of up to $\sim 200$ cells per scale height, which are unprecedented in Rad-HD VSI simulations, aiming to investigate the role of secondary instabilities on the VSI saturation in Paper II. The resulting computational overhead is limited by artificially reducing the speed of light, which in this context can be done as long as thermal relaxation timescales are unaffected by this operation, as we verified both analytically and numerically in this work.
Also advantageous in terms of computational efficiency is the good parallel scalability of our employed numerical code \citep{MelonFuksman2021}, which is the first Rad-HD method applied to the study of disk instabilities not to rely on globally implicit solvers.


This article is organized as follows. In Section \ref{S:NumericalMethod} we outline our employed numerical scheme and describe our disk models. In Section \ref{SS:VSIgrowth} we characterize the velocity and vertical shear distributions in the different growth and saturation phases of the VSI, while in Section \ref{S:AngMom} we analyze the resulting redistribution of angular momentum in relation with the small and large-scale flows in our simulations.
In Section \ref{S:ThermalEvolution} we explore the thermal evolution of our simulated disks, and in Section \ref{S:StabilityAnalysis} we analyze our obtained stability regions in terms of local criteria. Finally, in Section \ref{S:Discussion} we discuss our results, and in Section \ref{S:Conclusions} we summarize our conclusions.
Additional figures and calculations are included in the appendices.




 \section{Numerical method and disk models}\label{S:NumericalMethod}
 
 \subsection{Radiation hydrodynamics}\label{S:Equations}

 We modeled the gas-dust mixture in a circumstellar disk and its heating and cooling processes by means of the Rad-HD module \citep{MelonFuksman2021} implemented in the open-source \sftw{PLUTO} code \citep[version 4.4,][]{Mignone2007}. In the chosen configuration for this work, the following system of equations is solved:
\begin{equation}\label{Eq:RadHD}
 \begin{split}
\frac{\partial \rho}{\partial t} + \nabla \cdot 
\left(\rho \mathbf{v}\right) &= 0 \\
\frac{\partial( \rho \mathbf{v})}{\partial t} + \nabla \cdot 
\left(\rho \mathbf{v} \mathbf{v}\right)+\nabla p &= 
   \mathbf{G}     -\rho\nabla \Phi  \\
\frac{\partial\left( E+\rho\Phi\right)}{\partial t} + \nabla \cdot 
\left[(E+p+\rho\Phi) \mathbf{v}\right] &= c\,G^0  
  -\nabla\cdot\mathbf{F}_\mathrm{Irr} \\
\frac{1}{\hat{c}}\frac{\partial E_r}{\partial t}+\nabla\cdot \mathbf{F}_r &= -G^0 \\
\frac{1}{\hat{c}}\frac{\partial \mathbf{F}_r}{\partial t}+\nabla\cdot \mathbb{P}_r &= -\mathbf{G}\,,
\end{split}
\end{equation}
where $\rho$, $p$, and $\mathbf{v}$ are the gas density, pressure and velocity, while $E_r$, $\mathbf{F}_r$, and $\mathbb{P}_r$ are respectively the radiation energy, flux, and pressure tensor. We normalize the radiation flux by the speed of light $c$ in such a way that both $E_r$ and $\mathbf{F}_r$ are measured in energy density units. The gas energy density is defined as $E=\rho \epsilon + \frac{1}{2}\rho \mathbf{v}^2$ imposing an ideal equation of state to the internal energy density, namely $\rho \epsilon = \frac{p}{\Gamma - 1}$, where we set the heat capacity ratio as $\Gamma=1.41$ \citep{DeCampli1978}. This system is completely defined by imposing the M1 closure \citep{Levermore1984M1} to define the radiation pressure tensor as a function of $\mathbf{F}_r$ and $E_r$ as $P^{ij}_r=D^{ij}E_r$, where the Eddington tensor $D^{ij}$ is defined as $
D^{ij}=\frac{1-\xi}{2}\,\delta^{ij}+
\frac{3\xi-1}{2}n^in^j
$, 
with
$
\xi=\frac{3+4f^2}{5+2\sqrt{4-3f^2}}
$, 
where $\bm{n}=\mathbf{F}_r/\vert\vert\mathbf{F}_r\vert\vert$,
$f=\vert\vert\mathbf{F}_r\vert\vert/E_r$, and
$\delta^{ij}$ is the Kronecker delta.

We assume axisymmetry and solved these equations in spherical coordinates $(r,\theta)$, where $\theta$ is the polar angle, solving also for the azimuthal $\phi$-component of all vector fields. We include a gravitational potential modeling a star of mass $M_s$ located at the center of coordinates, that is, $\Phi=-M_s G/r$. We include stellar heating by computing after each HD step the stellar irradiation flux produced by the central star as 
\begin{equation}\label{Eq:Firrad}
    \mathbf{F}_\mathrm{Irr}(r,\theta) =
    \pi\left(\frac{R_s}{r}\right)^2
    \int_{\nu_\mathrm{min}}^{\nu_\mathrm{max}} \mathrm{d}\nu\,
    B_\nu(T_s)\,
    e^{-\tau_\nu(r,\theta)}\, \hat{\mathbf{r}} \,,
\end{equation}
where $T_s$ is the star temperature, $R_s$ is the star radius, 
$B_\nu(T)=(2 h \nu^3/c^2)/(e^{h\nu/k_B T}-1)$
is the Planck radiative intensity at a temperature $T$ and frequency $\nu$, and $h$ and $k_B$ are the Planck and Boltzmann constants, respectively. The gas temperature, assumed to be the same for dust grains, is computed according to the ideal gas law
$
T = \frac{\mu u}{k_\mathrm{B}}\frac{p}{\rho}
$, 
where $\mu=2.35$ and $u$ are respectively the assumed gas mean molecular weight and the atomic mass unit. The frequency-dependent optical depth $\tau_\nu(r,\theta)$ is computed via ray tracing after each density update. We include for this calculation the optical depth of the part of the disk left out of the simulation domain. When constructing initial conditions (Section \ref{SS:InitialConds}), we estimate this value following the prescription in \cite{MelonFuksman2022}. In the Rad-HD simulations, we retrieve it from the initial conditions.

Optical depths are computed considering only opacities due to dust absorption, which at optical and near-infrared wavelengths is dominated by small ($\lesssim 1$ $\mu$m) grains. This does not mean that there is not significant mass in larger grain sizes, but that larger sizes do not contribute to the total opacity, as specified in the next section. We neglect dust settling of small grains and assumed a perfect coupling between the dust and gas distributions with a constant dust-to-gas mass ratio $f_\mathrm{dg}$ of small grains, thus computing the dust density of small grains as $\rho_\mathrm{d} = f_\mathrm{dg}\, \rho$. We employ for this calculation the same absorption opacities as in \cite{Krieger2020,Krieger2022}, which are obtained for spherical grains composed of 62.5\% silicate and 37.5\% graphite with a density of $2.5$ g cm$^{-3}$ and a size distribution $\mathrm{d}n\sim a^{-3.5} \mathrm{d}a$ with radii in the range $a\in[5,250]$ nm. Opacity values are logarithmically sampled in the frequency range  $[\nu_\mathrm{min},\nu_\mathrm{max}]=[1.5\times 10^{11},6\times 10^{15}]$ Hz. We also use these opacities to compute the frequency-integrated radiation-matter interaction terms, which in the gas comoving frame have the expressions
\begin{equation}\label{Eq:Gcomov}
  \begin{split}
  \tilde{G}^0&=\kappa^d_P\,\rho_\mathrm{d}\left(\tilde{E}_r- a_R T^4\right) \\
  \tilde{\mathbf{G}}&=\chi^d_R\,\rho_\mathrm{d}\,\tilde{\mathbf{F}}_r\,,
  \end{split}
\end{equation}
where tildes indicate comoving quantities, while $a_r=4 \sigma_\mathrm{SB}/c$ is the radiation constant, $\sigma_\mathrm{SB}$ is the Stefan-Boltzmann constant, and $\kappa^d_P$ and $\chi^d_R$ are, respectively, the Planck- and Rosseland-averaged absorption and extinction coefficients. The laboratory frame coefficients $(G^0,\mathbf{G})$ are obtained via a Lorentz transformation of $(\tilde{G}^0,\tilde{\mathbf{G}})$, which introduces terms of orders $\mathbf{v}/c$ and $\mathbf{v}^2/c^2$ as detailed in \cite{MelonFuksman2021}. In the simulations presented in this work these terms are negligible, as is the contribution of $\mathbf{G}$ to the gas momentum, which has no role in the resulting gas dynamics. Throughout this work we neglect scattering, namely, we take $\chi^d_R=\kappa^d_R$, which is the Rosseland-averaged absorption opacity, although the derivations in Appendix \ref{A:CoolingTimeScale} consider a general $\chi^d_R$. Graphs of the frequency-dependent and frequency-averaged opacities used in this work are shown in \cite{MelonFuksman2022}.

Our resolution method relies on the reduced speed of light approximation (RSLA), in which the speed of light in the radiation block is reduced to an artificially low value $\hat{c}<c$ in order to reduce the scale disparity between gas and radiation characteristic speeds. This avoids the employment of the prohibitively small time steps that would be given by the Courant-Friedrichs-Lewy condition using the real speed of light, thus reducing the overall computational cost. The obtained solutions of the Rad-HD equations are in general unchanged for large enough $\hat{c}$ values, as we later verify in our setup.

\begin{figure*}[t!]
\centering
\includegraphics[width=\linewidth]{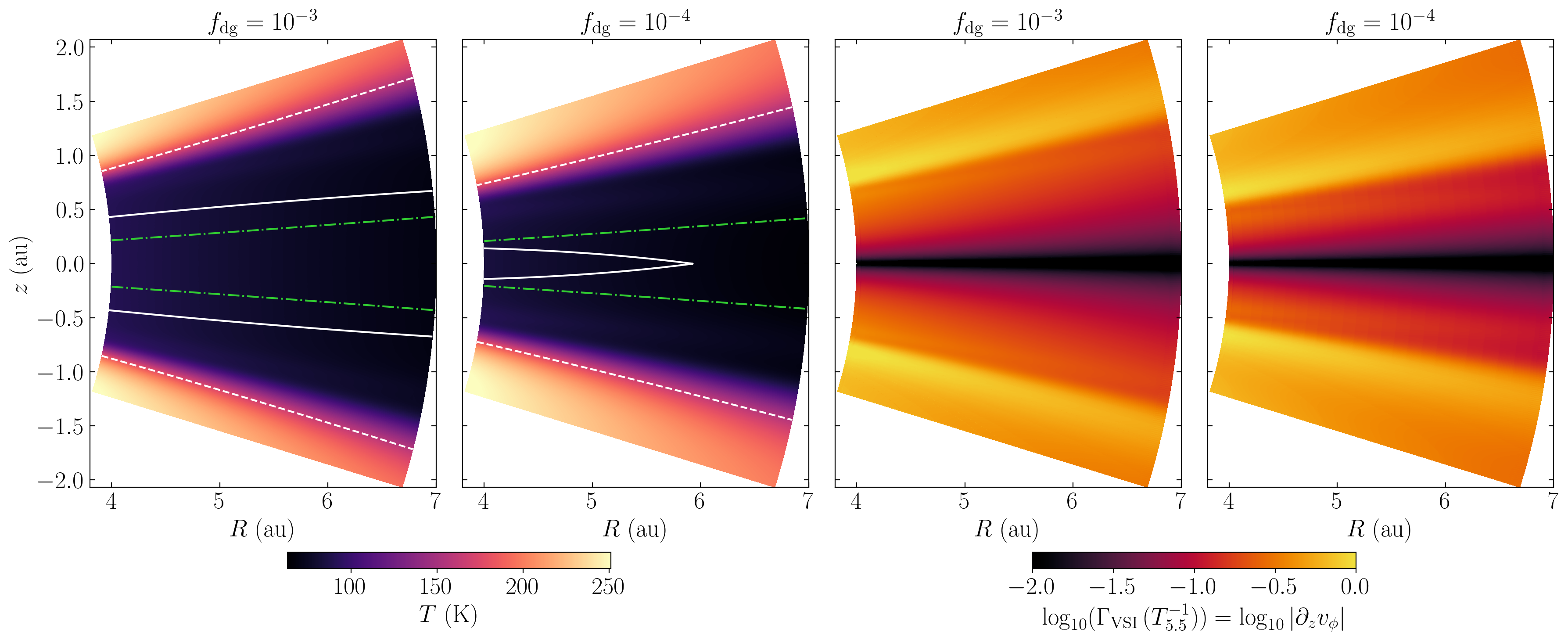}
\caption{
Initial temperature and vertical shear distributions. The latter also correspond to the VSI growth rates predicted by linear theory for instant cooling, shown here in units of the inverse orbital time at $5.5$ au ($T_{5.5}^{-1}$). Dashed, white lines indicate the $\tau_P(T_s)=1$ surface for stellar photons computed using Planck-averaged opacities, solid white lines show the $\tau_P=1$ surfaces for vertically traveling infrared photons, and green dash-dotted lines correspond to the $z=\pm H$ surfaces.}
\label{fig:hydrostaticTshear}
\end{figure*}

\begin{figure}
\centering
\includegraphics[width=\linewidth]{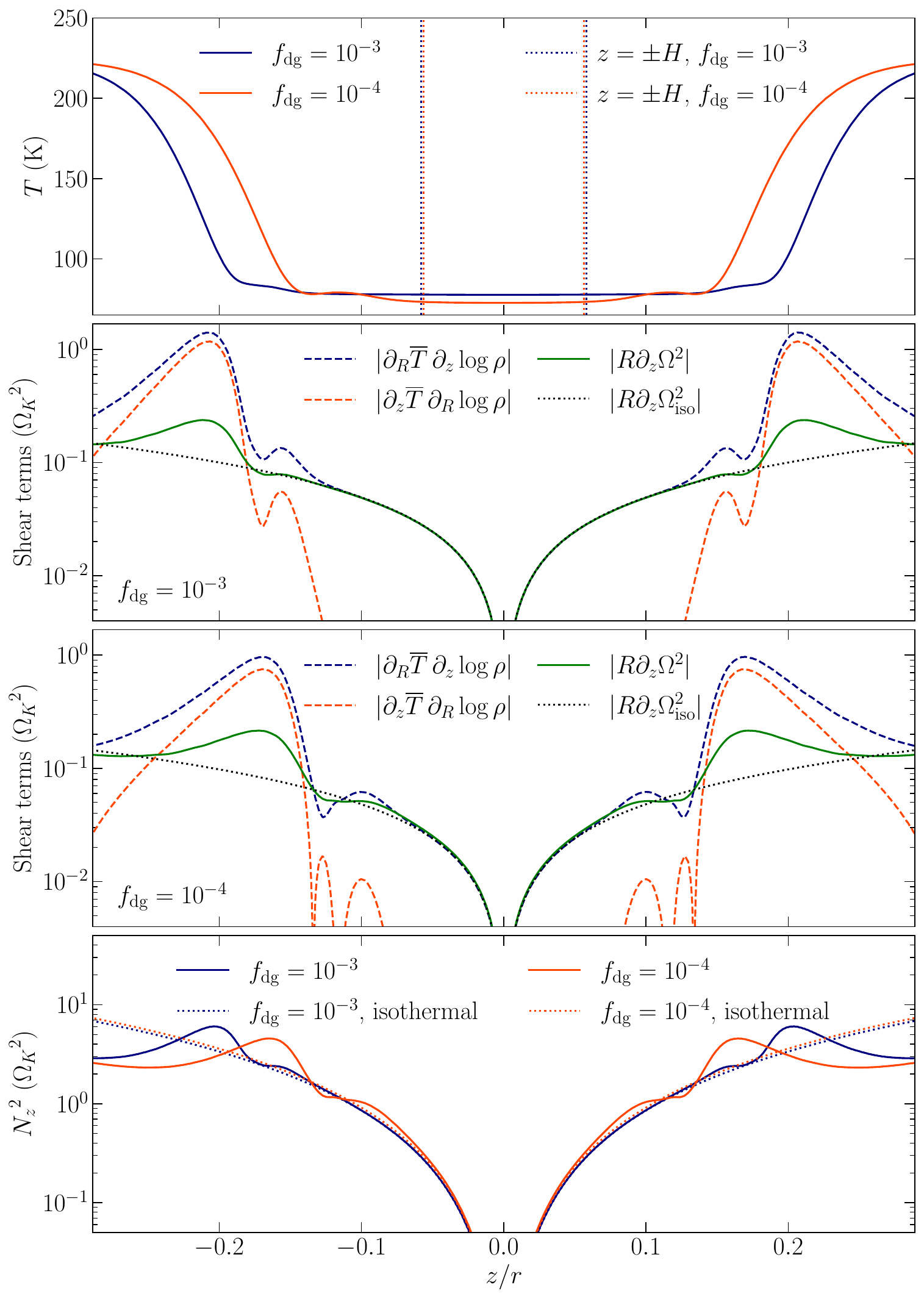}
\caption{
Vertical slices at $r=5.5$ au in the initial hydrostatic state.
Top panel: Temperature profiles. Middle panels: $R\partial_z\Omega^2$ profiles compared to the corresponding values for a vertically isothermal disk ($\Omega=\Omega_\mathrm{iso}$), and different terms contributing to the vertical shear in Eq. \eqref{Eq:ThermalWind} for $f_\mathrm{dg}=10^{-3}$ (second from the top) and $10^{-4}$ (third from the top) in units of the local squared Keplerian frequency $\Omega_K^2$. Bottom panel: Squared vertical Brunt-V\"ais\"al\"a frequency for both $f_\mathrm{dg}$ values in units of $\Omega_K^2$ (solid lines) compared to their values for vertically isothermal disks (dotted lines).
}
\label{fig:barocl_terms_T}
\end{figure}

\subsection{Initial conditions}\label{SS:InitialConds}

We considered disks initially in hydrostatic equilibrium, with temperature distributions resulting from the balance of heating due to stellar irradiation and cooling by infrared emission. We obtained our initial configurations by means of the same iterative procedure in \cite{MelonFuksman2022}, in which we alternated the computation of $(\rho,v_\phi)$ solving hydrostatic equilibrium equations and of $(T,E_r,\mathbf{F}_r)$ by means of our treatment of stellar and reprocessed radiation. The latter step was carried out by evolving the radiation fields in each iteration for a total time $t_\mathrm{iter}$ of $0.1$ orbits at $1$ au, which is enough to prevent artefacts induced by self-shadowing  \citep{MelonFuksman2022} and reach convergence. 

We constructed models of protoplanetary disks around a T Tauri star of mass $M_s=0.5$ $M_\odot$, effective temperature $T_s=4000$ K, and radius $R_s=2.5$ $R_\odot$, which leads to an effective luminosity of $\sim 1.4$ $L_\odot$. We assumed a gas column density of $\Sigma(r)=600\,\mathrm{g}\,\mathrm{cm}^{-2}\,(r/\mathrm{au})^{-1}$ which is kept constant throughout iterations. The hydrostatic and thermal equilibrium computations were carried out in a domain $(r,\theta)\in[0.4,100]\,\mathrm{au}\times[\pi/2-0.5,\pi/2+0.5]$ discretized with a resolution of $N_r\times N_\theta = 240\times 400$. The total gas mass in that region is thus $\sim 0.04 M_\odot$.  If a vertically integrated dust-to-gas mass ratio of approximately $0.01$ is assumed, the total dust mass is approximately on the upper end of the distribution of estimated dust masses as a function of stellar masses for disks observed with ALMA \citep{Manara2018}.

The only physical parameter that varies among our simulations is the assumed dust-to-gas ratio of small grains, $f_\mathrm{dg}$, whereas all other physical quantities are kept constant. We considered two values of $f_\mathrm{dg}$: a nominal case with $f_\mathrm{dg}=10^{-3}$ and a dust-depleted case with $f_\mathrm{dg}=10^{-4}$. If we assume a total dust-to-gas mass ratio of  $10^{-2}$ including all grain sizes, these values correspond respectively to maximum grain sizes of $19$ $\mu$m and $1.8$ mm, in which cases only $10\%$ and $1\%$ of the total dust mass is in the small-size population responsible for the heating and cooling processes in the disk.
The latter case can represent the late evolution of the disk after a few Myr \citep{Birnstiel2012} or simply a disk where significant coagulation and settling took place. As shown by the synthetic images in \cite{Dullemond2022}, that degree of dust depletion can still be consistent with the vertically extended appearance of scattered light images of protoplanetary disks \citep[see also][]{Pfeil2023}.

\subsection{Hydrostatic distributions}\label{SS:TempDistr}

In our Rad-HD simulations, we employed a higher-resolution grid than the one used to construct initial conditions. We defined this new grid in the domain of study $(r,\theta)\in[4,7]\,\mathrm{au}\times[\pi/2-0.3,\pi/2+0.3]$, in which we retrieved all initial distributions from the hydrostatic solution via bilinear interpolation. Inner regions of protoplanetary disks ($\lesssim 10$ au) are rather unexplored in numerical simulations \citep[see, however,][]{Stoll2016,Pfeil2021} as they are currently hard to resolve in mm observations \citep[e.g.,][]{Andrews2018}. However, our chosen disk region serves as a laboratory to study the thermal evolution produced by the VSI (Section \ref{S:ThermalEvolution}) in regions of different vertical optical depth, as this domain becomes either vertically optically thick or thin depending on the dust content (Fig. \ref{fig:hydrostaticTshear} and this section).
On the other hand, its reduced radial extent also allows us to resolve the secondary instabilities studied in Paper II, where we also investigate the stability of outer disk regions via local criteria.

The resulting initial temperature distributions for both $f_\mathrm{dg}$ values are shown in Fig. \ref{fig:hydrostaticTshear}, while vertical slices at $r=5.5$ au are shown in Fig. \ref{fig:barocl_terms_T}.
Close to the midplane, the disk is approximately vertically isothermal up to a few scale heights above the midplane ($\sim 2.5$ and $3.5$ for $f_\mathrm{dg}=10^{-3}$ and $10^{-4}$, respectively), with a temperature well described by a power law $T\propto R^{-q}$, where $q=0.49$ and $0.48$ for $f_\mathrm{dg}=10^{-3}$ and $10^{-4}$, respectively, where $R$ is the cylindrical radius. This corresponds to a disk aspect ratio $H/R=h_0\left(\frac{R}{4\,\mathrm{au}}\right)^{\frac{1-q}{2}}$, where $h_0=0.054$ and $0.052$ for $f_\mathrm{dg}=10^{-3}$ and $10^{-4}$, respectively, while $H$ is the pressure scale height measured at the midplane. The flaring index, $\frac{1-q}{2}$, is in every case positive and close to $1/4$.


At higher distances from the midplane, the temperature distributions transition from the midplane values of $65-90$ K to higher values up to $\sim 250$ K. This transition occurs near the location of the $\tau_\nu=1$ surfaces for photons emitted at the star with frequencies corresponding to maximum stellar luminosity, where the irradiation heating rate is maximal. As shown in Fig. \ref{fig:hydrostaticTshear}, this region is approximately located at the irradiation surface, which is defined through the condition $\tau_P(T_s)=1$, where $\tau_P(T_s)$ is the same optical depth computed with the Planck-averaged opacity at the star temperature. Throughout this work, we shall use the terms "upper" or "surface layers" and "middle layer" to refer to the disk regions above and below that surface, respectively. In the same figure it is shown that this temperature transition occurs a few scale heights above the region where the disk becomes optically thin for vertically traveling reprocessed photons, defined as the vertical $\tau_P=1$ surfaces computed with $\kappa_P^d(T)$ (Fig. \ref{fig:hydrostaticTshear}). In our models, the total vertical optical depth computed in this way from $z=0$ to infinity is $\tau_P\approx 10-20$ for $f_\mathrm{dg}=10^{-3}$ and $0.7-2$ for $f_\mathrm{dg}=10^{-4}$, in which case most of the domain is optically thin for vertically traveling photons. 

Above the irradiation surface, the disk is optically thin to radially traveling photons, and thus the temperature results from the balance of irradiation heating and optically thin cooling. Specifically, in terms of Equation \eqref{Eq:RadHD},
\begin{equation}\label{Eq:Tatm}
\begin{split}
    T &= \left(\frac{-\nabla\cdot\mathbf{F}_\mathrm{Irr}}{c
    \rho_d \kappa^d_P(T) a_R}\right)^{1/4}\\
    &\approx \left(\frac{\kappa^d_P(T_s)}{4\kappa^d_P(T)}\right)^{1/4}
    \left(\frac{R_s}{r}\right)^{1/2}
    e^{-\tau_P(T_s)/4}\, T_s\,,
\end{split}
\end{equation}
which is roughly proportional to $r^{-1/2}$ if opacities vary slowly with temperature and $\tau_P(T_s)\ll 1$. This explains the obtained midplane temperature power-law index $q\sim 0.5$ (steeper profiles would require the inclusion of an additional heat source, for instance, due to viscosity). Under this surface, the temperature is determined by the transport of reprocessed radiation. As described in \cite{MelonFuksman2022}, the fact that we are using a fluid approach to describe this process causes a slight deviation from isothermality right below the temperature transition ($z/r\sim 0.1-0.2$ in Fig. \ref{fig:barocl_terms_T}), which barely alters the resulting shear distribution with respect to an isothermal stratification (see Fig. \ref{fig:barocl_terms_T}).

\subsection{Vertical shear  distribution}\label{SS:VertShear}

When the disk is in hydrostatic equilibrium, an expression for the vertical shear can be obtained by writing the curl of the momentum equation (second line in Equation \eqref{Eq:RadHD}) in cylindrical coordinates $(R,\phi,z)$, which leads to the following relations:
\begin{equation}\label{Eq:ThermalWind}
\begin{split}
    \kappa_z^2 &\equiv R \partial_z \Omega^2 \\
    &= \frac{\left(\nabla p\times\nabla \rho\right)_\phi}{\rho^2}\\
    &= \left(\nabla\overline{T}\times\nabla \log\rho \right)_\phi\\
    &= \partial_z\overline{T}\partial_R\log\rho
    - \partial_R\overline{T}\partial_z\log\rho
    \,,
\end{split}
\end{equation}
where $\kappa^2_z=\frac{1}{R^3}\partial_z j_z^2$, with an analogous expression for the epicyclic frequency $\kappa^2_R=\frac{1}{R^3}\partial_R j_z^2$, quantifies the vertical variation of the specific angular momentum $j_z=R^2\Omega$, while $\Omega=v_\phi/R$ is the orbital angular frequency and $\overline{T}=p/\rho$. This equation, which is a form of the so-called thermal wind equation in geophysical scenarios, shows that a vertically varying rotation velocity occurs if the disk is baroclinic, namely, such that $\nabla p\times\nabla \rho\neq \mathbf{0}$. As evidenced by the last line, baroclinic stratifications naturally occur in vertically isothermal disks as long as $\partial_R T\neq 0$. For this reason, below the irradiation surface, $\kappa_z^2$ approximates with good accuracy its value for a vertically isothermal temperature distribution, as shown in Fig. \ref{fig:barocl_terms_T}. On the contrary, above the irradiation surface, the term proportional to $\partial_z\overline{T}$ becomes important, and the obtained vertical shear results from the balance of both terms on the right-hand side of Eq. \eqref{Eq:ThermalWind}, which have opposite signs. In particular, $\partial_z\overline{T}\partial_R\log\rho$ is positive above the midplane and negative below, and the opposite is true for $- \partial_R\overline{T}\partial_z\log\rho$. In every case, the term proportional to the radial temperature gradient dominates, and its effect is diminished by the opposite contribution of the vertical temperature gradient. At the region of maximal shear, located at the temperature transition region, both $|\partial_z\overline{T}\partial_R\log\rho|$ and $|\partial_R\overline{T}\partial_z\log\rho|$ are about $5$ times larger than the maximum $|\kappa_z^2|$, which reaches up to $2$ times its value for a vertically isothermal disk.

\begin{table*}[t!]
\centering
\caption{Varying parameters of the presented simulations and derived quantities.}
\label{tt:RadHD}
\begin{tabular}{
*{1}{p{0.10\linewidth}}
*{1}{p{0.04\linewidth}}
*{1}{p{0.10\linewidth}}
*{1}{p{0.05\linewidth}}
*{1}{p{0.05\linewidth}}
*{1}{p{0.06\linewidth}}
*{4}{p{0.09\linewidth}}}
\hline\hline
Label              & $f_\mathrm{dg}$ & $N_r\times N_\theta$ & $H/\Delta r$ & $\hat{c}/c$ 
& $t_f\,(T_{5.5})$ & $\Omega t_\mathrm{cool}(\frac{\Gamma H}{2})$ & $\Omega t_\mathrm{cool}(z_\mathrm{max})$ & $\frac{t_\mathrm{cool}}{t_\mathrm{crit}}(\frac{\Gamma H}{2})$ & $\frac{t_\mathrm{cool}}{t_\mathrm{crit}}(z_\mathrm{max})$  \\ \hline
\sftw{dg3c4\_256}   & $10^{-3}$       & $240\times 256$     & $25$ & $10^{-4}$ & $300$ &  $0.01$ & $10^{-4}$ & $0.15$ & $4\times 10^{-3}$  \\
\sftw{dg4c4\_256}   & $10^{-4}$       & $240\times 256$     & $25$ & $10^{-4}$ & $300$ & $0.13$ & $10^{-3}$ & $1.98$ & $4\times 10^{-2}$ \\
\sftw{dg3c2\_512}   & $10^{-3}$       & $480\times 512$     & $50$ & $10^{-2}$ & $300$ &  $0.01$ & $10^{-4}$ & $0.15$ & $4\times 10^{-3}$ \\
\sftw{dg3c3\_512}   & $10^{-3}$       & $480\times 512$     & $50$ & $10^{-3}$ & $300$ &  $0.01$ & $10^{-4}$ & $0.15$ & $4\times 10^{-3}$ \\
\sftw{dg3c4\_512}   & $10^{-3}$       & $480\times 512$     & $50$ & $10^{-4}$ & $3800$ &  $0.01$ & $10^{-4}$ & $0.15$ & $4\times 10^{-3}$ \\
\sftw{dg3c5\_512}   & $10^{-3}$       & $480\times 512$     & $50$ & $10^{-5}$ & $300$ &  $0.01$ & $10^{-4}$ & $0.15$ & $4\times 10^{-3}$ \\
\sftw{dg4c3\_512}   & $10^{-4}$       & $480\times 512$     & $50$ & $10^{-3}$ & $300$ & $0.13$ & $10^{-3}$ & $1.98$ & $4\times 10^{-2}$ \\
\sftw{dg4c4\_512}   & $10^{-4}$       & $480\times 512$     & $50$ & $10^{-4}$ & $300$ & $0.13$ & $10^{-3}$ & $1.98$ & $4\times 10^{-2}$ \\
\sftw{dg4c5\_512}   & $10^{-4}$       & $480\times 512$     & $50$ & $10^{-5}$ & $300$ & $0.13$ & $10^{-3}$ & $1.98$ & $4\times 10^{-2}$ \\
\sftw{dg3c4\_1024}  & $10^{-3}$       & $960\times 1024$    & $100$ & $10^{-4}$ & $1450$ &  $0.01$ & $10^{-4}$ & $0.15$ & $4\times 10^{-3}$ \\
\sftw{dg3c5\_1024}  & $10^{-3}$       & $960\times 1024$    & $100$ & $10^{-5}$ & $300$ &  $0.01$ & $10^{-4}$ & $0.15$ & $4\times 10^{-3}$ \\
\sftw{dg4c4\_1024}  & $10^{-4}$       & $960\times 1024$    & $100$ & $10^{-4}$ & $300$ & $0.13$ & $10^{-3}$ & $1.98$ & $4\times 10^{-2}$ \\
\sftw{dg4c5\_1024}  & $10^{-4}$       & $960\times 1024$    & $100$ & $10^{-5}$ & $300$ & $0.13$ & $10^{-3}$ & $1.98$ & $4\times 10^{-2}$ \\
\sftw{dg3c4\_2048}  & $10^{-3}$       & $1920\times 2048$   & $200$ & $10^{-4}$ & $300$ &  $0.01$ & $10^{-4}$ & $0.15$ & $4\times 10^{-3}$ \\
\sftw{dg4c4\_2048}  & $10^{-4}$       & $1920\times 2048$   & $200$ & $10^{-4}$ & $300$ & $0.13$ & $10^{-3}$ & $1.98$ & $4\times 10^{-2}$ \\\hline
\end{tabular}
\tablefoot{Dust-to-gas mass ratio of small grains ($f_\mathrm{dg}$), resolution ($N_r\times N_\theta$), approximate number of cells per scale height ($H/\Delta r$), ratio between the reduced and real values of the speed of light ($\hat{c}/c$), and total simulation time ($t_f$) in units of the Keplerian orbital time at $5.5$ au ($T_{5.5}$). Also shown for reference are the normalized cooling time $\Omega t_\mathrm{cool}$ and the ratio $\frac{t_\mathrm{cool}}{t_\mathrm{crit}}$ between the cooling time and the local critical value for instability (Equation \eqref{Eq:tcrit_loc}) computed in Section \ref{S:StabilityAnalysis}, both shown at the maximum height in the domain ($z/r\approx0.3$) and the height $z=\frac{\Gamma H}{2}$ determining the vertically global stability criterion in Equation \eqref{Eq:tcrit_glob}.}
\end{table*}

We can expect this larger vertical shear than isothermal disks to be translated into larger local linear growth rates and an overall stronger instability, as suggested by the relation $\alpha\propto q^2$ in \cite{Manger2021II}. For instant cooling, a local Boussinesq linear analysis predicts that the local growth rate of the fastest-growing mode is $\Gamma_\mathrm{VSI}=|\frac{\kappa_z^2}{2\Omega}|=|\partial_z v_\phi|$ \citep[][]{UrpinBrandenburg1998,Klahr2023a}. The 2D distribution of this quantity in our models is shown in Fig. \ref{fig:hydrostaticTshear}, where it can be seen that the fastest growth is expected above the irradiation surface. This does not necessarily mean larger unstable regions than in isothermal disks, since these not only depend on the vertical shear but also on the stabilizing vertical buoyancy, which is also affected by the temperature transition. We can quantify this in terms of the squared vertical Brunt-V\"ais\"al\"a frequency corresponding to vertical buoyant oscillations, $N_z^2=-\frac{1}{\rho \Gamma}\frac{\partial p}{\partial z}\frac{\partial \log (p/\rho^\Gamma)}{\partial z}$ \citep[e.g.,][]{Rudiger2002}. This quantity is larger than in isothermal disks at the region of maximal shear and smaller above (see Fig. \ref{fig:barocl_terms_T}), due to the larger and smaller vertical entropy gradient in these locations, respectively. The combined effect of these quantities on the disk stability is studied in Section \ref{S:StabilityAnalysis}. On the other hand, local analyses such as this are not enough to predict the nonlinear evolution of the instability, which is essentially nonlocal due to the vertical mass advection along several scale heights and the cooling due to radiative transfer over a broad range of lengthscales. The evolution and saturation of the VSI in its nonlinear regime are studied in the next section.


\subsection{Rad-HD simulations' setup}

 Starting from the described initial conditions, we ran Rad-HD simulations in order to study the growth and saturation of the VSI. A list of the presented runs and their corresponding parameters is shown in Table \ref{tt:RadHD}. Simulations are labeled as, \sftw{dgN$_\mathrm{dg}$cN$_c$\_N$_\theta$} where the numbers $N_\mathrm{dg}$ and $N_c$ indicate the employed $f_\mathrm{dg}$ and $\hat{c}/c$ values, respectively, while $N_\theta$ is the resolution in the $\theta$-direction. We employed four different resolutions, $N_r\times N_\theta=240\times 256$, $480\times 512$, $960\times 1024$, and $1920\times 2048$, which correspond to approximately $25$, $50$, $100$ and $200$ cells per scale height, respectively. This discretization is chosen in such a way that in all cases the cell aspect ratio $r \Delta \theta / \Delta r$ is almost exactly $1$.
 
 In every run considered in the next sections we used $\hat{c}/c=10^{-4}$. This value is still high enough to avoid introducing unphysical phenomena in our simulations, as justified in Appendices \ref{A:CoolingTimeScale} and \ref{A:RSLAtests} via analytical computations and testing with varying $\hat{c}/c$, respectively. In particular, it is shown there that the radiative cooling time is unaffected for high enough $\hat{c}/c$, which is crucial for this work given the dependence on the VSI on that timescale.
 
 The Rad-HD equations are solved using third-order Runge-Kutta time integration for the gas fields and the IMEX method derived from it in \cite{MelonFuksman2019} for the radiation fields. Gas and radiation fluxes are computed using the HLLC solvers in \cite{Toro} and \cite{MelonFuksman2019}, respectively. We employed in every case third-order WENO spatial reconstruction \citep{Yamaleev2009} with the modifications introduced in \cite{Mignone2014reconstruction} for spherical grids. In all cases, Dirichlet boundary conditions are imposed by fixing both radiation and HD fields in the ghost cells to their corresponding values in the initial hydrostatic configurations.

\section{VSI growth and saturation}\label{SS:VSIgrowth}

\subsection{Growth phases}\label{SS:VelEkin}
 
\begin{figure*}[t!]
\centering
\includegraphics[width=\linewidth]{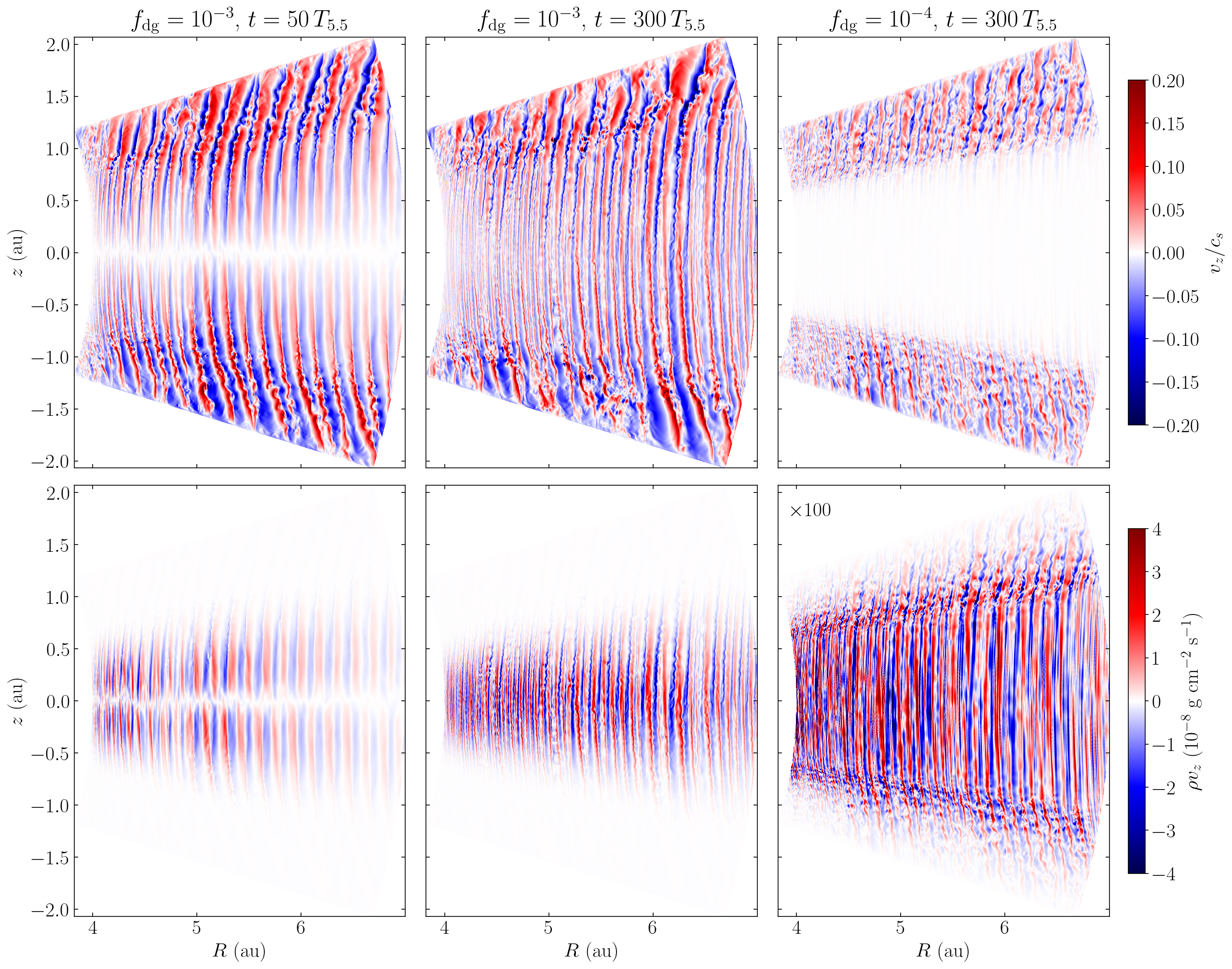}
\caption{ Vertical flows produced by the VSI. Top row: vertical velocity distributions normalized by the isothermal sound speed $c_s=\sqrt{p/\rho}$ in our highest-resolution simulations after 300 orbits (also shown after 50 orbits for $f_\mathrm{dg}=10^{-3}$). Bottom row: vertical mass flux in the same snapshots, increased by a factor of $100$ for $f_\mathrm{dg}=10^{-4}$.}
\label{fig:vel_2048}
\end{figure*}

\begin{figure}
\centering
\includegraphics[width=\linewidth]{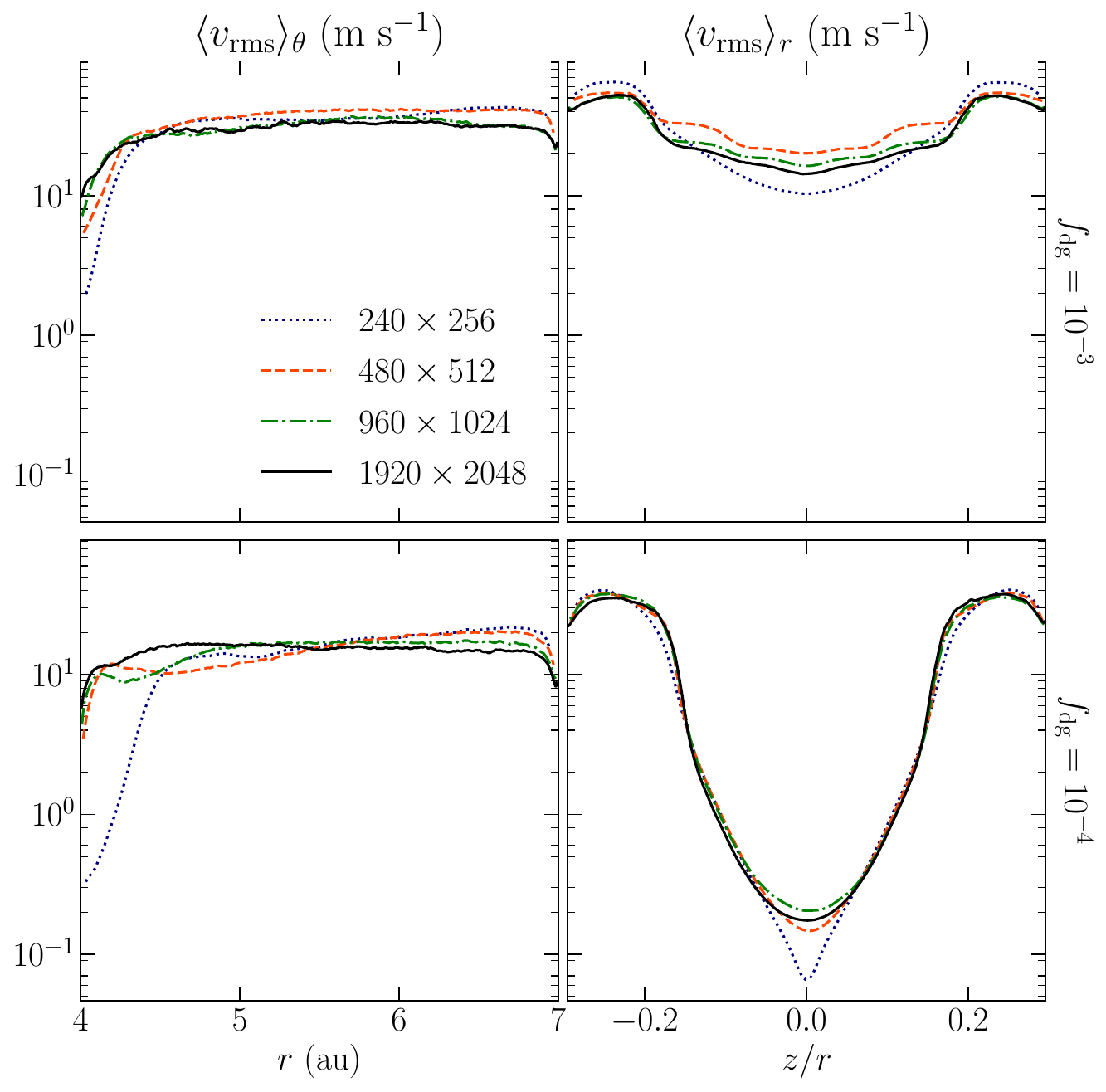}
\caption{
R.m.s. velocities computed between $150$ and $300$ orbits for different resolutions and dust content, averaged at constant $r$ (left) and constant $\theta$ (right).
}
\label{fig:vrms_1Davg}
\end{figure}

\begin{figure}
\centering
\includegraphics[width=\linewidth]{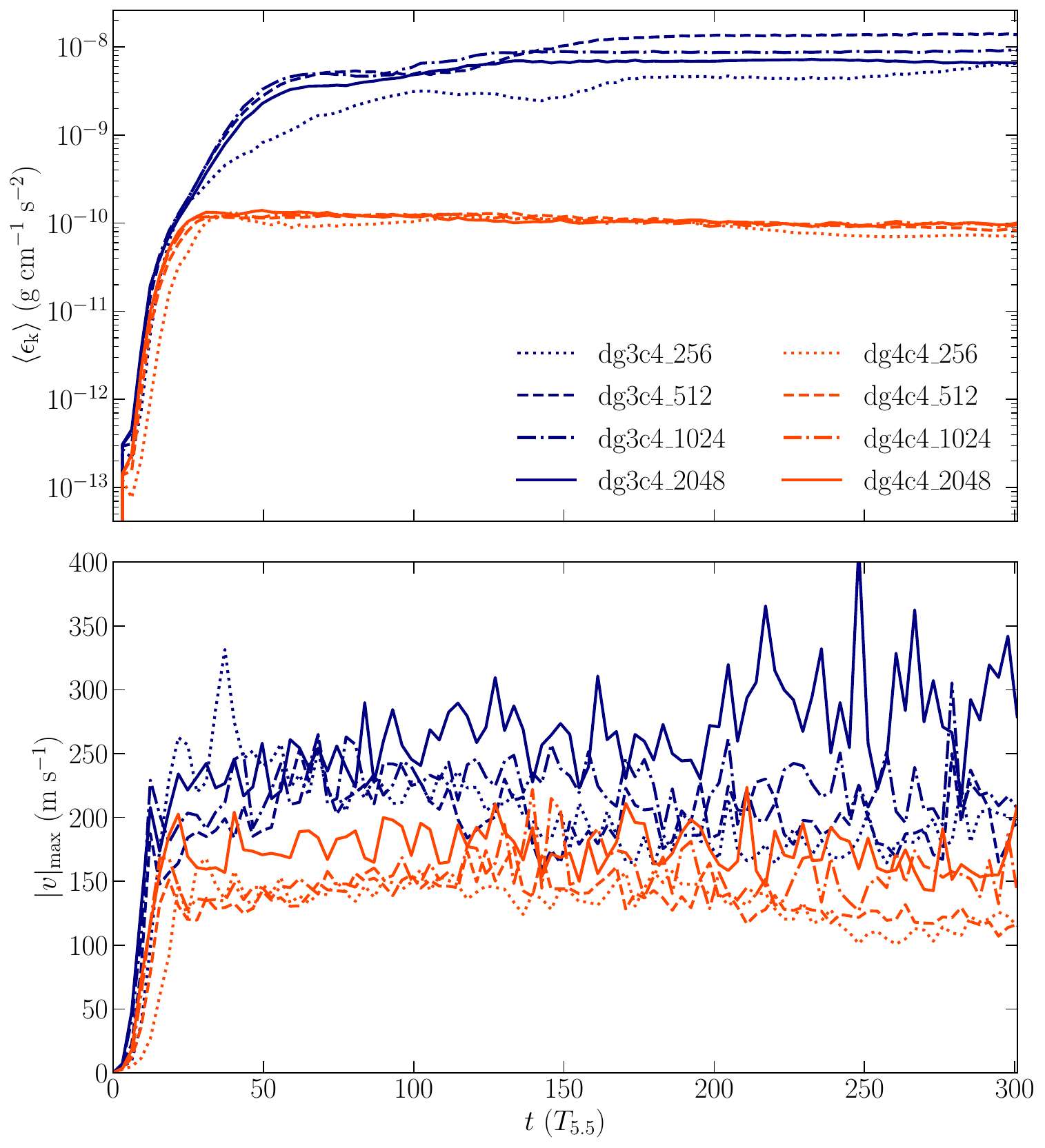}
\caption{Time evolution of the volume-averaged kinetic energy (top) and maximum gas velocity (bottom) in all runs with $\hat{c}/c=10^{-4}$.}
\label{fig:ekin_vmax_t}
\end{figure}

We begin by characterizing the flows produced by the VSI in its different growth stages. In all of our simulations, vertically elongated VSI modes can be first seen above the irradiation surface beginning at the smallest radii and occupying the entire radial domain within $\sim10$ orbits, after which they propagate toward the midplane and occupy the entire domain. This time sequence holds some similarity with the occurrence of surface modes in the vertically isothermal disk model by \cite{BarkerLatter2015}. Such modes, which dominate the early growth phases in that work, have maximum amplitude close to the vertical boundaries, where the local growth rate predicted by linear theory is maximum \citep[][]{UrpinBrandenburg1998,Klahr2023a}. In our disk models, the local growth rate is instead maximum right above the irradiation surface (Fig. \ref{fig:hydrostaticTshear}), and thus VSI modes first appear in those locations before occupying the entire irradiated regions. Thus, both phenomena might result from the time ordering determined by the local growth rate distribution.

The resulting vertical gas velocity and mass flux in our highest-resolution simulations are shown in Figure \ref{fig:vel_2048} after $50$ and $300$ orbits (we define one orbit as the rotation period $T_{5.5}$ at $5.5$ au).  For $f_\mathrm{dg}=10^{-3}$, as shown in the velocity snapshots at $t=50$ $T_{5.5}$, the instability first produces "finger" modes \citep{Nelson2013} characterized by an approximately vertically antisymmetric $v_z$, in which case no mass is transported through the midplane. After $\sim 100$ orbits, the velocity distribution transitions into vertically symmetric "body" modes occupying the entirety of the vertical domain, in which case mass is transported upward and downward at the midplane. On the contrary, for $f_\mathrm{dg}=10^{-4}$ we observe no transition between finger and body modes. Unlike in the previous case, the VSI is almost entirely suppressed at the middle layer, where meridional mass fluxes are $\sim 10^2$ times lower than for $f_\mathrm{dg}=10^{-3}$. A localized VSI suppression at the midplane is also observed in the $\beta$-cooling simulations by \cite{Fukuhara2023}. As later described in Section \ref{S:StabilityAnalysis}, this reduced activity results from the stabilizing effect of buoyancy for long enough cooling times. 
 
 Gas flows are in every case fastest at the disk upper layers, where the vertical shear is maximum and the 
 local instability criterion is met in all cases (Section \ref{S:StabilityAnalysis}).
 The gas r.m.s. velocity in the saturated phase is maximum above the irradiation surface (see Fig. \ref{fig:vrms_1Davg}), reaching at the highest resolution up to $\sim50$ and $40$ m s$^{-1}$, corresponding to Mach numbers $\mathcal{M}\approx 0.07$ and $0.05$, for $f_\mathrm{dg}=10^{-3}$ and $10^{-4}$, respectively. At that resolution, the VSI modes produce maximum vertical velocities of $\sim 200$ m s$^{-1}$ ($\mathcal{M}\sim 0.24$) and $\sim100$ m s$^{-1}$ ($\mathcal{M}\sim 0.12$) for $f_\mathrm{dg}=10^{-3}$ and $10^{-4}$, respectively. At the middle layer, r.m.s. velocities are about $3$ times lower than at the upper layers for $f_\mathrm{dg}=10^{-3}$, whereas for $f_\mathrm{dg}=10^{-4}$ a reduction of $2$ orders of magnitude occurs due to the suppression of the VSI in that region. Some level of convergence can be seen in the r.m.s. velocities at different resolutions in Fig. \ref{fig:vrms_1Davg}, but this is not true for other quantities. 
 As discussed in the next sections, significant changes in the gas flow are seen when increasing the resolution, and likely our highest resolution is not enough to reach convergence. In particular, the width of the body modes consistently decreases with increasing resolution, as shown in Fig. \ref{fig:vel_resolution}.
 
The different growth phases of the instability can be determined by evaluating the time evolution of the volume-averaged meridional kinetic energy, $\langle\epsilon_k\rangle=\langle \frac{\rho}{2}(v_r^2+v_\theta^2)\rangle$, shown in Fig. \ref{fig:ekin_vmax_t} for varying dust content and resolution. For $f_\mathrm{dg}=10^{-3}$, two distinct growth phases can be seen: after $\sim 50-100$ orbits, the kinetic energy plateaus following the saturation of the finger modes, after which it starts increasing again as body modes start growing and it stalls when these saturate. Similar behavior is seen, for example, in \cite{Stoll2014}, \cite{Manger2021II}, and \cite{FloresRivera2020}. The saturation of the body modes occurs at $t\sim 125-160$ orbits in all cases except for $N_\theta=256$, in which case body modes take up to $\sim 400$ orbits to grow due to the higher numerical diffusion. On the contrary, for $f_\mathrm{dg}=10^{-4}$, we observe a single growth phase, as no clear transition occurs between dominant modes due to the reduced VSI activity close to the midplane.

\begin{figure*}[t!]
\centering
\includegraphics[width=\linewidth]{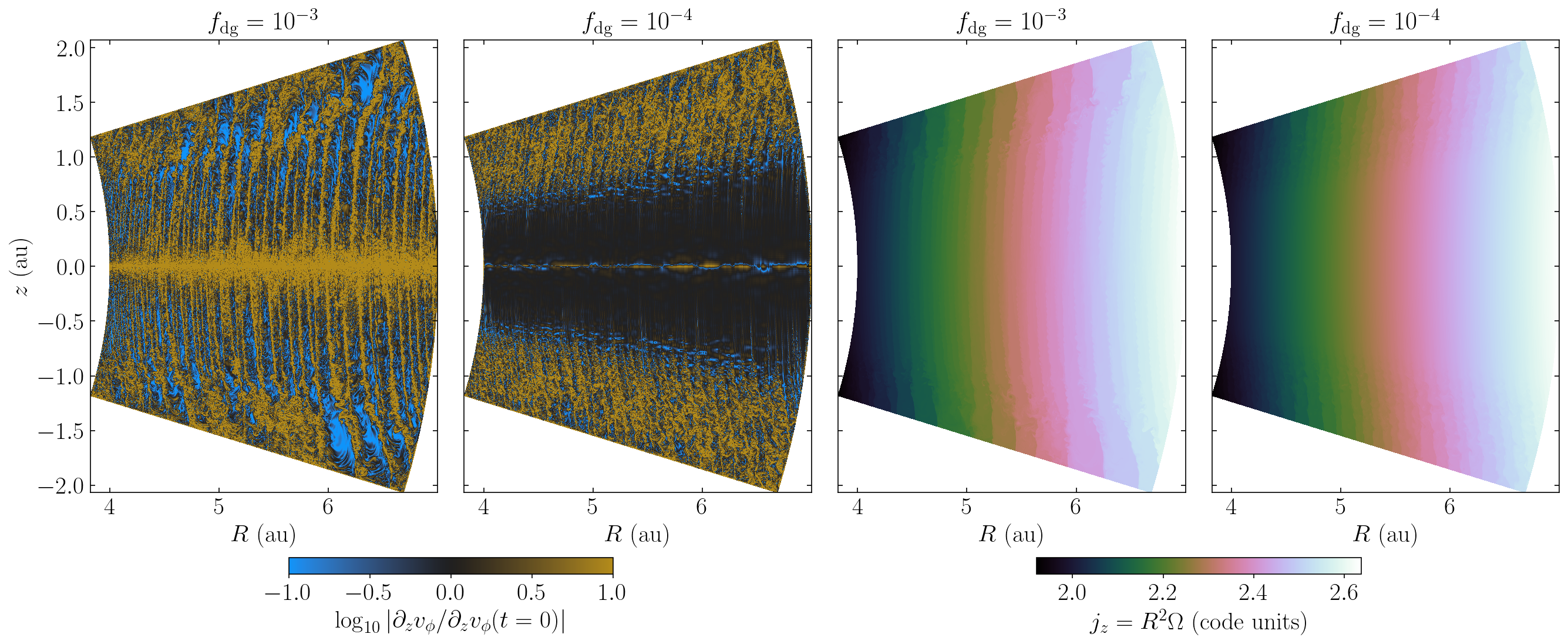}
\caption{Vertical shear and angular momentum. Left panels: comparison between the vertical shear at $t=0$ and $t=300$ orbits in the highest-resolution runs. Right panels: corresponding specific angular momentum distributions after 300 orbits.}
\label{fig:sh_sh0_j}
\end{figure*}

It can be seen in Fig. \ref{fig:ekin_vmax_t} that the saturation value of $\langle\epsilon_k\rangle$ is smaller for increasing resolution for $f_\mathrm{dg}=10^{-3}$. This result highlights a general feature of global VSI models discussed, for example, in \cite{BarkerLatter2015}, which is the fact that the smallest scales in which the VSI can develop are limited by the viscous length \citep{LatterPapaloizou2018}, which in inviscid models such as ours is given entirely by numerical diffusion. On the contrary, the saturated value of $\langle\epsilon_k\rangle$ for $f_\mathrm{dg}=10^{-4}$ reaches convergence for $N_\theta\geq 512$ after $\sim 30$ orbits, when the saturation of the VSI modes occurs above the irradiation surface. However, changes in the gas flow can still be seen for increasing resolution, not only in the wavelength of the body modes but also in the increased growth of parasitic Kelvin-Helmholtz (KH) eddies, visible in Fig. \ref{fig:vel_2048} as ripples in between the vertical flows and further studied in Paper II. As discussed in that work, the saturation of the VSI may result from the transfer of kinetic energy from the VSI flows into KH eddies.
This is consistent with the earlier VSI saturation for increasing resolution, and may also explain the decreasing saturated value of $\langle\epsilon_k\rangle$ with resolution for $f_\mathrm{dg}=10^{-3}$ if the Kelvin-Helmholtz instability (KHI) limits the maximum energy of the saturated body modes. This effect may be enhanced by the favored growth of the KHI when numerical diffusion is reduced, and also by the increase with resolution of the number of KH-unstable interfaces between vertical flows, given that larger resolutions lead to smaller VSI wavelengths (Figs. \ref{fig:vel_resolution} and \ref{fig:kH_r}).
On top of this, some KH eddies are subject to baroclinic amplification, as also analyzed in Paper II. The fast vortices produced by this process are responsible for the maximum velocities in Figure \ref{fig:ekin_vmax_t}, which explains why the maximum velocities increase with resolution despite the overall decrease of the kinetic energy. 

 \subsection{Vertical shear evolution}\label{SS:Shear}
 

As the VSI modes grow, the produced vertical flows tend to remove the initially unstable velocity stratification that originated the instability. Even though the initial baroclinic state of the disk is maintained by the balance of stellar irradiation and radiative cooling, the vertical shear is no longer determined by Equation \eqref{Eq:ThermalWind} after a few orbits, since it is affected by the departure of the gas velocity from its hydrostatic equilibrium distribution.

A comparison of the vertical shear distribution in the initial state and after $300$ orbits is shown in Fig. \ref{fig:sh_sh0_j} next to the corresponding distributions at that time of the specific angular momentum, $j_z=R^2\Omega$. In these snapshots it can be seen that the VSI flows generate regions of approximately uniform $j_z$, where consequently the vertical shear is reduced with respect to its initial state, as $\partial_z v_\phi = \frac{1}{R} \partial_z j_z$. As explored in Paper II, these bands begin forming due to the angular momentum redistribution produced by the motion of VSI-destabilized gas parcels across constant-$j_z$ surfaces \citep[see also][]{JamesKahn1970}. The subsequent reduction of the vertical shear causes the baroclinic torque to prevail over the centrifugal torque. Once the vertically moving flows are connected in upper disk regions, angular momentum can be transported in the sense of circulation favored by the baroclinic torque, until uniform-$j_z$ regions are formed \citep[see Paper II and][]{Klahr2023b}.

The formation of reduced shear bands is harder to see for increasing resolution, due to the increased development of KH eddies (see Paper II), which produce perturbations in the shear distribution. Conversely, the radial shear is increased with respect to the quasi-Keplerian initial state (not shown here), since $v_\phi=j_z/R\sim R^{-1/2}$ for a quasi-Keplerian flow and $\sim R^{-1}$ for constant $j_z$.
In each case, the shear in all directions is always maximum in between the uniform-$j_z$ zones. 

\begin{figure}[t]
\centering
\includegraphics[width=\linewidth]{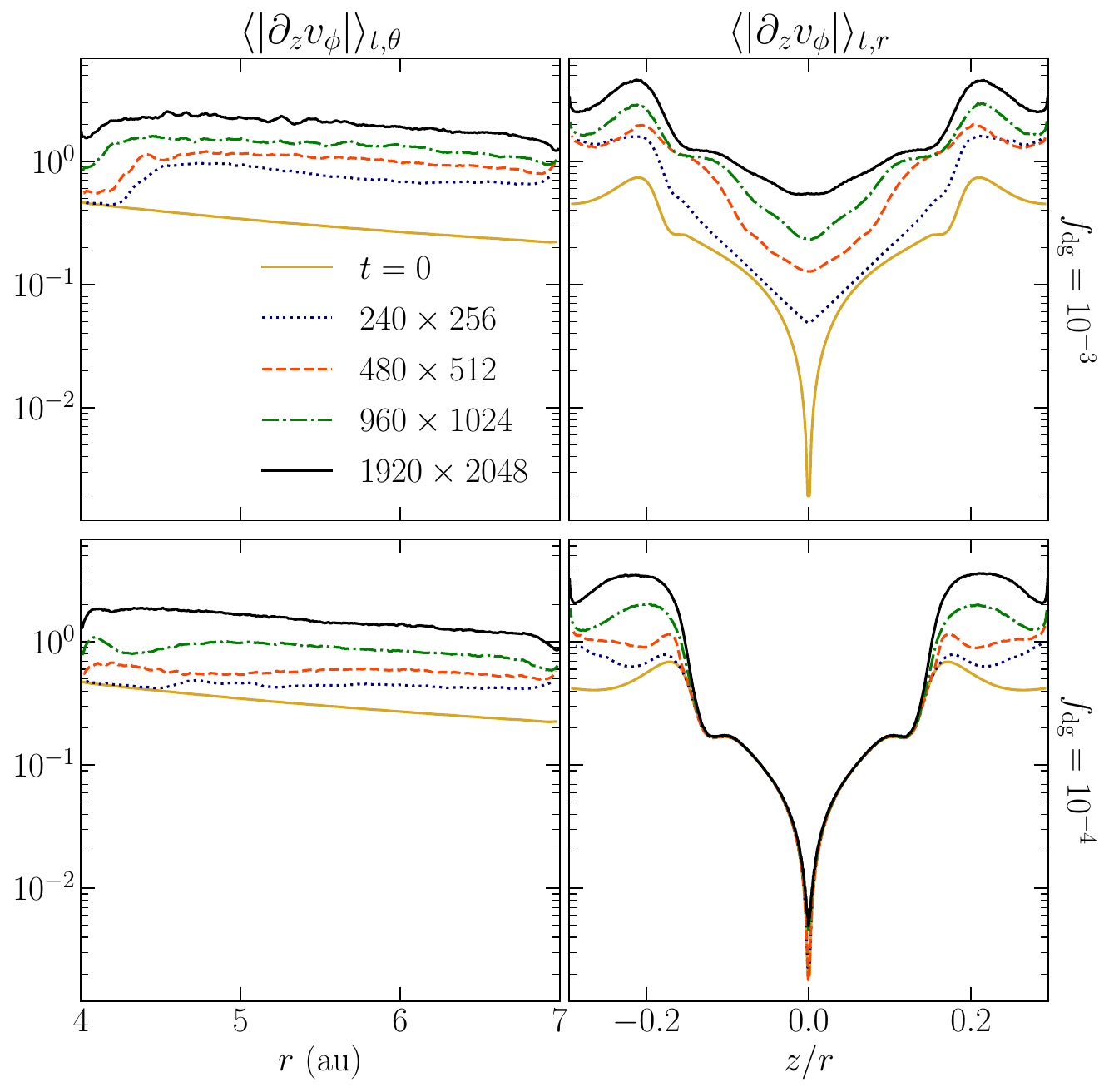}
\caption{Same as Fig. \ref{fig:vrms_1Davg}, but showing instead constant-$r$ and constant-$\theta$ averages of the averaged absolute value of the vertical shear between $150$ and $300$ orbits. The corresponding values at $t=0$ are shown for comparison.}
\label{fig:vshear_1Davgs_res}
\end{figure}

Overall, the increase of the shear in between uniform-$j_z$ regions, together with the extra shear produced by KH eddies, result in an increase of the time- and volume-averaged absolute value of the vertical shear with respect to the initial state in all regions where the VSI is active, as shown in Fig. \ref{fig:vshear_1Davgs_res}. For $f_\mathrm{dg}=10^{-4}$, this occurs everywhere except in the VSI-inactive middle region, where the shear remains unchanged. The growth of this quantity increases with resolution, almost by one order of magnitude in our highest-resolution runs. This effect is likely enhanced by the favored growth of small-scale structures such as KHI eddies as numerical diffusion is reduced (Paper II). Conversely, equivalent averages of the signed $\partial_z v_\phi$ remain unchanged with respect to the initial state. Shear fluctuations therefore largely exceed their mean value resulting from hydrostatic equilibrium a result of the nonlinear evolution of the VSI. As shown by the decrease with resolution of both the saturated $\langle \epsilon_k\rangle$ values and the components of the Reynolds stress tensor (Section \ref{SS:ReynoldsStress}), the increasing amplitude of the shear fluctuations with resolution is not translated into an increase of the VSI strength, which is instead enforced by the disk's baroclinicity and limited by the VSI saturation mechanism (see Paper II).

\section{Angular momentum transport}\label{S:AngMom}

\subsection{Reynolds stresses}\label{SS:ReynoldsStress}

\begin{figure}[t!]
\centering
\includegraphics[width=\linewidth]{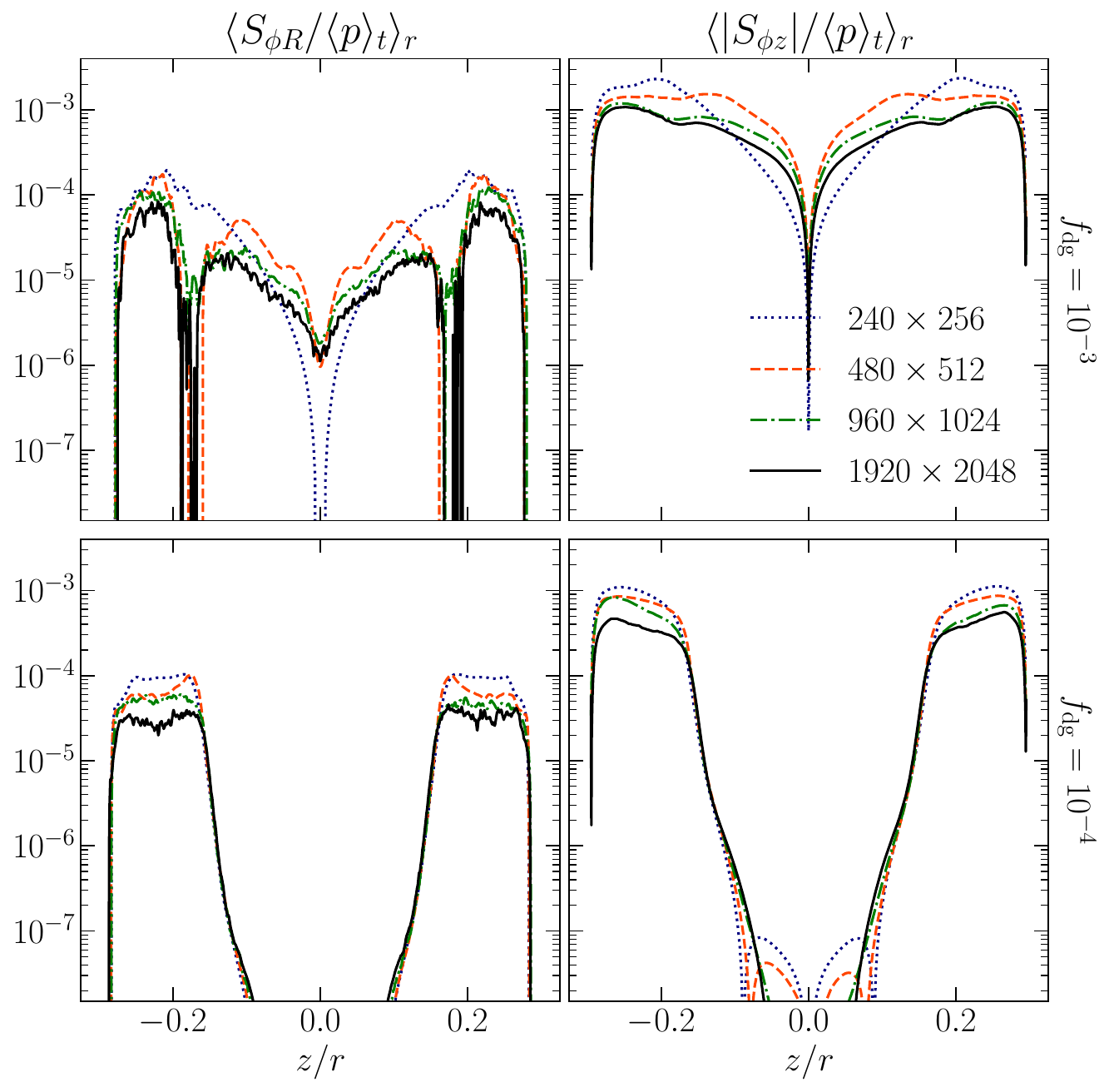}
\caption{Reynolds stress cylindrical components normalized by the average pressure, averaged over $r$ at constant $\theta$ for varying resolution and dust content.}
\label{fig:Sphi_prs_1Davgs_res}
\end{figure}

We now turn to the transport of angular momentum in the disk, which depends on the time-averaged Reynolds stress produced by the VSI. Since our disk model is axisymmetric, we replaced the $\phi$-averages normally employed in this calculation with time averages between $150$ and $300$ orbits (after the saturation of the body modes for $N_\theta\geq 512$), taking them as representative of an ensemble average over fluctuations. We thus computed the $\phi$-row of the Reynolds stress tensor as 
\begin{equation}
    \mathbf{S}_\phi = \langle \delta v_\phi   \delta(\rho\mathbf{v}) \rangle_t
                    = \langle \rho v_\phi \mathbf{v} \rangle_t -
                    \langle v_\phi \rangle_t \langle \rho \mathbf{v} \rangle_t\,,
\end{equation}
where $\langle\cdot\rangle_t$ indicates time average and $\delta$ denotes fluctuations with respect to the mean. The components of this tensor can be turned into stress-to-pressure ratios which can be used in effective 1D disk evolution models \citep[e.g.,][]{Burn2022}. However, as with other forms of disk turbulence \citep{KlahrBodenheimer2003}, the VSI-induced angular momentum transport cannot be modeled by means of an isotropic viscosity coefficient $\nu = \alpha c_s^2 / \Omega_K$ \citep{ShakuraSunyaev}, since the vertical stresses produced by the VSI are orders of magnitude larger than the radial stresses, as shown in \cite{Stoll2017} and next in this section. Despite this, the $R$-component of $\mathbf{S}_\phi$ can still be used to compute the flux of vertically integrated mass and angular momentum, and therefore it can still be parameterized as an $\alpha$-viscosity model implementable in 1D models. This can be seen by taking the azimuthal average of the angular momentum conservation equation as done, for example, in \cite{BalbusPapaloizou1999},
which in axisymmetric disks can be replaced with the mentioned average over fluctuations. This results in the following expression:
\begin{equation}\label{Eq:dtangmom}
    \partial_t(\langle \rho\rangle j_z) +
    \langle \rho \mathbf{v}\rangle \cdot \nabla j_z +
    \nabla \cdot (R \mathbf{S}_\phi) = 0\,,
\end{equation}
where $\langle\cdot\rangle$ denotes the chosen averaging procedure, and where we assumed $\langle\delta v_\phi \delta \rho\rangle=0$. The second and third terms in this equation account for the transport of angular momentum due to gas advection and VSI-induced mixing, respectively.
In quasi steady-state accretion, the projection of the average mass flux perpendicular to the surfaces of constant specific angular momentum is thus determined by the divergence of $R \mathbf{S}_\phi$. Since such surfaces are approximately vertical close to the midplane, Equation \eqref{Eq:dtangmom} can be used to compute the vertically integrated mass flux $F_\Sigma$ via an integration in $z$. In absence of disk winds, in which case both $\langle \rho v_z\rangle$ and $S_{\phi z}$ are null at $z=\pm \infty$, this results in
\begin{equation}\label{Eq:1DFSigma}
    F_\Sigma \equiv \int_{-\infty}^{\infty} \langle \rho v_R\rangle dz
    \approx
    -\frac{2}{R^{1/2}}\frac{\partial}{\partial R}
    \left( \frac{R^{1/2}}{\Omega}
    \int_{-\infty}^{\infty} S_{\phi R} dz
    \right)
\end{equation}
\citep[see, e.g.,][]{Jacquet2013}, where $\Omega$ is evaluated at the midplane. 
In turn, the column density evolves according to
\begin{equation}\label{Eq:1DSigma}
    \partial_t \Sigma + \frac{1}{R} \frac{\partial(R F_\Sigma)}{\partial R}=0\,,
\end{equation}
and so the radial transport of vertically averaged mass and angular momentum in steady state depends solely on $S_{\phi R}$.

\begin{figure}[t!]
\centering
\includegraphics[width=\linewidth]{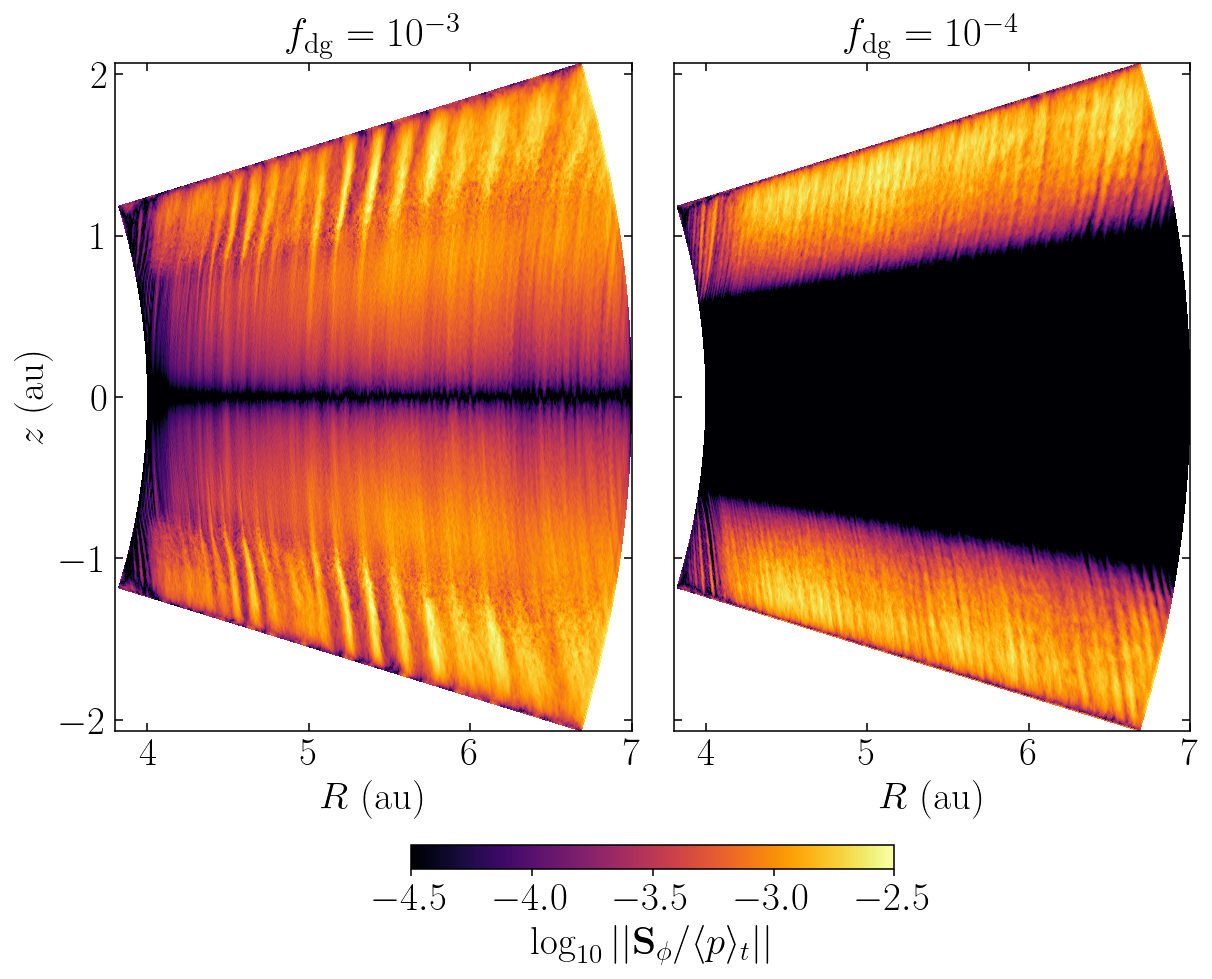}
\caption{Norm of the $\phi$-row of the Reynolds stress tensor normalized by the average pressure at the highest resolution.}
\label{fig:logSphi_p}
\end{figure}

In Fig. \ref{fig:Sphi_prs_1Davgs_res} we show pressure-weighted radial averages of $S_{\phi R}/\langle p\rangle_t$ and $|S_{\phi z}|/\langle p\rangle_t$. In every case $|S_{\phi z}|/\langle p\rangle_t$ is between $1$ and $2$ orders of magnitude larger than $S_{\phi R}/\langle p\rangle_t$, which evidences that the mixing of angular momentum in the disks occurs predominantly in the vertical direction, as obtained in \cite{Stoll2017}. In particular, $S_{\phi z}$ is positive for $z>0$ and negative for $z<0$, which means that angular momentum is transported away from the midplane. In the 2D profiles of $||\mathbf{S}_\phi/\langle p\rangle_t||$ in Fig. \ref{fig:logSphi_p}, we note that this quantity is maximum in between the approximately uniform-$j_z$ bands described in Section \ref{SS:Shear} (we can see this because some body modes do not significantly migrate radially throughout the snapshots used to compute time averages). It is then likely that the mixing of angular momentum in the interfaces between such bands, enhanced by the triggering of the KHI in those locations, is what causes them to merge and grow over time (Paper II), which also explains why the radial wavelength of the saturated modes is generally largest at the disk surface layers (see Appendix \ref{A:WavelengthEstimation}), where the mixing is maximal and the bands can merge most efficiently.

 We can see in Fig. \ref{fig:Sphi_prs_1Davgs_res} that stresses are maximum above the irradiation surface in all cases, where shear is maximum. For $f_\mathrm{dg}=10^{-4}$, stresses are null close to the midplane. The obtained values decrease with increasing resolution as in \cite{Stoll2014} and \cite{Manger2021II}, although they show early signs of convergence for the two largest resolutions. It is however unclear whether convergence would be reached for even higher resolution, and, even in that case, whether that would still happen in 3D. The observed stress reduction with resolution is possibly a consequence of the kinetic energy loss to KH eddies (Paper II), which limits the growth of the VSI modes and therefore also their produced transport of angular momentum.

To define an effective Reynolds $\alpha$ value that can be translated into effective eddy viscosity models of the form $S_{\phi R} = - \nu_R \rho R \frac{\partial \Omega}{\partial R}$ (i.e., the $\phi R$-component of the strain rate tensor of a fluid with viscosity $\nu_R$), while reproducing the usual parameterization $S_{\phi R}=\alpha p$ for thin, vertically isothermal disks, we define 
\begin{equation}\label{Eq:alphaR}
    \alpha_R = -\frac{3}{2} \frac{S_{\phi R}}{p} \Omega_K \left(R\,\frac{\partial \Omega}{\partial R}\right)^{-1}\,,
\end{equation}
 or equivalently, $\nu_R=\frac{2}{3}\alpha_R c_s^2 \Omega_K^{-1}$ \citep[see, e.g.,][]{ArmitageLectureNotes2022}, where $\Omega_K(R)$ is the Keplerian angular velocity (the extra $\frac{2}{3}$ factor accounts for the fact that $R \frac{\partial \Omega}{\partial R}\approx -\frac{3}{2} \Omega_K$ in thin, vertically isothermal disks). We have verified that the graphs of $\alpha_R$ defined in this way are practically indistinguishable from those of $S_{\phi R}/\langle p\rangle_t$,
 and so the $S_{\phi R}/\langle p\rangle_t$ averages
 shown in Fig. \ref{fig:Sphi_prs_1Davgs_res} can be interpreted as $\alpha_R$. To obtain an effective 1D viscous evolution model similar to, e.g., \cite{LyndenBellPringle1974}, we can then replace $S_{\phi R}= \alpha_R \langle p\rangle_t$ in Equation \eqref{Eq:1DFSigma}, which leads to
 \begin{equation}\label{Eq:AccretionFlux}
     F_\Sigma \approx 
    -\frac{2}{R^{1/2}}\frac{\partial}{\partial R}
    \left( R^{1/2} \langle \alpha_R \rangle_p \frac{c_s^2}{\Omega}
    \Sigma \right),
 \end{equation}
 where $\langle \alpha_R \rangle_p$ is the pressure-weighted $z$-average of $\alpha_R$ (the corresponding expression in \cite{Jacquet2013} has a multiplying factor of $3$ instead of $2$ due to their different definition of $\alpha$). In these expressions we assumed a vertically uniform squared isothermal sound speed $c_s^2=p/\rho$, which is a valid approximation given the temperature distribution in the disk middle layer. 
 
\begin{figure}[t!]
\centering
\includegraphics[width=\linewidth]{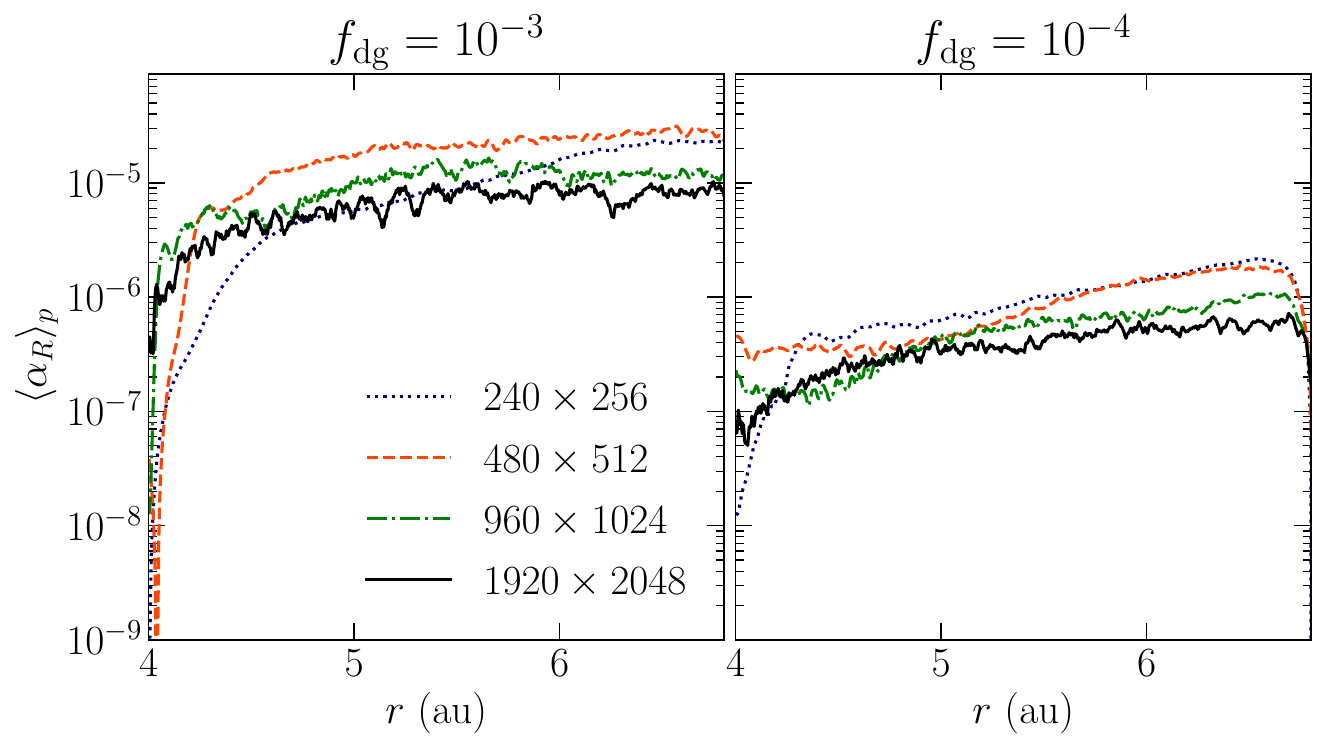}
\caption{Pressure-weighted $z$-averages of $\alpha_R$ computed for different resolutions and $f_\mathrm{dg}$ as in Equation \eqref{Eq:alphaR}.}
\label{fig:alphaR_zavgs}
\end{figure}

\begin{figure}[t!]
\centering
\includegraphics[width=\linewidth]{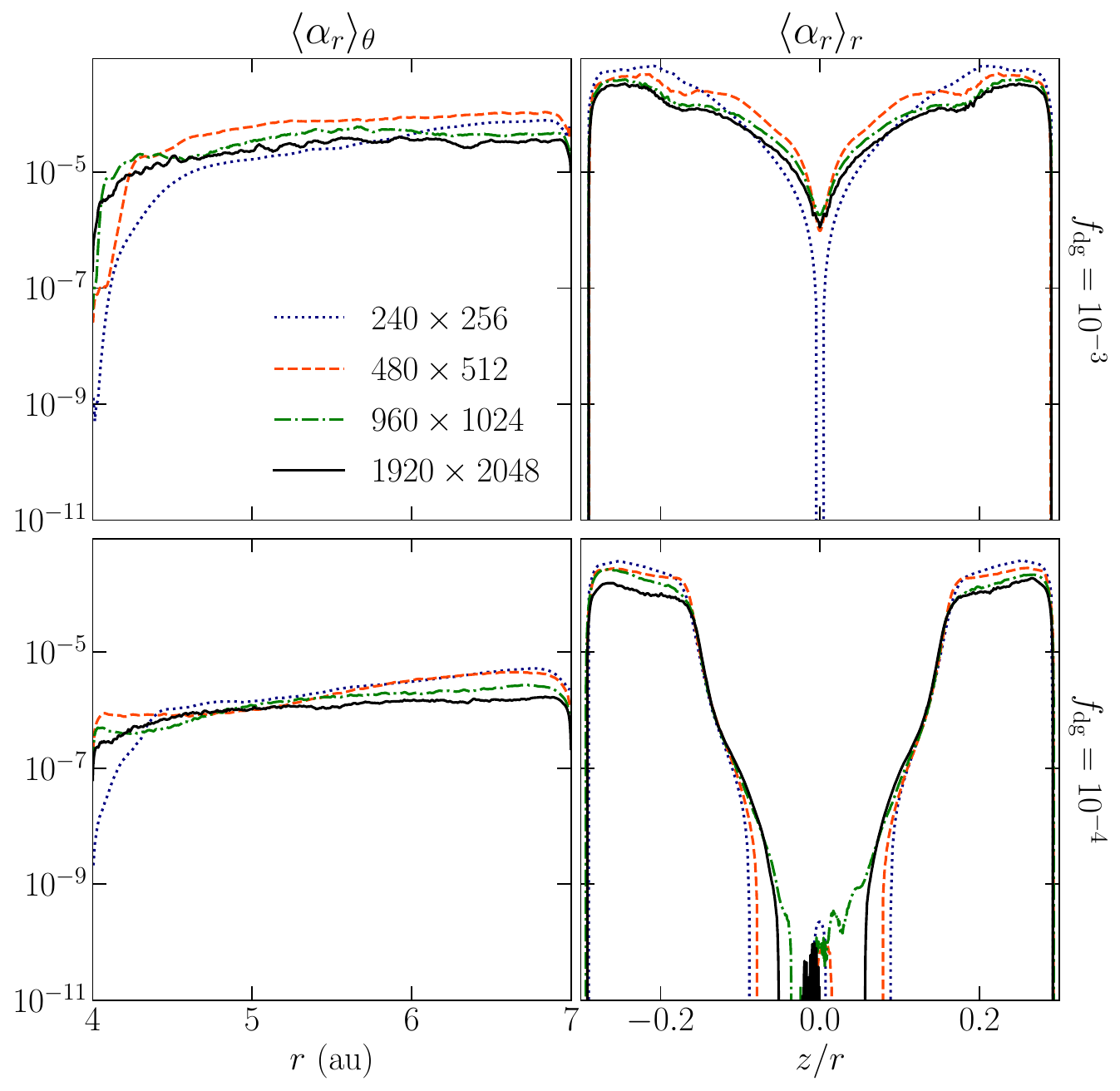}
\caption{$\alpha_r$ values for varying resolution and dust content computed as $S_{\phi r}/\langle p\rangle_t$, averaged at constant $r$ (left) and constant $\theta$ (right).}
\label{fig:alpha_r_1Davgs_res}
\end{figure}

 Values of $\langle \alpha_R \rangle_p$ computed via interpolation along constant-$R$ lines are shown in Fig. \ref{fig:alphaR_zavgs}.
 These are in the range $\sim10^{-6}-10^{-5}$ and $\sim10^{-7}-10^{-6}$ for $f_\mathrm{dg}=10^{-3}$ and $f_\mathrm{dg}=10^{-4}$. These values cannot be compared with other works, since most commonly the spherical $\alpha_r=S_{\phi r}/\langle p\rangle_t$ is computed instead of the cylindrical $\alpha_R$.
 We thus show 1D averages of $\alpha_r=$ in Fig. \ref{fig:alpha_r_1Davgs_res}, which shows that $\alpha_r\sim10^{-5}-10^{-4}$ and $\sim10^{-7}-10^{-6}$ for $f_\mathrm{dg}=10^{-3}$ and $f_\mathrm{dg}=10^{-4}$, respectively. These values are respectively about $4$ and $2$ times larger than the $\alpha_R$ averages in Fig. \ref{fig:alphaR_zavgs}. This difference despite the small $z/r$ values is somewhat surprising, but it simply results from the anisotropy of the angular momentum transport, i.e., from the fact that $|S_{\phi z}|\gg |S_{\phi R}|$. Thus, care must be taken in effective vertically integrated models, in which case $\langle \alpha_R\rangle_p$ should be used instead of the averaged $\alpha_r$
 (see Equation \eqref{Eq:AccretionFlux}).
 
\begin{figure}[t!]
\centering
\includegraphics[width=\linewidth]{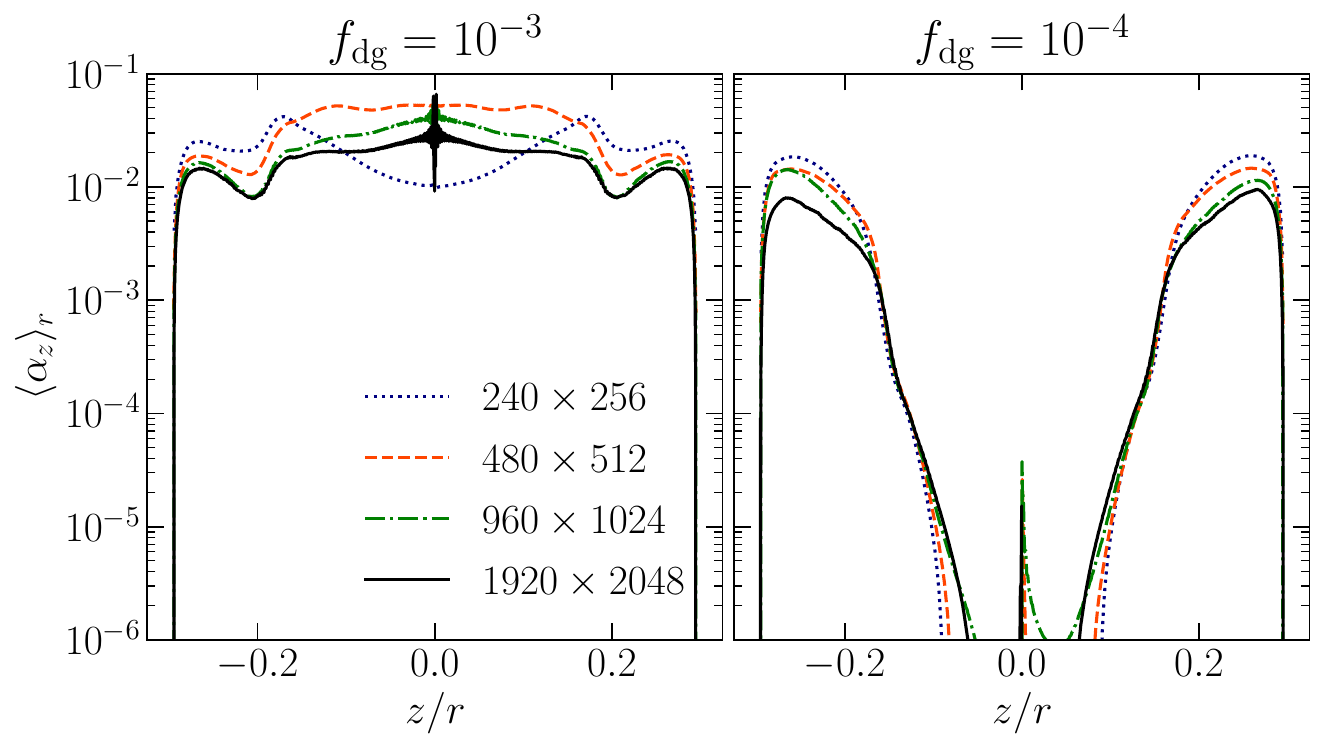}
\caption{Averages at constant $\theta$ of $\alpha_z$ computed as in Equation \eqref{Eq:alphaz} for varying resolution and $f_\mathrm{dg}$.}
\label{fig:alphaz_thavgs}
\end{figure}
 
The vertical component of the Reynolds stress can also be parameterized via an effective $\alpha_z$ number. This value is likely representative of the dust vertical diffusivity $D$ via $D=\alpha_z H^2 \Omega$ given that, unlike in the radial direction, turbulent mixing transports angular momentum and dust particles in the same direction (i.e., away from the midplane), following both the negative angular momentum and dust density gradients. The $\phi z$-component of the strain rate tensor in cylindrical coordinates is of the form $S_{\phi z}=-\nu_z \rho R \frac{\partial \Omega}{\partial z}$, and so in order to parameterize the eddy viscosity as $\nu_z=\alpha_z c_s^2\Omega_K^{-1}$ we define
 \begin{equation}\label{Eq:alphaz}
     \alpha_z = -\frac{S_{\phi z}}{p} \Omega_K
     \left(R\, \frac{\partial \Omega }{\partial z} \right)^{-1}\,.
 \end{equation}
 Even though $\frac{\partial \Omega }{\partial z}$ changes with time, we used for this calculation its initial equilibrium value, as that is the one that would be used in a viscous prescription. Resulting pressure-weighted $\alpha_z$ averages along constant-$\theta$ lines are shown in Fig. \ref{fig:alphaz_thavgs}. In every case, our obtained $\alpha_z$ values are between $2$ and $4$ orders of magnitude larger than $\alpha_R$. For $f_\mathrm{dg}=10^{-4}$, $\alpha_z$ decreases from values up to $10^{-2}$ to almost-zero at the midplane, where the instability is suppressed. On the other hand, for $f_\mathrm{dg}=10^{-3}$ at our highest resolution, we obtain $\alpha_z$ values that consistently stay between $1-2\times10^{-2}$, being approximately constant and equal to $2\times 10^{-2}$ close to the midplane, and thus the vertical mixing can be effectively modeled with a constant $\alpha_z$ of that value. A similar constancy of $\alpha_z$ is seen in \cite{Dullemond2022} and likely also in \cite{Stoll2017}, as suggested by their obtained linearity of $S_{\phi z}/\rho$ with $z$ for $z/H\lesssim 1$.
\begin{figure}
\centering
\includegraphics[width=\linewidth]{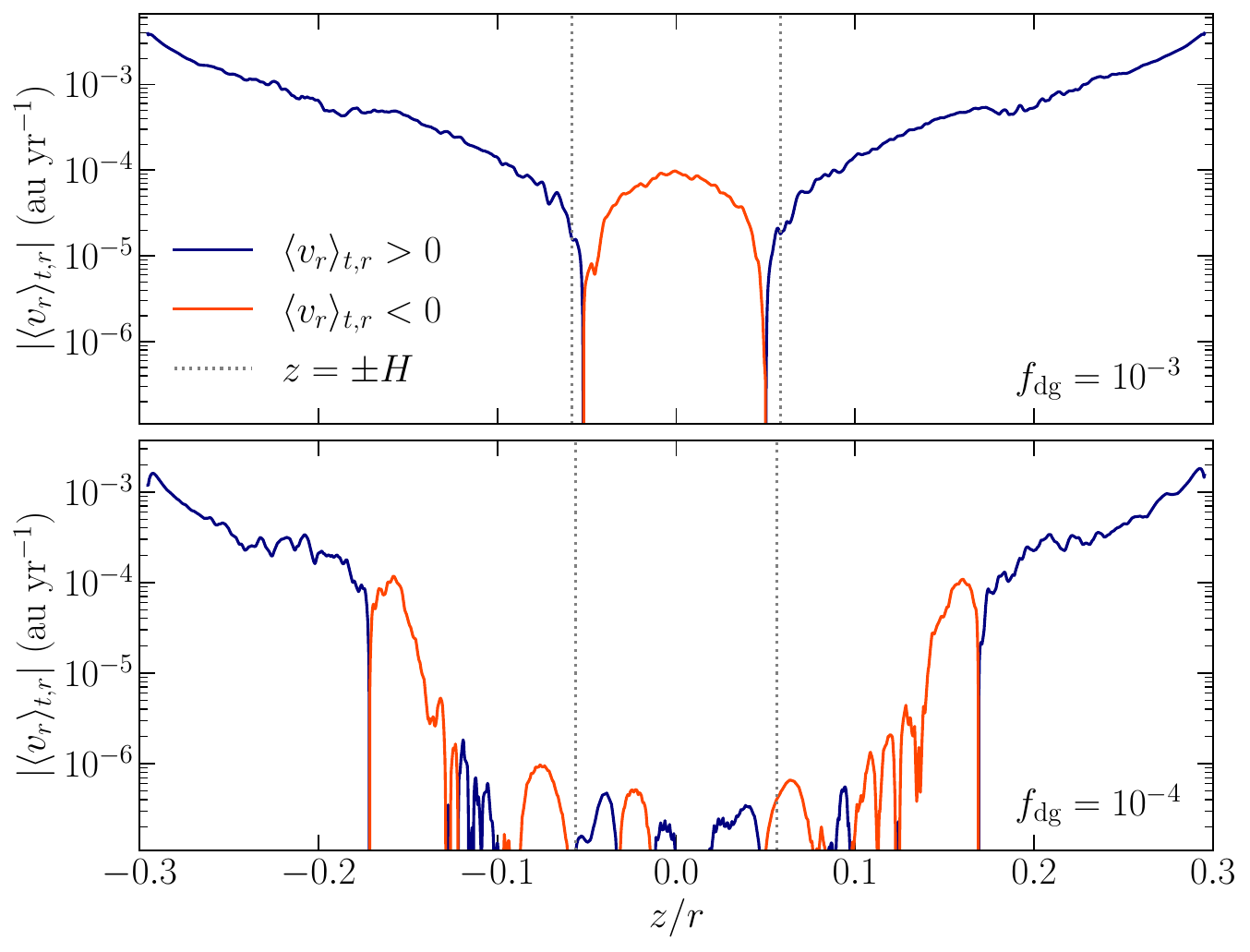}
\caption{Absolute value of the average radial velocity at $r=5.5$ au in our highest-resolution simulations. Blue and orange lines correspond respectively to outward and inward velocities, respectively. The pressure scale height is indicated for reference (gray dotted line).}
\label{fig:velr_1Davgs_res}
\end{figure}

\subsection{Meridional circulation}\label{SS:MeridionalCirculation}
 
We now turn to the average mass fluxes in our simulations. In all runs with $f_\mathrm{dg}=10^{-3}$, we obtain an average mass inflow onto the star close to the midplane and an outflow in upper layers (Fig. \ref{fig:velr_1Davgs_res}), also seen in \cite{Stoll2016,Stoll2017,Manger2018,Pfeil2021}. Above $|z|\approx H$, the gas moves away from the star with average radial velocities between $10^{-5}$ and $10^{-3}$ au yr$^{-1}$ (Fig. \ref{fig:velr_1Davgs_res}). This so-called meridional circulation results from the vertical transport of angular momentum away from the midplane as determined by Equation \eqref{Eq:dtangmom}, which causes gas flows at the midplane to fall to lower orbits around the star and surface gas layers to move to upper orbits due to their angular momentum excess.
The same phenomenon occurs in isotropic $\alpha$-viscosity models \citep[e.g.,][]{Urpin1984,Jacquet2013,Kley1992,Rafikov2017}, 
 with the difference that in those cases the circulation pattern is inverted, and the radial velocity as a function of $z$ is shaped as a concave downward parabola. Outward flows produced by meridional circulation can lead to large-scale outward transport of dust particles \citep[][]{TakeuchiLin2002,Stoll2016}, which has been proposed \citep[e.g.,][]{Ciesla2007} to contribute to the puzzling observed presence of high-temperature minerals in outer regions of protoplanetary disks \citep{Olofsson2009} and our own solar system (\cite{Brownlee2006}, \cite{Bryson2021}, and references therein) that likely formed in hotter, inner regions.



\begin{figure}
\centering
\includegraphics[width=\linewidth]{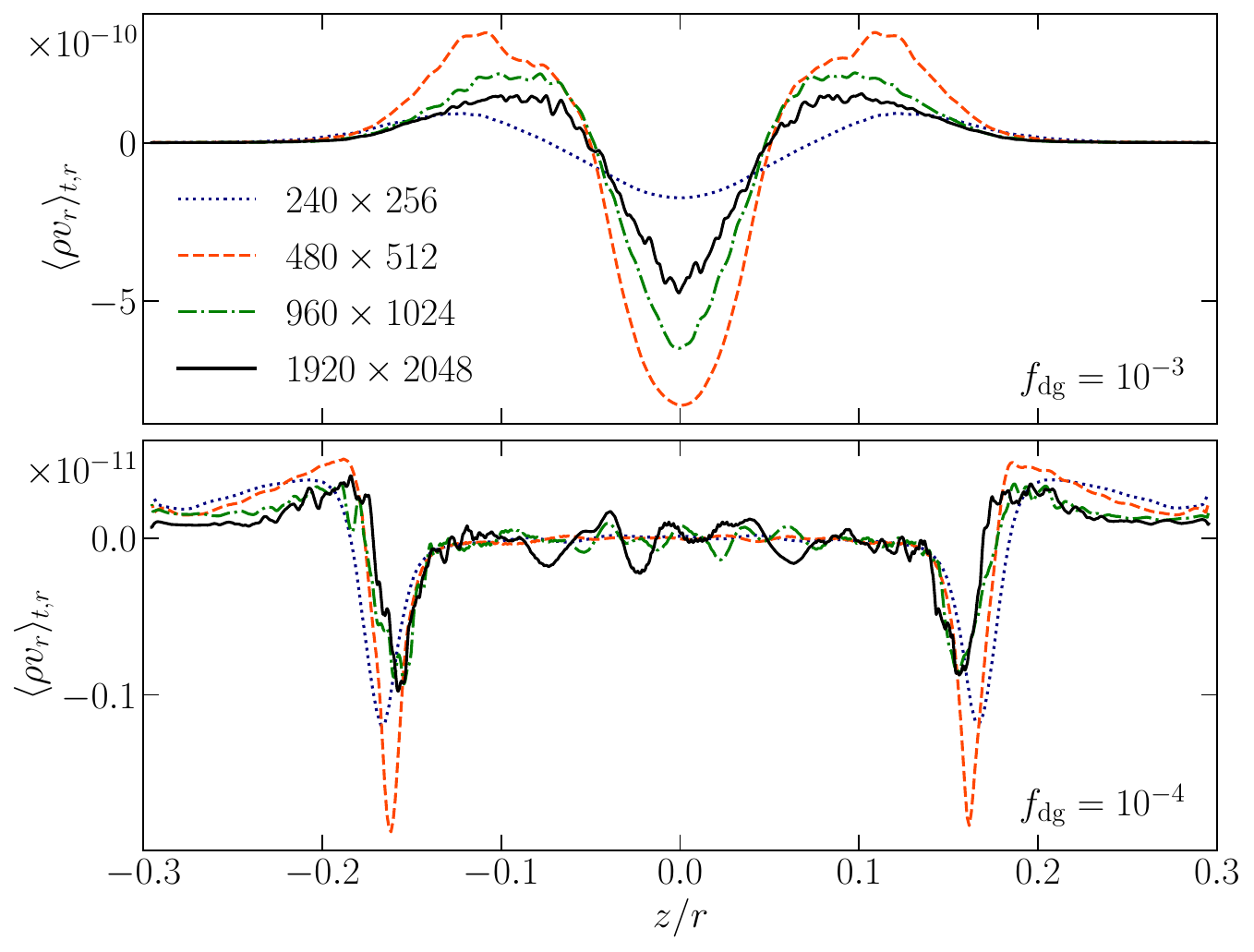}
\caption{Average mass flux in the radial direction at $r=5.5$ au for varying resolution and dust content.}
\label{fig:mflux_1Davgs_res}
\end{figure}
 
A meridional circulation can also be seen in the dust-depleted case, albeit with much smaller mass fluxes, as shown in Fig. \ref{fig:mflux_1Davgs_res}. In that case, the gas flux is negative close to the irradiation surface and positive for higher altitudes, while at the midplane it oscillates around zero. This occurs because a net upward vertical angular momentum transport is only produced at the VSI-active disk surface layers, as shown, e.g., in Figs. \ref{fig:Sphi_prs_1Davgs_res} and \ref{fig:logSphi_p}. Therefore, outward flows can still occur in upper layers even if the VSI is suppressed at the disk middle layer. Disk regions with an unstable surface layer and a VSI-inactive middle layer can be expected for sufficient depletion of small dust (Section \ref{S:StabilityAnalysis} and Paper II).



\begin{figure}
\centering
\includegraphics[width=\linewidth]{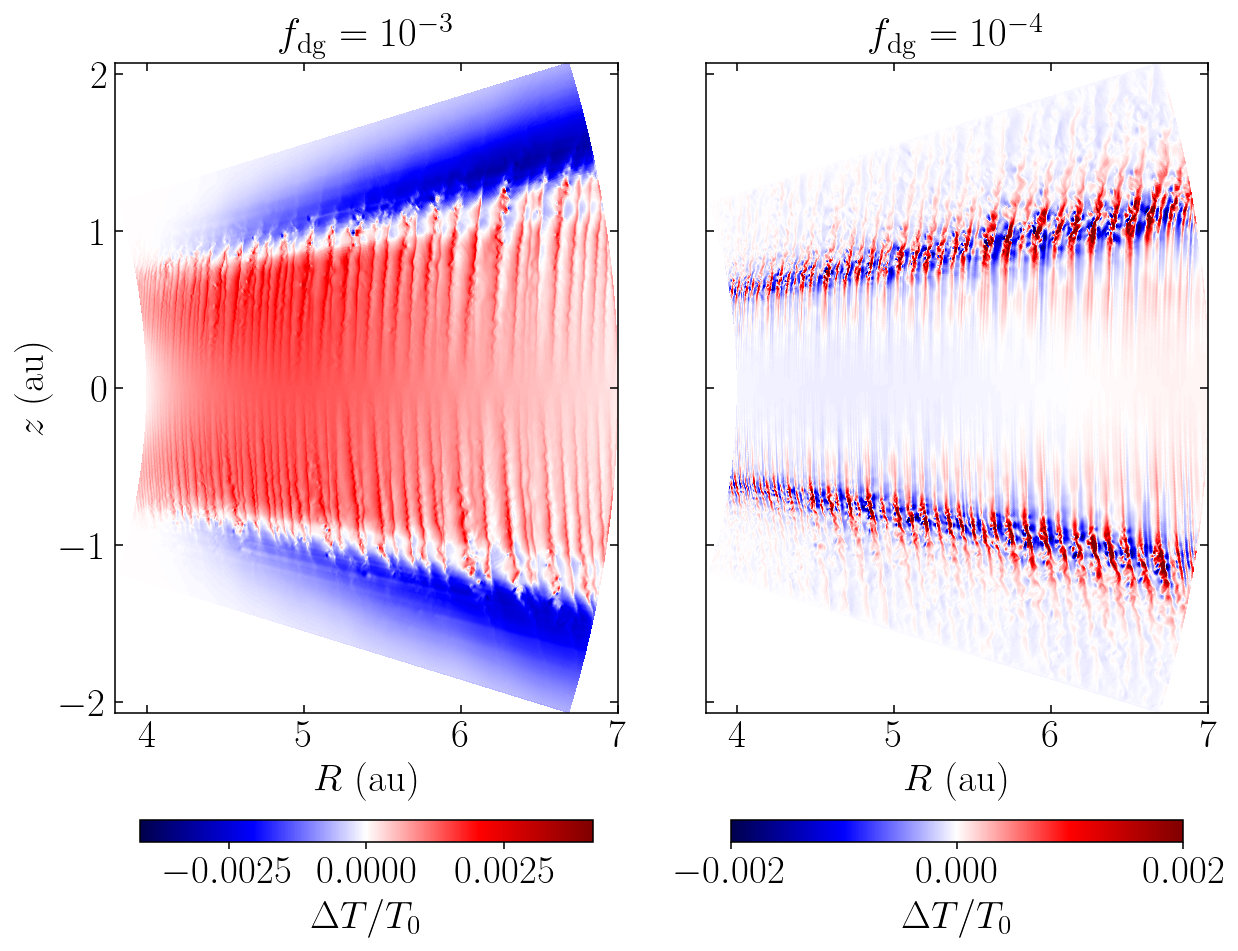}
\caption{Relative temperature variation after $300$ orbits at the maximum employed resolution.}
\label{fig:dT_T0}
\end{figure}

\section{Thermal evolution}\label{S:ThermalEvolution}

Unlike in isothermal and $\beta$-cooling simulations, the average temperature in our simulations is not forced to remain close to its initial value, which allows us
to quantify the temperature perturbations produced by the VSI. The ratio $\Delta T/T_0=(T-T_0)/T_0$ comparing the temperature distributions $T$ and $T_0$, computed respectively at $t=300$ and $4$ orbits (the latter being after the initial system relaxation but before VSI modes grow), is shown in Fig. \ref{fig:dT_T0} for the maximum-resolution runs. 

\begin{figure}
\centering
\includegraphics[width=\linewidth]{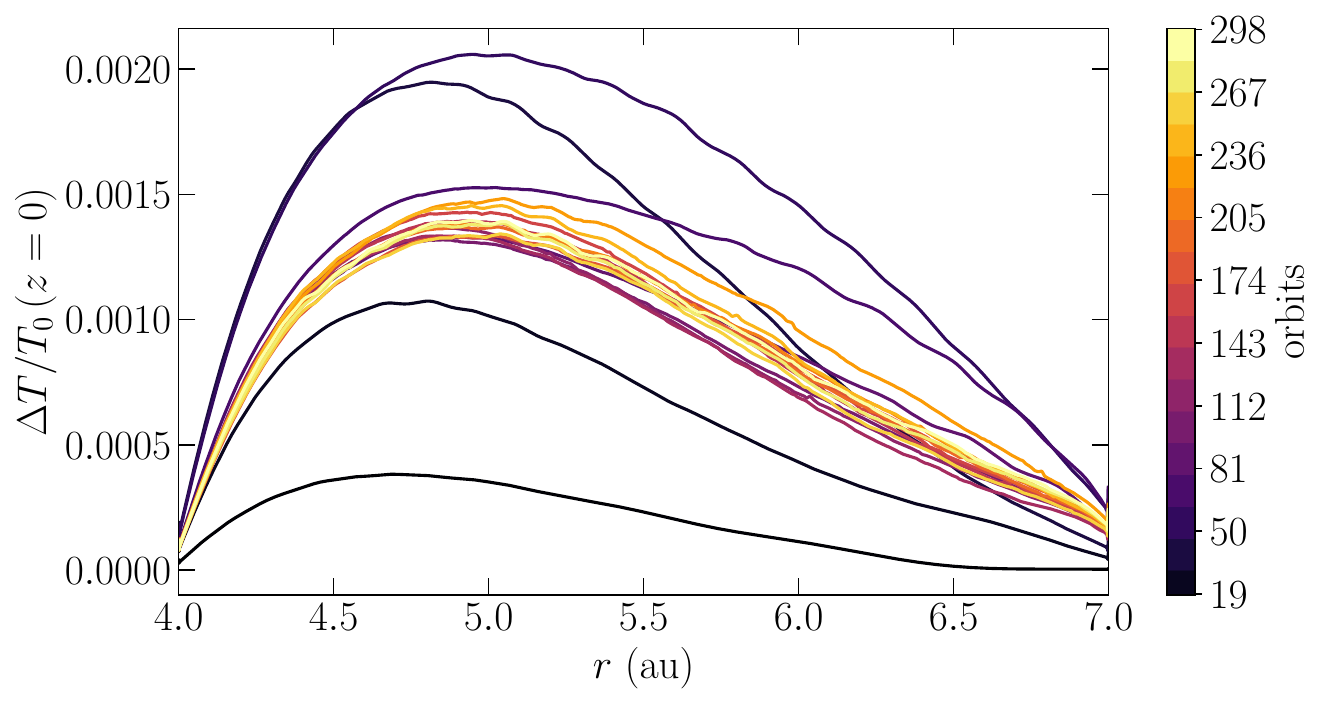}
\caption{Midplane relative temperature perturbation at different times in run \sftw{dg3c4\_2048}.}
\label{fig:dT_T0_z0_t}
\end{figure}

In all of our simulations, the temperature perturbations produced by the VSI are rather negligible. For $f_\mathrm{dg}=10^{-3}$, some heating occurs at the disk middle layer due to the radiative dissipation of the mechanical energy of the VSI modes.
Thermal equilibrium is reached when radiative cooling balances the VSI heating rate. In our highest-resolution run, this occurs at approximately $250$ orbits, i.e., about $100$ orbits after the saturation of the body modes, as shown, for instance, by the midplane temperature perturbation profiles in Fig. \ref{fig:dT_T0_z0_t}. The temperature perturbations are maximum at the maximum-shear layers in between uniform-$j_z$ regions (see Fig. \ref{fig:sh_sh0_j}), with maximum $\Delta T/T_0\sim 5\times 10^{-3}$ in such locations at the maximum resolution. This pattern is in part produced by the maximum dissipation occurring in such regions, as the heating rate is proportional to $\alpha_R$ (Equation \eqref{Eq:Qplus} later in this section), which reaches a maximum in between uniform-$j_z$ bands. On the other hand, the observed temperature perturbations are also affected by the transport of gas along the vertically oriented entropy gradient, given that also vertical velocities are maximum in those regions (Fig. \ref{fig:vel_2048}).
The energy dissipation in those locations may also be connected to the turbulent heating produced by the KHI at the regions of maximum shear and even by the baroclinic deceleration of KH eddies, as discussed in Paper II.

Above the irradiation surface, vertically elongated temperature perturbations
cannot longer be seen due to the shorter thermal relaxation timescale in that region (Section \ref{S:StabilityAnalysis}). Instead, we obtain an overall temperature decrease that becomes larger with radius. This is caused by the slight vertical expansion of the disk due to the heating of the middle layer, which leads to a density increase in the surface layers and thus to additional shadowing of stellar irradiation. Also visible in those regions are faint "shadows" of the reduced density regions produced by small-scale amplified eddies (Paper II).

\begin{figure}
\centering
\includegraphics[width=\linewidth]{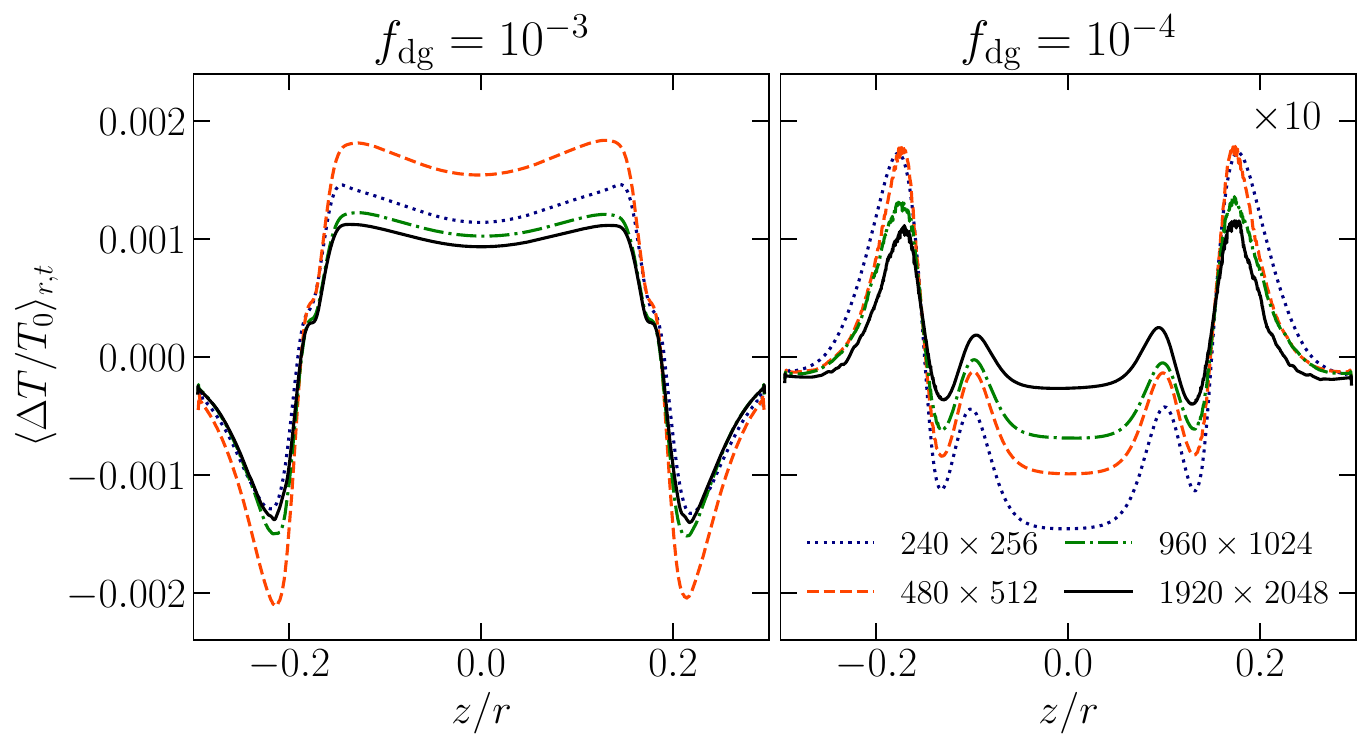}
\caption{Radially averaged relative temperature variation for all employed resolutions and $f_\mathrm{dg}$ values. Displayed values have been averaged in time between $285$ and $300$ orbits to get rid of fluctuations. The values for $f_\mathrm{dg}=10^{-4}$ have been multiplied by $10$ to facilitate the comparison.}
\label{fig:dT_T0_ravgs_res}
\end{figure}

\begin{figure}
\centering
\includegraphics[width=\linewidth]{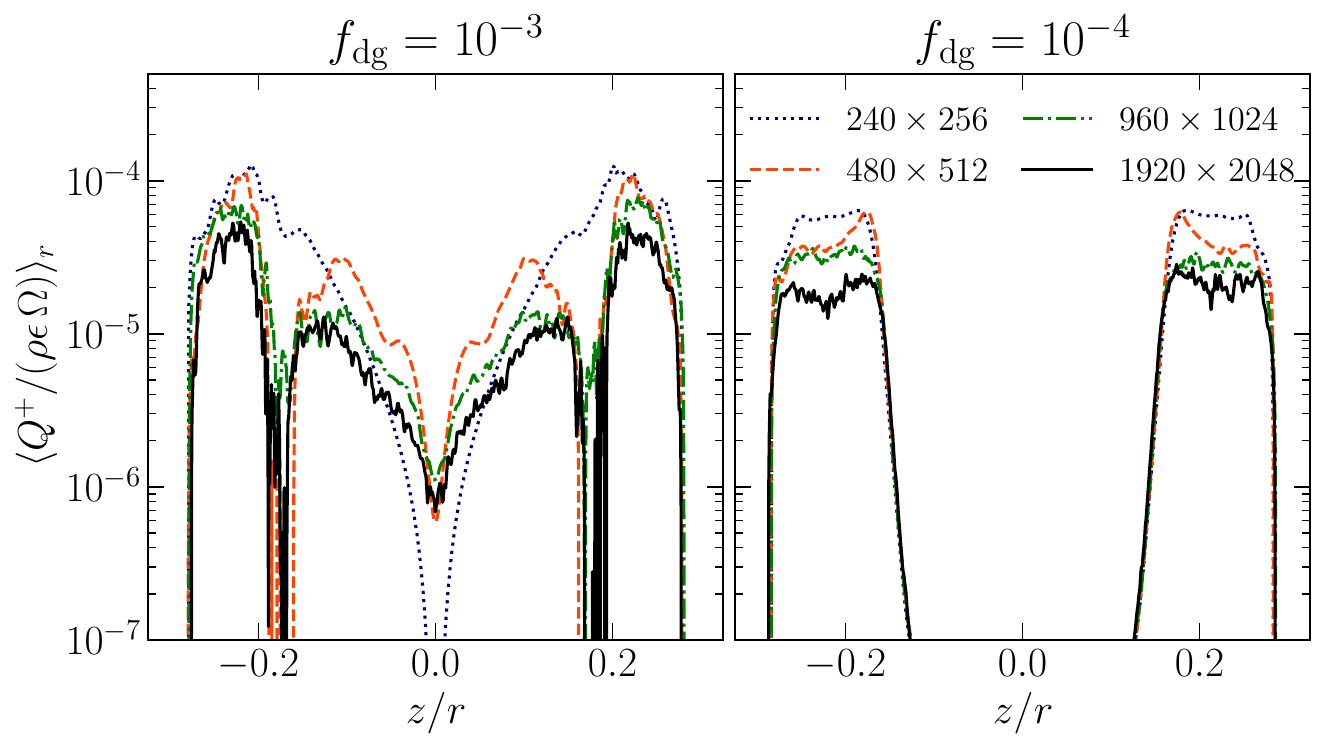}
\caption{Radial average of the heating rate produced by the VSI in units of $\Omega$ for all employed resolutions and $f_\mathrm{dg}$ values.}
\label{fig:Qplus}
\end{figure}

For $f_\mathrm{dg}=10^{-4}$, temperature perturbations are confined to a region near the irradiation surface, where both positive and negative temperature perturbations are produced by the vertical transport of entropy across the temperature transition (Fig. \ref{fig:hydrostaticTshear}). Unlike in the previous case, the average heating produced by the VSI only becomes evident when radially averaging the values of $\Delta T/T_0$, as shown in Fig. \ref{fig:dT_T0_ravgs_res}. In that case, the VSI produces above the irradiation surface a maximum temperature increase about $10$ times smaller than the maximum increase for $f_\mathrm{dg}=10^{-3}$ in the middle layer.
We also see a slight decrease in the midplane temperature due to the disk's reconfiguration over time that is not compensated by VSI heating, which however is reduced with increasing resolution.
The overall smaller average $\Delta T/T_0$ in this case results both from the weaker VSI activity in the entire domain (e.g., Figs. \ref{fig:vrms_1Davg} and \ref{fig:Sphi_prs_1Davgs_res}) and the lower optical depth in the region where heating takes place (Figs. \ref{fig:hydrostaticTshear} and \ref{fig:Qplus}). Conversely, the relatively larger heating of the middle layer for $f_\mathrm{dg}=10^{-3}$ results from the fact that the disk is both VSI-active in that region and vertically optically thick.

By Reynolds-averaging the HD equations, the average local heating power density can be computed as
\begin{equation}\label{Eq:Qplus}
\begin{split}
    Q^+ &= -S_{\phi R} R \partial_R \Omega
           -S_{\phi z} R \partial_z \Omega\\
        &= \nu_R \rho\, (R \partial_R \Omega)^2
        + \nu_z \rho\, (R \partial_z \Omega)^2\,,
\end{split}
\end{equation}
where $\nu_R$ and $\nu_z$ are the effective eddy viscosities computed in the previous section. We have verified that the term proportional to $\nu_z$ in this expression can be safely neglected, as it can be shown taking $\Omega\approx \Omega_K$ that the ratio between the $z$- and $R$-contributions to $Q^+$ is approximately $\frac{1}{36}\frac{\alpha_z}{\alpha_R}\left(\frac{z}{R}\right)^2$, which is $\ll 1$ despite $\alpha_z\gg \alpha_R$. Resulting $r$-averages of the heating rate $Q^+/(\rho\epsilon)$ in units of $\Omega$ are shown in Fig. \ref{fig:Qplus}. For $f_\mathrm{dg}=10^{-3}$, this quantity grows with height and is minimum at the midplane, where the disk's baroclinicity is zero and consequently the VSI driving is minimum (Fig. \ref{fig:Sphi_prs_1Davgs_res}). This explains the observed temperature variation, which is locally minimum at the midplane, reaches a maximum a few scale heights above the midplane, and is finally overtaken by the balance of irradiation and cooling at the optically thin surface layers. For $f_\mathrm{dg}=10^{-4}$, the heating rate is only nonzero at the VSI-active surface layers, which is not enough to heat the midplane.

 Due to the vertical dependence of $\alpha_R$, our obtained heating distributions resemble those in the simulations of MRI-heated disks by \citep{Flock2013}, which are also locally minimum at the midplane and peak a few scale heights above it.
 This observation contradicts the usual assumption resulting from $\alpha$-viscosity models that turbulent heating is maximum at the midplane, which is commonly used to compute temperature distributions via $1$D radiative transfer \citep[e.g.,][]{Nakamoto1994,Pfeil2019}. The vertical temperature distributions resulting from this method peak at the midplane \citep[e.g.,][]{Mori2019}, as opposed to our obtained distributions (Fig. \ref{fig:dT_T0_ravgs_res}). 

\section{Stability analysis}\label{S:StabilityAnalysis}



\begin{figure}[t!]
\centering
\includegraphics[width=\linewidth]{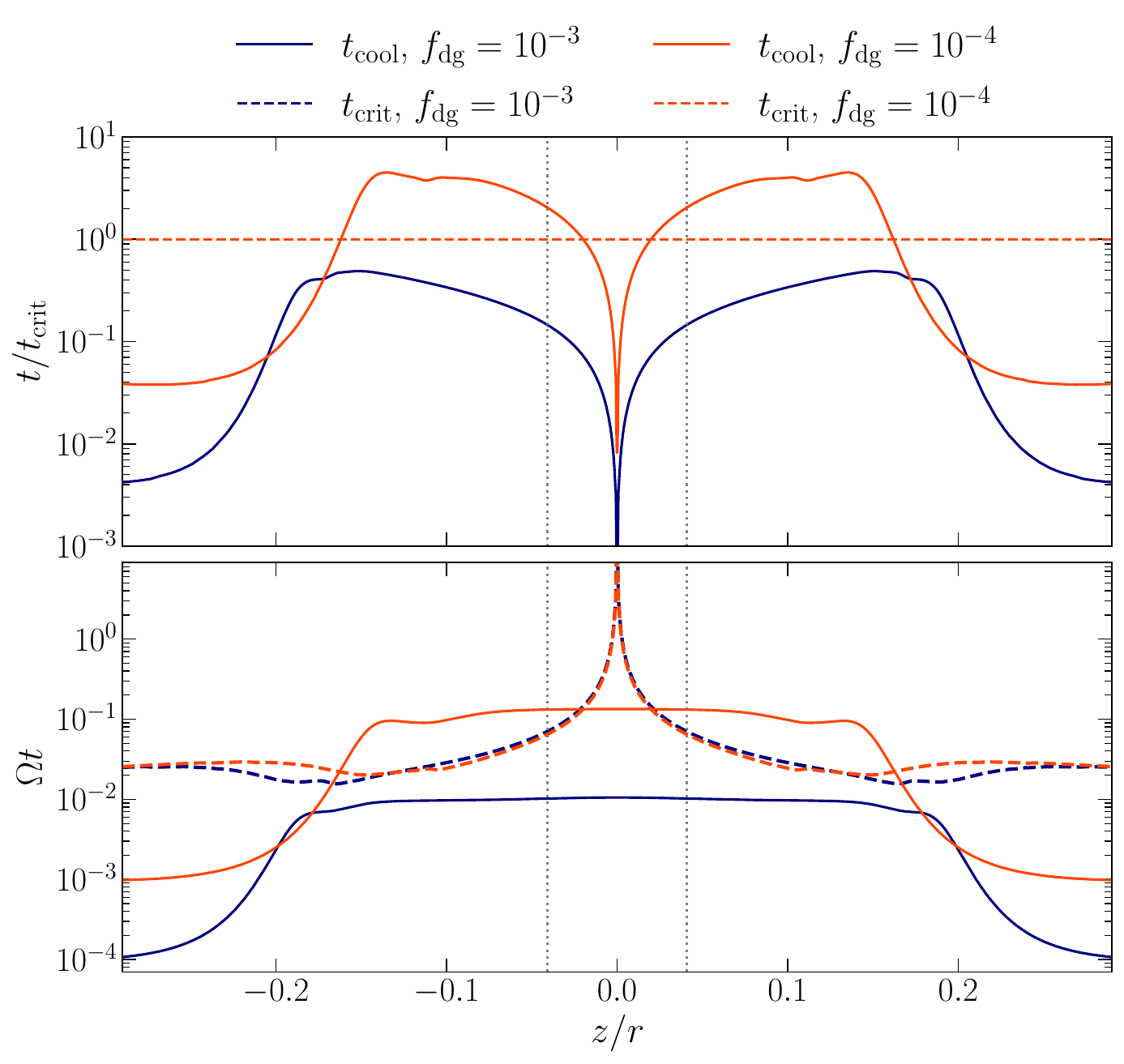}
\caption{Radiative cooling times normalized by the critical cooling time $t_\mathrm{crit}$ (top) and the orbital frequency (bottom) at $r=5.5$ au for both $f_\mathrm{dg}$ values.
Also shown are the corresponding values for $t=t_\mathrm{crit}$ and the locations $z=\pm \Gamma H/2$ (gray dotted lines) at which the local $t_\mathrm{crit}$ coincides with its value in the global stability criterion (Equation \eqref{Eq:tcrit_glob}).}
\label{fig:tcool_tcrit}
\end{figure}

In this section we analyze the applicability of local and global stability criteria in our disk models. Via a heuristic analysis, \cite{Urpin2003} proposed a local Richardson-like criterion in which the VSI is subject to the instability condition
\begin{equation}\label{Eq:tcrit_loc}
t_\mathrm{rel}\lesssim\frac{|\partial_z v_\phi|}{N_z^2}\,,
\end{equation}
where $t_\mathrm{rel}$ is the local thermal relaxation timescale, while $N_z$ is the vertical Brunt-V\"ais\"al\"a frequency (see Fig. \ref{fig:barocl_terms_T}).
A linear analysis in \cite{Klahr2023a} for arbitrary cooling times similar to that in \cite{Goldreich1967} shows that the linear growth rate decreases as $\sim 1/t_\mathrm{rel}$ for $t_\mathrm{rel}<t_\mathrm{crit}$, where the critical cooling time $t_\mathrm{crit}$ is defined as the right-hand side of Equation \eqref{Eq:tcrit_loc}.
On the other hand, the vertically global instability criterion by \cite{LinYoudin2015} reads
\begin{equation}\label{Eq:tcrit_glob}
  t_\mathrm{rel}<\frac{|q|}{\Gamma-1}
  \frac{H}{R}\,
  \Omega_K^{-1}\,,
\end{equation}
where $H$ and $\Gamma$ are respectively the gas pressure scale height and heat capacity ratio. This expression, which corresponds to Equation \eqref{Eq:tcrit_loc} evaluated at a height $z=\Gamma H/2$, is derived in \cite{LinYoudin2015} under the assumptions that (I) the disk is vertically isothermal with a temperature distribution depending on the cylindrical radius as $T\propto R^q$, and (II) that $\beta=t_\mathrm{rel}/\Omega_K$ is vertically uniform, where $\Omega_K$ is the Keplerian frequency. 
Most stability studies found in the literature \citep[e.g.,][]{Flock2017RadHydro,Dullemond2022} rely on this criterion, some of them relying on its local application to produce stability maps \citep[e.g.,][]{Malygin2017,Pfeil2019}. However, the cited validity conditions cease to be satisfied above the irradiation surface, where both the temperature and the relaxation timescale abruptly vary (Figs. \ref{fig:hydrostaticTshear} and \ref{fig:tcool_tcrit}) and the vertical shear does no longer solely depend on the radial temperature gradient (Section \ref{SS:VertShear}). We therefore analyze the disk stability in terms of the local critical time $t_\mathrm{crit}$ defined as the right-hand side of Equation \eqref{Eq:tcrit_loc} and later discuss in Section \ref{SS:DiscStability} the local applicability of Equation \eqref{Eq:tcrit_glob}.
 
 In the simulations presented in this work, the thermal relaxation time $t_\mathrm{rel}$ is entirely determined by radiative transfer, and equals\footnote{In reality $t_\mathrm{rel}=\Gamma t_\mathrm{cool}$ if one includes the effect of PdV work in the relaxation process \citep[see, e.g.,][]{Goldreich1967,Lesur2022PPVII,Klahr2023a}, but this widely used approximation is accurate enough for the purpose of a comparison with $t_\mathrm{crit}$, especially considering that the suppression of the VSI for $t_\mathrm{rel}>t_\mathrm{crit}$ as a function of $t_\mathrm{rel}$ is not abrupt \citep{LinYoudin2015,Manger2021II}.} the local radiative cooling time, $t_\mathrm{cool}$. We estimate this timescale in Appendix \ref{A:CoolingTimeScale} via a local lineal analysis assuming small temperature perturbations with a given wavelength $\lambda$ on a uniform background, obtaining
\begin{equation}\label{Eq:tcool_ms}
    t_\mathrm{cool} = t_\mathrm{cool,thin}+
    t_\mathrm{cool,thick}
\end{equation}
with
\begin{equation}\label{Eq:tcool_thin_thick}
\begin{split}
    t_\mathrm{cool,thin} &= 
    \frac{\lambda_P \zeta}{ c} \\
    t_\mathrm{cool,thick} &= 
    \frac{3\zeta}{c\, \lambda_R k^2}\\
    \zeta &= \frac{\rho \epsilon}{4\, a_R T^4+ b_P(a_R T^4-E_r)}
    \,,
\end{split}
\end{equation}
 where $|k|=2\pi/\lambda$, $\lambda_P=(\rho_d\kappa^d_P)^{-1}$ and $\lambda_R=(\rho_d\kappa^d_R)^{-1}$ are respectively the Planck- and Rosseland-averaged mean free paths, and $b_P=\frac{d \log \kappa^d_P}{d \log T}$. This expression evidences a dependence of the cooling mechanism on $\lambda$: temperature perturbations occurring over much smaller lengthscales than $\lambda_R$ and $\lambda_P$ ($\lambda_P \lambda_R k^2 \gg 1$) get damped at the scale-independent rate $t_\mathrm{cool,thin}^{-1}$ corresponding to thermal emission in an optically thin medium. On the other hand, optically thick perturbations ($\lambda_P \lambda_R k^2 \ll 1$) cool down through radiative diffusion at a rate $t_\mathrm{cool,thick}^{-1}$, which increases proportionally to $k^2$. The adiabatic limit $t_\mathrm{cool}\rightarrow\infty$ is thus recovered for $\lambda_R k\rightarrow 0$. For constant opacities, Equation \eqref{Eq:tcool_ms} tends to the corresponding expression in \cite{LinYoudin2015}. In Appendix \ref{A:CoolingTimeScale} it is shown that this expression is exact for any treatment of gray radiative transfer as long as a local first-order perturbation analysis is applicable, but different cooling time prescriptions used in the literature \citep[e.g.,][]{Stoll2016,Malygin2017,Dullemond2022} only differ slightly and should yield equivalent results. In the same appendix, we show that the resulting cooling times are not affected by the RSLA for large enough $\hat{c}$, in particular for our employed value of $\hat{c}=10^{-4}$.
 
 To estimate the characteristic wavelength of the VSI flows, we computed the distance between consecutive sign changes of $v_\theta$ at fixed $\theta$ values and extrapolate the resulting $k$ distributions to the entire domain. This computation is detailed in Appendix \ref{A:WavelengthEstimation}, where it is shown that $k H=\mathcal{O}(10)$ and $\lambda_P \lambda_R k^2\gg 1$ everywhere, i.e., induced temperature perturbations of the same wavelength as the velocity flows are optically thin. This is true at all times, and since the optically thin cooling timescale is wavelength-independent, the estimated $t_\mathrm{cool}$ values are unchanged through time. This validates the employment of $\beta$-cooling prescriptions in even higher-resolution axisymmetric simulations or 3D simulations using the same disk models, with the difference that such an approach cannot reproduce the (rather negligible) disk's thermal evolution (Section \ref{S:ThermalEvolution}). Further differences can be expected if large-scale optically thick structures are eventually formed (e.g., vortices in 3D).
 
 Resulting slices of $t_\mathrm{cool}/t_\mathrm{crit}$ and $t_\mathrm{cool}\Omega$ at $r=5.5$ au are shown in Fig. \ref{fig:tcool_tcrit} for each $f_\mathrm{dg}$. In all cases, the cooling times are almost constant below the irradiation surface, which would validate in our model the application of a vertically global criterion with uniform cooling time (Equation \eqref{Eq:tcrit_glob}) in that region. Above that surface, $t_\mathrm{cool}$ decreases up to $2$ orders of magnitude despite the surface layers having much smaller optical depths. The reason for this is that the ratio $\rho \epsilon/a_R T^4$, which stems from the balance between the gas internal energy and the dust emissivity (Appendix \ref{A:CoolingTimeScale}), decreases much faster with height than the increase in $\lambda_P$ (Equation \eqref{Eq:tcool_thin_thick}). For $f_\mathrm{dg}=10^{-3}$, the resulting $t_\mathrm{cool}$ is everywhere below $t_\mathrm{crit}$, which is consistent with the obtained instability in the whole domain.
 
 For $f_\mathrm{dg}=10^{-4}$, the local instability criterion is met at the surface layers, which explains our obtained localized VSI activity in those regions. On the other hand, the increase of $t_\mathrm{cool}$ below the irradiation surface forms two layers above the midplane where $t_\mathrm{cool}>t_\mathrm{crit}$, 
 which explains our obtained VSI suppression in the middle layer. A similar phenomenon is seen in the $\beta$-cooling HD simulations by \cite{Fukuhara2023}, who obtained that a stable midplane layer is formed in cases where a local application of Equation \eqref{Eq:tcrit_glob} predicts stability up to heights larger than $\sim H$. In our case, Equation \ref{Eq:tcrit_loc} predicts stability up to $\sim 3 H$, and thus our observed VSI suppression is consistent with that work.
 
 Close to the midplane, the instability condition is again met. The reason for this is that both the vertical shear and the squared vertical buoyancy frequency vanish with height, the first as $R\partial_z \Omega \propto z$ and the second as $N_z^2 \propto z^2$, and so the local critical cooling time close to the midplane diverges as $1/|z|$. 
 According to our local analysis, we should then expect a localized increased VSI growth in the unstable region close to the midplane. However, in Fig. \ref{fig:vel_2048} we instead see slow ($v_z \ll c_s$) vertical modes with similar amplitude in the midplane region and in the inactive zones above and below. 
 One possible explanation for this is the fact that the local linear VSI growth rate equals the local vertical shear \citep[e.g.,][]{UrpinBrandenburg1998,Klahr2023a}, which is very small close to the midplane. However, under this criterion, and provided the numerical dissipation in our code is low enough, we should still see some growth occurring after several hundreds of orbits, as it can be seen in the local growth rate distribution shown in Fig. \ref{fig:hydrostaticTshear}. Another explanation can be given if the maximum amplitude of the saturated VSI perturbations is limited by the local vertical shear. A hint in this direction is given by the vertically global model by \cite{LinYoudin2015}, which shows that the VSI growth rates are reduced if the maximum height of the domain, and therefore also the $z$-averaged vertical shear, are decreased. This hypothesis is also supported by the approximate relation $\langle \alpha_r\rangle_\rho \propto q^2$ in the global vertically isothermal model by \cite{Manger2021II}, which shows that the radial Reynolds stress in the saturated stage increases for larger average vertical shear. Lastly, resolution might also play a role in limiting VSI growth close to the midplane. The linearly unstable directions, strictly confined between the constant-$R$ and constant-$j_z$ surfaces \citep[e.g.,][]{Knobloch1982,Klahr2023b}, are for small $|z|$ only slightly slanted with respect to the vertical direction, and thus large resolutions might be needed to reproduce linear growth in such directions. This can be investigated in high-resolution box simulations of the midplane region.

 However useful, this picture has the limitation that VSI modes are nonlocal. Thus, another possibility is that the amplitude of the flows at the middle layer does not only depend on the in situ driving of the VSI but also on the excitation of inertial modes in unstable layers
 extending to otherwise inactive regions, as it likely occurs in the T states in \cite{Fukuhara2023}. This, together with the fact that VSI modes can in general occupy regions of largely different cooling times, makes it difficult to apply local stability criteria in more complex $t_\mathrm{cool}$ distributions to predict the saturated behavior of the VSI, for instance if the unstable region in this example is enlarged. Still, our analysis shows that evaluating which specific regions satisfy Equation \ref{Eq:tcrit_loc} gives a good insight into their expected stability \citep[see also][]{Pfeil2021}.

\section{Discussion}\label{S:Discussion}

\subsection{Angular momentum distribution}\label{SS:DiscAngMom}

The formation of bands of approximately uniform $j_z$ is likely enhanced by the imposed axisymmetry of our models. In 3D simulations, we would expect constant-$j_z$ zones to be unstable to non-axisymmetric modes \citep[e.g.,][]{PapaloizouPringle1984}, in which case it remains to be seen if the VSI can still enforce the formation of reduced shear bands. Surprisingly, this is likely the case in a number of published works: in the 3D simulations by \cite{Richard2016}, the vertical component of the vorticity, $\omega_z$, proportional to $\partial_R j_z$ for axisymmetric flows, peaks in between VSI modes, indicating abrupt jumps in $j_z$. In between those jumps, $\omega_z/\Omega$ is reduced with respect to its Keplerian value of $1/2$, although this reduction is below the value $\Delta \omega_z/\Omega=-1/2$ corresponding to constant $j_z$. More evidence showing abrupt jumps in $j_z$ is given by the simulations by \cite{Manger2018}, in which the azimuthally averaged critical function $\mathcal{L}$, approximately proportional to $1/\partial_R j_z$, has peaks of similar width as the VSI modes. On top of this, the 3D Rad-HD simulations of irradiated disks by \cite{Flock2020} exhibit a formation of approximately uniform-$j_z$ bands which is most predominant in the maximum-shear upper layers, as shown in Appendix \ref{A:AddFigures} (Fig. \ref{fig:angmom3D}). It is then possible that non-axisymmetric modes, while preventing the formation of uniform-$j_z$ zones, may still permit the creation of regions with reduced $\nabla j_z$ (see Paper II), although further thorough investigation is required to test this hypothesis.

\subsection{Reynolds stress comparison}\label{SS:DiscReyn}

 Our obtained vertically averaged $\alpha_r$ values for the nominal dust-to-gas ratio $f_\mathrm{dg}=10^{-3}$, where the instability is not suppressed close to the midplane, are slightly lower than the values $\alpha_r\approx 10^{-4}$ reported in \cite{Flock2020} with $H/R=0.09-0.14$ and \cite{Manger2021II} and \cite{BarrazaAlfaro2021} with $H/R=0.1$. On the other hand, \cite{Pfeil2021} and \cite{Manger2020I} obtained $\alpha_r\sim 10^{-5}$ respectively in 2D and 3D simulations with $H/R\sim 0.05$. Considering the dependence of $\alpha_r$ with $H/R$ studied in \cite{Manger2020I,Manger2021II}, who propose $\alpha_r \propto q^2 (H/R)^{2.6}$, our $\alpha_r$ values obtained with $H/R\sim 0.05-0.06$ are consistent with those obtained in 3D models.  The consistency with vertically isothermal simulations is likely a result of the pressure weighting of $\alpha_r$, which highlights its value at the vertically isothermal middle layer. Additionally, the consistency with 3D models points to the fact that the saturated flows produced by the VSI maintain some degree of axisymmetry, as shown, for example, in the high-resolution simulations in \cite{Flock2020,BarrazaAlfaro2021} \citep[see, however,][]{Richard2016,Manger2018}, and thus it is likely that the mixing of angular momentum in 3D is predominantly axisymmetric. We defer such verification to follow-up works on this topic. Still, differences in the spatial distribution of averaged velocities and stresses in 2D and 3D simulations are seen in \cite{Stoll2014,Stoll2017,Pfeil2021}.

 \subsection{Meridional circulation and accretion}\label{SS:DiscMeridCirc}

  Using the obtained mean gas flows, we can compute the total mass flux delivered to the star by integrating $\langle \rho \mathbf{v}\rangle_t$ onto any surface crossing our domain vertically. When doing so integrating the $r$-flux along constant-$r$ surfaces, we obtain
 that the positive and negative contributions cancel each other out almost exactly. More precisely, the $\theta$-averaged radial mass flux takes small positive values between $2$ and $3$ orders of magnitude smaller than the maximum values shown in Fig. \ref{fig:mflux_1Davgs_res}.
 This occurs in contradiction with Equation \eqref{Eq:AccretionFlux}, which for our simulation parameters predicts that
 the positivity of $\langle\alpha_R\rangle_p$ and its approximate constancy (Fig. \ref{fig:alphaR_zavgs}) are enough to produce an average mass inflow.
 The reason for this is that our imposed boundary conditions do not allow for either outflow or inflow of mass, and thus the inward-directed gas flow is forced at the inner radial boundary to turn away from the midplane and merge with the outward-directed meridional flow, while the inverse process occurs at the outer radial boundary. This increases the outward flux, creating an equilibrium configuration in which outgoing and ingoing flows balance each other while no mass escapes the domain. The extent to which this affects the resulting gas structure in our simulations and all mentioned works is unknown. This drawback could be solved by including the inner rim in the domain and allowing gas to escape the radial boundary toward the star, ideally including magnetic effects in well-ionized regions \citep{Flock2017InnerRim}. 
 
The ideal MHD simulations of MRI turbulence by \cite{Fromang2011} and \cite{Flock2011} show similar downward-concave parabola patterns as in Figure \ref{fig:velr_1Davgs_res}, with the difference that, unlike both in \cite{Flock2011} and the present paper, in \cite{Fromang2011} the radial flux remains everywhere positive and there is no average inflow close to the midplane. Given the expected quenching of the MRI by nonideal effects in several disk regions \citep[e.g.,][]{Lesur2022PPVII}, global nonideal MHD simulations are required to better characterize the radial transport of solids in protoplanetary disks.

 \subsection{Turbulent heating}\label{SS:DiscThermalEvolution}

 In our nominal model, heating due to turbulent viscosity is only marginal in comparison with stellar heating, and the average temperature increase produced by the VSI is at most on the order of $0.1$ K. This is comparable to the prediction by \citep[][]{Pfeil2019}, obtained via a $1$D vertical diffusion model, for our obtained $\langle\alpha_R\rangle_p$ values of $\mathcal{O}(10^{-5})$. If we assume a constant $\alpha_R$ of that magnitude, we can expect from that work that the VSI can only produce a significant temperature increase well inside of $1$ au. However, is it unclear whether the VSI can still produce such $\alpha_R$ values in such dense regions, as the denser the disk is, the more the maximum VSI-unstable wavelengths predicted by linear theory get shifted to lower values, as determined by Equation \eqref{Eq:tcool_ms}. In our disk model, as detailed in Paper II, the VSI should be suppressed at the midplane for our maximum average wavenumbers ($kH\approx 70$, see Fig. \ref{fig:kH_r}) inside of $\sim 2$ au, where only smaller-wavelength modes can grow. Given our obtained trend of decreasing VSI strength with decreasing dominant wavelength (see Figs. \ref{fig:Sphi_prs_1Davgs_res} and \ref{fig:Qplus}), we can expect in that region a suppression of the VSI at the midplane due to the reduction of the maximum VSI-unstable wavelengths. It therefore appears unlikely that the VSI can in general lead to any significant heating, since that would simultaneously require (I) large enough vertical optical depths to retain the produced heat and (II) modes with small radial wavelengths producing larger levels of turbulence than predicted by simulations. This hypothesis can be further studied in high-resolution Rad-HD simulations of the inner disk inside of $\sim 1$ au, keeping in mind that close enough to the star the MRI is expected to become predominant and suppress the VSI \citep{Lesur2022PPVII}.

 The model implemented in \cite{Pfeil2019} also predicts that larger $\alpha_R$ values of $\sim 10^{-4}-10^{-3}$ can lead to an increase of up to a few tens of K outside of $1$ au. Such high $\alpha_R$ values can only be expected for steeper temperature profiles than in this work \citep[$T\propto R^{-1}$, ][]{Pfeil2021,Manger2021II} or for large aspect ratios at a few au of up to $0.1$ \citep{Manger2020I}, both of which are hard to justify in passive irradiated disks. In this context, it is also unclear whether steeper temperature gradients, predicted by models of viscous-dominated heating in optically thick accretion disks \citep[$q= -\frac{3}{4}$,][]{Pringle1981}, can be maintained solely by VSI-induced turbulence. Moreover, the cited $\alpha$-disk model in \cite{Pfeil2019} assumes maximum heating at the midplane, whereas our maximum heating regions occur in upper regions of lower optical depth, and thus the temperature increase in that work can be regarded as an upper limit in the case of VSI-induced heating. Further explorations are however required to test whether any heating can be produced by the VSI under a plausible choice of parameters.
 

The fact that the temperature distributions in our nominal disk are at all times minimum close to the $r$-boundaries (Fig. \ref{fig:dT_T0_z0_t}) evidences that the obtained temperature increase is not only limited by vertical diffusion but also by radiative cooling through the radial boundaries. This is a numerical effect produced by the employed Dirichlet boundary conditions, which leads to outflow of radiation as soon as the temperature inside of the domain exceeds that at the boundary. Larger domains could be used in the future to explore whether this effect limits the maximum temperature perturbations.

Determining the equilibrium temperatures produced by
the VSI is also made difficult by the lack of convergence of the temperature distributions with resolution due to the reduction of the VSI strength, as shown in Fig. \ref{fig:dT_T0_ravgs_res}. However, the average temperatures obtained with our two highest resolutions are rather similar, with differences of $\sim 10\%$. Whether such resolutions are enough to accurately predict the temperature increase produced by the VSI can be explored in the future via even higher-resolution simulations. However, given our obtained trend of decreasing heating with increasing resolution, this effect should not alter the conclusion that the heating produced by the VSI is most likely negligible.

\subsection{Local and vertically global stability criteria}\label{SS:DiscStability}

In Fig. \ref{fig:tcool_tcrit} it can be seen that the validity conditions of the global stability criterion by \cite{LinYoudin2015}, namely that the disk is vertically isothermal and that the cooling time is vertically uniform, are approximately met in the entire extent of the disk middle layer. Furthermore, the locations $z=\pm \Gamma H/2$ at which $t_\mathrm{crit}$ equals its value in the global stability criterion (Equation \eqref{Eq:tcrit_glob}) are well inside the middle layer (Fig. \ref{fig:tcool_tcrit}). This suggests that we can expect the global stability criterion to be applicable in the disk's middle layer, with a stability condition depending on the critical time $t^\mathrm{glob}_\mathrm{crit}=t_\mathrm{crit}(\Gamma H/2)$. This is in fact verified in our disk models, since $t_\mathrm{cool}$ is smaller and larger than $t^\mathrm{glob}_\mathrm{crit}$ for $f_\mathrm{dg}=10^{-3}$ and $10^{-4}$, respectively (see Fig. \ref{fig:tcool_tcrit}). Further evidence for this is given by the vertically isothermal $\beta$-cooling simulations by \cite{Fukuhara2023}, which only show a stable middle layer if the height $\Delta L_s/2$ of the stability region predicted by the local application of Equation \eqref{Eq:tcrit_glob} is at least $\sim H$. Given the functional form of $t_\mathrm{cool}(z)$ in that work, $\Delta L_s/2$ must be larger than $z=\Gamma H/2$ to verify $t_\mathrm{cool}\gg t_\mathrm{crit}$ at $z=\pm \Gamma H/2$. Thus, the midplane stability for $\Delta L_s/2\gtrsim H$ in that work is consistent with the vertically global criterion by \cite{LinYoudin2015}, even though, unlike these authors, \cite{Fukuhara2023} did not consider a vertically uniform cooling time. These results indicate that the cited vertically global criterion should suffice to predict the stability of regions below the irradiation surface. 

The vertical constancy of $t_\mathrm{cool}$ in the middle layer is not guaranteed if the radial optical depth of the VSI modes is large enough that the cooling timescale is determined by radiative diffusion (Equation \eqref{Eq:tcool_ms}), as shown in Fig. 2 of \cite{Flock2017RadHydro}. It remains to be explored if global stability criteria are still applicable to such cases.

Lastly, we have not considered in this work the effect on the thermal relaxation timescale of the dust-gas thermal equilibration time \citep[see, e.g.,][]{Barranco2018}. This becomes important at the disk upper layers, where the dust and gas densities are low enough that $t_\mathrm{rel}$ is increased due to their infrequent collisions, with a stabilizing effect in those regions \citep{Pfeil2021}. On the other hand, $t_\mathrm{rel}$ may be reduced by molecular emissivity depending on the chemical composition, favoring the instability. These effects can be studied in future simulations decoupling the gas and dust temperatures, as done in \cite{Muley2023}. The impact of these phenomena on the disk stability is explored in terms of both local and global criteria in Paper II.
\section{Conclusions}\label{S:Conclusions}

In this work we studied the linear and nonlinear evolution of the VSI in protoplanetary disks, focusing on the transport of angular momentum, the produced temperature perturbations, and the applicability of local stability conditions. We employed for this axisymmetric two-moment radiation-hydrodynamical simulations with realistic temperature stratifications and cooling times, both of which result from the balance of stellar irradiation and radiative cooling, which are treated in our code using realistic tabulated dust opacities. We assumed typical parameters for disks around T Tauri stars considering two different values of the dust-to-gas mass ratio of small grains, namely a nominal case with $f_\mathrm{dg}=10^{-3}$ and a dust-depleted case with $f_\mathrm{dg}=10^{-4}$, making our studied disk region between $4$ and $7$ au vertically optically thick and thin, respectively.
Our main findings can be summarized as follows:

\begin{enumerate}
\setlength\itemsep{1em}
    \item The temperature gradient in our disk models leads to a localized increase of the vertical shear in the temperature transition region between the midplane and the atmosphere. The radial and vertical components of the temperature gradient tend to increase and decrease the overall vertical shear, respectively. At its peak, the resulting vertical shear reaches up to $2$ times its value in equivalent vertically isothermal models, which is translated into similarly larger local VSI growth rates than predicted by linear theory for vertically isothermal disks. The stabilizing buoyancy frequency is also maximum in that region, while above it decreases below its vertically isothermal value.
    \item In our nominal case, we observe two distinct growth stages corresponding to the growth and saturation of finger and body modes occupying the entire domain. Instead, in the dust-depleted case, the VSI is suppressed at the disk middle layer, and body modes never grow. This behavior results in a single growth phase of VSI modes localized at the disk surface layers.
    \item In every case, the VSI reduces the vertical shear inside of bands of approximately uniform specific angular momentum. Conversely, shear is maximal at the interfaces between these bands. Even though this process is likely enhanced by the enforced axisymmetry, bands of reduced shear can also be seen in 3D simulations, suggesting that their formation mechanism may in some cases predominate over their destruction by non-axisymmetric modes.
    \item The time- and volume-averaged absolute value of the vertical shear increases with resolution as a result of the growth of small-scale velocity fluctuations, exceeding the mean shear value by almost one order of magnitude in our highest-resolution runs. This is not translated into an increase of the VSI strength, which is instead enforced by the disk’s baroclinicity and limited by the VSI saturation mechanism.
    \item The Reynolds stress is in every case maximum at the maximum shear regions above the irradiation surfaces and in between the approximately uniform-$j_z$ bands, with $z$ components between $1$ and $2$ orders of magnitude larger than the $R$ components. The pressure-weighted $\alpha_r$ values of $\sim 10^{-5}$ obtained in the nominal case coincide with reported values for $3$D vertically isothermal simulations. This suggests that, even in 3D, the radial mixing of angular momentum may be predominantly axisymmetric.
    \item The anisotropy of the Reynolds stress tensor causes differences between its radial components in cylindrical and spherical coordinates, with $\alpha_R<\alpha_r$ by factors of $\sim 4$ and $\sim 2$ in the nominal and dust-depleted cases, respectively. Determining whether this is a general trend for different disk parameters requires further testing. While $\alpha_R$ should be used to compute the vertically integrated mass flux, $\alpha_r$ is usually reported in the literature.
    \item As in previous works \cite[e.g.,][]{Stoll2016}, the anisotropic angular momentum transport produced by the VSI produces a meridional circulation with inflow at the midplane and outflow in upper layers. This pattern only occurs at the disk's surface layer in the dust-depleted case. This circulation does not lead to accretion in our simulations due to our imposed no-outflow and no-inflow boundary conditions, which likely increases the resulting outflow velocities with respect to their values if mass could escape the domain. 
    \item The turbulent heating produced by the VSI in our models causes negligible temperature increases of $0.1\%$ and $0.01\%$ in the nominal and dust-depleted models, respectively. In the nominal case, the energy dissipated by the VSI modes heats the disk middle layer with a heating rate distribution peaking a few ($\sim 3$) scale heights above the midplane, contrarily to the heating distributions in $\alpha$-viscosity models, which peak at the midplane. We conclude that the VSI is in general unlikely to produce any significant temperature increase, as 
    that would either require it to produce larger heating rates than expected in dense disk regions inside of $\sim 1$ au, where largely optically thick radial wavelengths cease to be VSI-unstable, or unrealistically large radial temperature gradients or aspect ratios leading to larger average $\alpha_R$ values ($10^{-4}-10^{-3}$) than can be obtained in models of passive disks. Further explorations of the parameter space via high-resolution Rad-HD simulations of inner disk regions should either deny or confirm this hypothesis.
    \item The Richardson-like local criterion by \cite{Urpin2003}  explains the instability regions observed in our simulations, showing that, despite VSI modes are vertically global, a local analysis can provide good insight into the stability of specific disk regions. On the other hand, the global criterion in \cite{LinYoudin2015} derived for vertically isothermal disks with uniform cooling time can still be applied to the vertically isothermal disk middle layer, which justifies its general application to assess the stability close to the midplane as long as VSI modes are radially optically thin.
    \item We verified via analytical computations and numerical testing that the RSLA is safely applicable to this problem, since for high enough $\hat{c}<c$ the cooling times are not affected by the artificial reduction of the speed of light. This is important due to the fact that the VSI is sensitive in both its linear and nonlinear regimes to the precise value of $t_\mathrm{cool}$ if this value is similar or larger than $t_\mathrm{crit}$. This also shows that IMEX schemes for Rad-HD, with generally good scalability properties in comparison with globally implicit methods, can be applied to this problem without introducing unphysical phenomena, which is rather useful given the high resolutions required to resolve the VSI modes. 
\end{enumerate}

Our stability analysis does not contemplate effects due to nonzero thermal coupling timescales between dust grains and gas particles and gas thermal emissivity. These can affect the stability of the low-density disk surface layers, in particular in outer disk regions currently observable with ALMA \citep[e.g.,][]{Andrews2018}. These effects are included in Paper II to study the stability of such regions.




\begin{acknowledgements}
The research of J.D.M.F. and H.K. is supported by the German Science Foundation (DFG) under the priority program SPP 1992: "Exoplanet Diversity" under contract KL 1469/16-1/2. We thank our collaboration partners on this project in Kiel, Sebastian Wolf and Anton Krieger, under contract WO 857/17-1/2, for providing us with the employed tabulated opacity coefficients. M.F. acknowledges funding from the European Research Council (ERC) under the European Union’s Horizon 2020 research and innovation program (grant agreement No. 757957). All numerical simulations were run on the ISAAC and VERA clusters of the MPIA and the COBRA cluster of the Max Planck Society, all of these hosted at the Max-Planck Computing and Data Facility in Garching (Germany). We thank the anonymous referee, whose comments helped us improve the quality of this work.
\end{acknowledgements}

\bibliographystyle{aa}
\bibliography{refs}

\begin{thebibliography}{95}
\expandafter\ifx\csname natexlab\endcsname\relax\def\natexlab#1{#1}\fi

\bibitem[{{Al-Mohy} \& {Higham}(2010)}]{matexp}
{Al-Mohy}, A.~H. \& {Higham}, N.~J. 2010, SIAM Journal on Matrix Analysis and
  Applications, 31, 970

\bibitem[{{Andrews} {et~al.}(2018){Andrews}, {Huang}, {P{\'e}rez}, {Isella},
  {Dullemond}, {Kurtovic}, {Guzm{\'a}n}, {Carpenter}, {Wilner}, {Zhang}, {Zhu},
  {Birnstiel}, {Bai}, {Benisty}, {Hughes}, {{\"O}berg}, \&
  {Ricci}}]{Andrews2018}
{Andrews}, S.~M., {Huang}, J., {P{\'e}rez}, L.~M., {et~al.} 2018, \apjl, 869,
  L41

\bibitem[{{Armitage}(2022)}]{ArmitageLectureNotes2022}
{Armitage}, P.~J. 2022, arXiv e-prints, arXiv:2201.07262

\bibitem[{{Bai}(2015)}]{Bai2015}
{Bai}, X.-N. 2015, \apj, 798, 84

\bibitem[{{Bai} \& {Stone}(2013)}]{BaiStone2013}
{Bai}, X.-N. \& {Stone}, J.~M. 2013, \apj, 769, 76

\bibitem[{{Balbus} \& {Hawley}(1991)}]{BalbusHawley1991}
{Balbus}, S.~A. \& {Hawley}, J.~F. 1991, \apj, 376, 214

\bibitem[{{Balbus} \& {Papaloizou}(1999)}]{BalbusPapaloizou1999}
{Balbus}, S.~A. \& {Papaloizou}, J. C.~B. 1999, \apj, 521, 650

\bibitem[{{Barker} \& {Latter}(2015)}]{BarkerLatter2015}
{Barker}, A.~J. \& {Latter}, H.~N. 2015, \mnras, 450, 21

\bibitem[{{Barranco} {et~al.}(2018){Barranco}, {Pei}, \&
  {Marcus}}]{Barranco2018}
{Barranco}, J.~A., {Pei}, S., \& {Marcus}, P.~S. 2018, \apj, 869, 127

\bibitem[{{Barraza-Alfaro} {et~al.}(2021){Barraza-Alfaro}, {Flock}, {Marino},
  \& {P{\'e}rez}}]{BarrazaAlfaro2021}
{Barraza-Alfaro}, M., {Flock}, M., {Marino}, S., \& {P{\'e}rez}, S. 2021, \aap,
  653, A113

\bibitem[{{B{\'e}thune} {et~al.}(2017){B{\'e}thune}, {Lesur}, \&
  {Ferreira}}]{Bethune2017}
{B{\'e}thune}, W., {Lesur}, G., \& {Ferreira}, J. 2017, \aap, 600, A75

\bibitem[{{Birnstiel} {et~al.}(2012){Birnstiel}, {Klahr}, \&
  {Ercolano}}]{Birnstiel2012}
{Birnstiel}, T., {Klahr}, H., \& {Ercolano}, B. 2012, \aap, 539, A148

\bibitem[{{Brownlee} {et~al.}(2006){Brownlee}, {Tsou}, {Al{\'e}on},
  {Alexander}, {Araki}, {Bajt}, {Baratta}, {Bastien}, {Bland}, {Bleuet},
  {Borg}, {Bradley}, {Brearley}, {Brenker}, {Brennan}, {Bridges}, {Browning},
  {Brucato}, {Bullock}, {Burchell}, {Busemann}, {Butterworth}, {Chaussidon},
  {Cheuvront}, {Chi}, {Cintala}, {Clark}, {Clemett}, {Cody}, {Colangeli},
  {Cooper}, {Cordier}, {Daghlian}, {Dai}, {D'Hendecourt}, {Djouadi},
  {Dominguez}, {Duxbury}, {Dworkin}, {Ebel}, {Economou}, {Fakra}, {Fairey},
  {Fallon}, {Ferrini}, {Ferroir}, {Fleckenstein}, {Floss}, {Flynn}, {Franchi},
  {Fries}, {Gainsforth}, {Gallien}, {Genge}, {Gilles}, {Gillet}, {Gilmour},
  {Glavin}, {Gounelle}, {Grady}, {Graham}, {Grant}, {Green}, {Grossemy},
  {Grossman}, {Grossman}, {Guan}, {Hagiya}, {Harvey}, {Heck}, {Herzog},
  {Hoppe}, {H{\"o}rz}, {Huth}, {Hutcheon}, {Ignatyev}, {Ishii}, {Ito}, {Jacob},
  {Jacobsen}, {Jacobsen}, {Jones}, {Joswiak}, {Jurewicz}, {Kearsley}, {Keller},
  {Khodja}, {Kilcoyne}, {Kissel}, {Krot}, {Langenhorst}, {Lanzirotti}, {Le},
  {Leshin}, {Leitner}, {Lemelle}, {Leroux}, {Liu}, {Luening}, {Lyon},
  {MacPherson}, {Marcus}, {Marhas}, {Marty}, {Matrajt}, {McKeegan}, {Meibom},
  {Mennella}, {Messenger}, {Messenger}, {Mikouchi}, {Mostefaoui}, {Nakamura},
  {Nakano}, {Newville}, {Nittler}, {Ohnishi}, {Ohsumi}, {Okudaira},
  {Papanastassiou}, {Palma}, {Palumbo}, {Pepin}, {Perkins}, {Perronnet},
  {Pianetta}, {Rao}, {Rietmeijer}, {Robert}, {Rost}, {Rotundi}, {Ryan},
  {Sandford}, {Schwandt}, {See}, {Schlutter}, {Sheffield-Parker},
  {Simionovici}, {Simon}, {Sitnitsky}, {Snead}, {Spencer}, {Stadermann},
  {Steele}, {Stephan}, {Stroud}, {Susini}, {Sutton}, {Suzuki}, {Taheri},
  {Taylor}, {Teslich}, {Tomeoka}, {Tomioka}, {Toppani}, {Trigo-Rodr{\'\i}guez},
  {Troadec}, {Tsuchiyama}, {Tuzzolino}, {Tyliszczak}, {Uesugi}, {Velbel},
  {Vellenga}, {Vicenzi}, {Vincze}, {Warren}, {Weber}, {Weisberg}, {Westphal},
  {Wirick}, {Wooden}, {Wopenka}, {Wozniakiewicz}, {Wright}, {Yabuta}, {Yano},
  {Young}, {Zare}, {Zega}, {Ziegler}, {Zimmerman}, {Zinner}, \&
  {Zolensky}}]{Brownlee2006}
{Brownlee}, D., {Tsou}, P., {Al{\'e}on}, J., {et~al.} 2006, Science, 314, 1711

\bibitem[{{Bryson} \& {Brennecka}(2021)}]{Bryson2021}
{Bryson}, J. F.~J. \& {Brennecka}, G.~A. 2021, \apj, 912, 163

\bibitem[{{Burn} {et~al.}(2022){Burn}, {Emsenhuber}, {Weder}, {V{\"o}lkel},
  {Klahr}, {Birnstiel}, {Ercolano}, \& {Mordasini}}]{Burn2022}
{Burn}, R., {Emsenhuber}, A., {Weder}, J., {et~al.} 2022, \aap, 666, A73

\bibitem[{{Chandrasekhar}(1961)}]{Chandrasekhar1961}
{Chandrasekhar}, S. 1961, {Hydrodynamic and hydromagnetic stability} (Clarendon
  Press, Oxford)

\bibitem[{{Ciesla}(2007)}]{Ciesla2007}
{Ciesla}, F.~J. 2007, Science, 318, 613

\bibitem[{{Cui} \& {Bai}(2022)}]{CuiBai2022}
{Cui}, C. \& {Bai}, X.-N. 2022, \mnras, 516, 4660

\bibitem[{{Decampli} {et~al.}(1978){Decampli}, {Cameron}, {Bodenheimer}, \&
  {Black}}]{DeCampli1978}
{Decampli}, W.~M., {Cameron}, A.~G.~W., {Bodenheimer}, P., \& {Black}, D.~C.
  1978, \apj, 223, 854

\bibitem[{{Dullemond} {et~al.}(2022){Dullemond}, {Ziampras}, {Ostertag}, \&
  {Dominik}}]{Dullemond2022}
{Dullemond}, C.~P., {Ziampras}, A., {Ostertag}, D., \& {Dominik}, C. 2022,
  \aap, 668, A105

\bibitem[{{Flock} {et~al.}(2011){Flock}, {Dzyurkevich}, {Klahr}, {Turner}, \&
  {Henning}}]{Flock2011}
{Flock}, M., {Dzyurkevich}, N., {Klahr}, H., {Turner}, N.~J., \& {Henning}, T.
  2011, \apj, 735, 122

\bibitem[{{Flock} {et~al.}(2013){Flock}, {Fromang}, {Gonz{\'a}lez}, \&
  {Commer{\c{c}}on}}]{Flock2013}
{Flock}, M., {Fromang}, S., {Gonz{\'a}lez}, M., \& {Commer{\c{c}}on}, B. 2013,
  \aap, 560, A43

\bibitem[{{Flock} {et~al.}(2017{\natexlab{a}}){Flock}, {Fromang}, {Turner}, \&
  {Benisty}}]{Flock2017InnerRim}
{Flock}, M., {Fromang}, S., {Turner}, N.~J., \& {Benisty}, M.
  2017{\natexlab{a}}, \apj, 835, 230

\bibitem[{{Flock} {et~al.}(2017{\natexlab{b}}){Flock}, {Nelson}, {Turner},
  {Bertrang}, {Carrasco-Gonz{\'a}lez}, {Henning}, {Lyra}, \&
  {Teague}}]{Flock2017RadHydro}
{Flock}, M., {Nelson}, R.~P., {Turner}, N.~J., {et~al.} 2017{\natexlab{b}},
  \apj, 850, 131

\bibitem[{{Flock} {et~al.}(2020){Flock}, {Turner}, {Nelson}, {Lyra}, {Manger},
  \& {Klahr}}]{Flock2020}
{Flock}, M., {Turner}, N.~J., {Nelson}, R.~P., {et~al.} 2020, \apj, 897, 155

\bibitem[{{Flores-Rivera} {et~al.}(2020){Flores-Rivera}, {Flock}, \&
  {Nakatani}}]{FloresRivera2020}
{Flores-Rivera}, L., {Flock}, M., \& {Nakatani}, R. 2020, \aap, 644, A50

\bibitem[{{Fromang} {et~al.}(2011){Fromang}, {Lyra}, \& {Masset}}]{Fromang2011}
{Fromang}, S., {Lyra}, W., \& {Masset}, F. 2011, \aap, 534, A107

\bibitem[{{Fukuhara} {et~al.}(2023){Fukuhara}, {Okuzumi}, \&
  {Ono}}]{Fukuhara2023}
{Fukuhara}, Y., {Okuzumi}, S., \& {Ono}, T. 2023, \pasj, 75, 233

\bibitem[{{Gerbig} {et~al.}(2020){Gerbig}, {Murray-Clay}, {Klahr}, \&
  {Baehr}}]{Gerbig2020}
{Gerbig}, K., {Murray-Clay}, R.~A., {Klahr}, H., \& {Baehr}, H. 2020, \apj,
  895, 91

\bibitem[{{Goldreich} \& {Schubert}(1967)}]{Goldreich1967}
{Goldreich}, P. \& {Schubert}, G. 1967, \apj, 150, 571

\bibitem[{{Gressel} {et~al.}(2020){Gressel}, {Ramsey}, {Brinch}, {Nelson},
  {Turner}, \& {Bruderer}}]{Gressel2020}
{Gressel}, O., {Ramsey}, J.~P., {Brinch}, C., {et~al.} 2020, \apj, 896, 126

\bibitem[{{Hartmann} {et~al.}(1998){Hartmann}, {Calvet}, {Gullbring}, \&
  {D'Alessio}}]{Hartmann1998}
{Hartmann}, L., {Calvet}, N., {Gullbring}, E., \& {D'Alessio}, P. 1998, \apj,
  495, 385

\bibitem[{{Jacquet}(2013)}]{Jacquet2013}
{Jacquet}, E. 2013, \aap, 551, A75

\bibitem[{{James} \& {Kahn}(1970)}]{JamesKahn1970}
{James}, H.~A. \& {Kahn}, F.~D. 1970, \aap, 5, 232

\bibitem[{{Johansen} {et~al.}(2014){Johansen}, {Blum}, {Tanaka}, {Ormel},
  {Bizzarro}, \& {Rickman}}]{Johansen2014}
{Johansen}, A., {Blum}, J., {Tanaka}, H., {et~al.} 2014, in Protostars and
  Planets VI, ed. H.~{Beuther}, R.~S. {Klessen}, C.~P. {Dullemond}, \&
  T.~{Henning}, 547

\bibitem[{{Johansen} {et~al.}(2007){Johansen}, {Oishi}, {Mac Low}, {Klahr},
  {Henning}, \& {Youdin}}]{Johansen2007}
{Johansen}, A., {Oishi}, J.~S., {Mac Low}, M.-M., {et~al.} 2007, \nat, 448,
  1022

\bibitem[{{Klahr}(2023)}]{Klahr2023a}
{Klahr}, H. 2023, \aap, submitted

\bibitem[{{Klahr} {et~al.}(2023){Klahr}, {Baehr}, \& {Melon
  Fuksman}}]{Klahr2023b}
{Klahr}, H., {Baehr}, H., \& {Melon Fuksman}, J.~D. 2023, arXiv e-prints,
  arXiv:2305.08165

\bibitem[{{Klahr} \& {Schreiber}(2020)}]{Klahr2020}
{Klahr}, H. \& {Schreiber}, A. 2020, \apj, 901, 54

\bibitem[{{Klahr} \& {Bodenheimer}(2003)}]{KlahrBodenheimer2003}
{Klahr}, H.~H. \& {Bodenheimer}, P. 2003, \apj, 582, 869

\bibitem[{{Kley} \& {Lin}(1992)}]{Kley1992}
{Kley}, W. \& {Lin}, D.~N.~C. 1992, \apj, 397, 600

\bibitem[{{Knobloch} \& {Spruit}(1982)}]{Knobloch1982}
{Knobloch}, E. \& {Spruit}, H.~C. 1982, \aap, 113, 261

\bibitem[{{Krieger} \& {Wolf}(2020)}]{Krieger2020}
{Krieger}, A. \& {Wolf}, S. 2020, \aap, 635, A148

\bibitem[{{Krieger} \& {Wolf}(2022)}]{Krieger2022}
{Krieger}, A. \& {Wolf}, S. 2022, \aap, 662, A99

\bibitem[{{Latter} \& {Papaloizou}(2018)}]{LatterPapaloizou2018}
{Latter}, H.~N. \& {Papaloizou}, J. 2018, \mnras, 474, 3110

\bibitem[{{Lesur} {et~al.}(2023){Lesur}, {Flock}, {Ercolano}, {Lin}, {Yang},
  {Barranco}, {Benitez-Llambay}, {Goodman}, {Johansen}, {Klahr}, {Laibe},
  {Lyra}, {Marcus}, {Nelson}, {Squire}, {Simon}, {Turner}, {Umurhan}, \&
  {Youdin}}]{Lesur2022PPVII}
{Lesur}, G., {Flock}, M., {Ercolano}, B., {et~al.} 2023, ASP Conference Series,
  534, 465

\bibitem[{{Levermore}(1984)}]{Levermore1984M1}
{Levermore}, C.~D. 1984, \jqsrt, 31, 149

\bibitem[{{Lin}(2019)}]{Lin2019}
{Lin}, M.-K. 2019, \mnras, 485, 5221

\bibitem[{{Lin} \& {Youdin}(2015)}]{LinYoudin2015}
{Lin}, M.-K. \& {Youdin}, A.~N. 2015, \apj, 811, 17

\bibitem[{{Lynden-Bell} \& {Pringle}(1974)}]{LyndenBellPringle1974}
{Lynden-Bell}, D. \& {Pringle}, J.~E. 1974, \mnras, 168, 603

\bibitem[{{Lyra} \& {Umurhan}(2019)}]{LyraUmurhan2019}
{Lyra}, W. \& {Umurhan}, O.~M. 2019, \pasp, 131, 072001

\bibitem[{{Malygin} {et~al.}(2017){Malygin}, {Klahr}, {Semenov}, {Henning}, \&
  {Dullemond}}]{Malygin2017}
{Malygin}, M.~G., {Klahr}, H., {Semenov}, D., {Henning}, T., \& {Dullemond},
  C.~P. 2017, \aap, 605, A30

\bibitem[{{Manara} {et~al.}(2018){Manara}, {Morbidelli}, \&
  {Guillot}}]{Manara2018}
{Manara}, C.~F., {Morbidelli}, A., \& {Guillot}, T. 2018, \aap, 618, L3

\bibitem[{{Manara} {et~al.}(2016){Manara}, {Rosotti}, {Testi}, {Natta},
  {Alcal{\'a}}, {Williams}, {Ansdell}, {Miotello}, {van der Marel}, {Tazzari},
  {Carpenter}, {Guidi}, {Mathews}, {Oliveira}, {Prusti}, \& {van
  Dishoeck}}]{Manara2016}
{Manara}, C.~F., {Rosotti}, G., {Testi}, L., {et~al.} 2016, \aap, 591, L3

\bibitem[{{Manger} \& {Klahr}(2018)}]{Manger2018}
{Manger}, N. \& {Klahr}, H. 2018, \mnras, 480, 2125

\bibitem[{{Manger} {et~al.}(2020){Manger}, {Klahr}, {Kley}, \&
  {Flock}}]{Manger2020I}
{Manger}, N., {Klahr}, H., {Kley}, W., \& {Flock}, M. 2020, \mnras, 499, 1841

\bibitem[{{Manger} {et~al.}(2021){Manger}, {Pfeil}, \& {Klahr}}]{Manger2021II}
{Manger}, N., {Pfeil}, T., \& {Klahr}, H. 2021, \mnras, 508, 5402

\bibitem[{{Marcus} {et~al.}(2015){Marcus}, {Pei}, {Jiang}, {Barranco},
  {Hassanzadeh}, \& {Lecoanet}}]{Marcus2015}
{Marcus}, P.~S., {Pei}, S., {Jiang}, C.-H., {et~al.} 2015, \apj, 808, 87

\bibitem[{{Melon Fuksman} {et~al.}(2023){Melon Fuksman}, {Flock}, \&
  {Klahr}}]{MelonFuksman2023}
{Melon Fuksman}, J.~D., {Flock}, M., \& {Klahr}, H. 2023, \aap, in press

\bibitem[{{Melon Fuksman} \& {Klahr}(2022)}]{MelonFuksman2022}
{Melon Fuksman}, J.~D. \& {Klahr}, H. 2022, \apj, 936, 16

\bibitem[{{Melon Fuksman} {et~al.}(2021){Melon Fuksman}, {Klahr}, {Flock}, \&
  {Mignone}}]{MelonFuksman2021}
{Melon Fuksman}, J.~D., {Klahr}, H., {Flock}, M., \& {Mignone}, A. 2021, \apj,
  906, 78

\bibitem[{{Melon Fuksman} \& {Mignone}(2019)}]{MelonFuksman2019}
{Melon Fuksman}, J.~D. \& {Mignone}, A. 2019, \apjs, 242, 20

\bibitem[{{Mignone}(2014)}]{Mignone2014reconstruction}
{Mignone}, A. 2014, Journal of Computational Physics, 270, 784

\bibitem[{{Mignone} {et~al.}(2007){Mignone}, {Bodo}, {Massaglia}, {Matsakos},
  {Tesileanu}, {Zanni}, \& {Ferrari}}]{Mignone2007}
{Mignone}, A., {Bodo}, G., {Massaglia}, S., {et~al.} 2007, \apjs, 170, 228

\bibitem[{{Mori} {et~al.}(2019){Mori}, {Bai}, \& {Okuzumi}}]{Mori2019}
{Mori}, S., {Bai}, X.-N., \& {Okuzumi}, S. 2019, \apj, 872, 98

\bibitem[{{Muley} {et~al.}(2023){Muley}, {Melon Fuksman}, \&
  {Klahr}}]{Muley2023}
{Muley}, D., {Melon Fuksman}, J.~D., \& {Klahr}, H. 2023, \aap

\bibitem[{{Nakamoto} \& {Nakagawa}(1994)}]{Nakamoto1994}
{Nakamoto}, T. \& {Nakagawa}, Y. 1994, \apj, 421, 640

\bibitem[{{Nelson} {et~al.}(2013){Nelson}, {Gressel}, \&
  {Umurhan}}]{Nelson2013}
{Nelson}, R.~P., {Gressel}, O., \& {Umurhan}, O.~M. 2013, \mnras, 435, 2610

\bibitem[{{Olofsson} {et~al.}(2009){Olofsson}, {Augereau}, {van Dishoeck},
  {Mer{\'\i}n}, {Lahuis}, {Kessler-Silacci}, {Dullemond}, {Oliveira}, {Blake},
  {Boogert}, {Brown}, {Evans}, {Geers}, {Knez}, {Monin}, \&
  {Pontoppidan}}]{Olofsson2009}
{Olofsson}, J., {Augereau}, J.~C., {van Dishoeck}, E.~F., {et~al.} 2009, \aap,
  507, 327

\bibitem[{{Ormel} \& {Cuzzi}(2007)}]{Ormel2007}
{Ormel}, C.~W. \& {Cuzzi}, J.~N. 2007, \aap, 466, 413

\bibitem[{{Papaloizou} \& {Pringle}(1984)}]{PapaloizouPringle1984}
{Papaloizou}, J.~C.~B. \& {Pringle}, J.~E. 1984, \mnras, 208, 721

\bibitem[{{Pfeil} {et~al.}(2023){Pfeil}, {Birnstiel}, \& {Klahr}}]{Pfeil2023}
{Pfeil}, T., {Birnstiel}, T., \& {Klahr}, H. 2023, arXiv e-prints,
  arXiv:2310.07332

\bibitem[{{Pfeil} \& {Klahr}(2019)}]{Pfeil2019}
{Pfeil}, T. \& {Klahr}, H. 2019, \apj, 871, 150

\bibitem[{{Pfeil} \& {Klahr}(2021)}]{Pfeil2021}
{Pfeil}, T. \& {Klahr}, H. 2021, \apj, 915, 130

\bibitem[{{Philippov} \& {Rafikov}(2017)}]{Rafikov2017}
{Philippov}, A.~A. \& {Rafikov}, R.~R. 2017, \apj, 837, 101

\bibitem[{{Pringle}(1981)}]{Pringle1981}
{Pringle}, J.~E. 1981, \araa, 19, 137

\bibitem[{{Richard} {et~al.}(2016){Richard}, {Nelson}, \&
  {Umurhan}}]{Richard2016}
{Richard}, S., {Nelson}, R.~P., \& {Umurhan}, O.~M. 2016, \mnras, 456, 3571

\bibitem[{{R{\"u}diger} {et~al.}(2002){R{\"u}diger}, {Arlt}, \&
  {Shalybkov}}]{Rudiger2002}
{R{\"u}diger}, G., {Arlt}, R., \& {Shalybkov}, D. 2002, \aap, 391, 781

\bibitem[{{Sch{\"a}fer} {et~al.}(2020){Sch{\"a}fer}, {Johansen}, \&
  {Banerjee}}]{schaefer2020}
{Sch{\"a}fer}, U., {Johansen}, A., \& {Banerjee}, R. 2020, \aap, 635, A190

\bibitem[{{Shakura} \& {Sunyaev}(1973)}]{ShakuraSunyaev}
{Shakura}, N.~I. \& {Sunyaev}, R.~A. 1973, in IAU Symposium, Vol.~55, X- and
  Gamma-Ray Astronomy, ed. H.~{Bradt} \& R.~{Giacconi}, 155

\bibitem[{{Simon} {et~al.}(2018){Simon}, {Bai}, {Flaherty}, \&
  {Hughes}}]{Simon2018}
{Simon}, J.~B., {Bai}, X.-N., {Flaherty}, K.~M., \& {Hughes}, A.~M. 2018, \apj,
  865, 10

\bibitem[{{Skinner} \& {Ostriker}(2013)}]{Skinner2013}
{Skinner}, M.~A. \& {Ostriker}, E.~C. 2013, \apjs, 206, 21

\bibitem[{{Stoll} \& {Kley}(2014)}]{Stoll2014}
{Stoll}, M. H.~R. \& {Kley}, W. 2014, \aap, 572, A77

\bibitem[{{Stoll} \& {Kley}(2016)}]{Stoll2016}
{Stoll}, M. H.~R. \& {Kley}, W. 2016, \aap, 594, A57

\bibitem[{{Stoll} {et~al.}(2017){Stoll}, {Kley}, \& {Picogna}}]{Stoll2017}
{Stoll}, M. H.~R., {Kley}, W., \& {Picogna}, G. 2017, \aap, 599, L6

\bibitem[{{Svanberg} {et~al.}(2022){Svanberg}, {Cui}, \&
  {Latter}}]{Svanberg2022}
{Svanberg}, E., {Cui}, C., \& {Latter}, H.~N. 2022, \mnras, 514, 4581

\bibitem[{{Takeuchi} \& {Lin}(2002)}]{TakeuchiLin2002}
{Takeuchi}, T. \& {Lin}, D.~N.~C. 2002, \apj, 581, 1344

\bibitem[{Toro(2009)}]{Toro}
Toro, E.~F. 2009, Riemann Solvers and Numerical Methods for Fluid Dynamics
  (Springer Berlin Heidelberg)

\bibitem[{{Townsend}(1958)}]{Townsend1958}
{Townsend}, A.~A. 1958, Journal of Fluid Mechanics, 4, 361

\bibitem[{{Urpin}(2003)}]{Urpin2003}
{Urpin}, V. 2003, \aap, 404, 397

\bibitem[{{Urpin} \& {Brandenburg}(1998)}]{UrpinBrandenburg1998}
{Urpin}, V. \& {Brandenburg}, A. 1998, \mnras, 294, 399

\bibitem[{{Urpin}(1984)}]{Urpin1984}
{Urpin}, V.~A. 1984, \sovast, 28, 50

\bibitem[{Virtanen {et~al.}(2020)Virtanen, Gommers, Oliphant, Haberland, Reddy,
  Cournapeau, Burovski, Peterson, Weckesser, Bright, {van der Walt}, Brett,
  Wilson, Millman, Mayorov, Nelson, Jones, Kern, Larson, Carey, Polat, Feng,
  Moore, {VanderPlas}, Laxalde, Perktold, Cimrman, Henriksen, Quintero, Harris,
  Archibald, Ribeiro, Pedregosa, {van Mulbregt}, \& {SciPy 1.0
  Contributors}}]{SciPy}
Virtanen, P., Gommers, R., Oliphant, T.~E., {et~al.} 2020, Nature Methods, 17,
  261

\bibitem[{{Voelkel} {et~al.}(2022){Voelkel}, {Klahr}, {Mordasini}, \&
  {Emsenhuber}}]{Voelkel2022}
{Voelkel}, O., {Klahr}, H., {Mordasini}, C., \& {Emsenhuber}, A. 2022, \aap,
  666, A90

\bibitem[{{Yamaleev} \& {Carpenter}(2009)}]{Yamaleev2009}
{Yamaleev}, N.~K. \& {Carpenter}, M.~H. 2009, Journal of Computational Physics,
  228, 4248

\end{thebibliography}

\begin{appendix}


\section{Additional figures}\label{A:AddFigures}

\begin{figure}[t!]
\centering
\includegraphics[width=\linewidth]{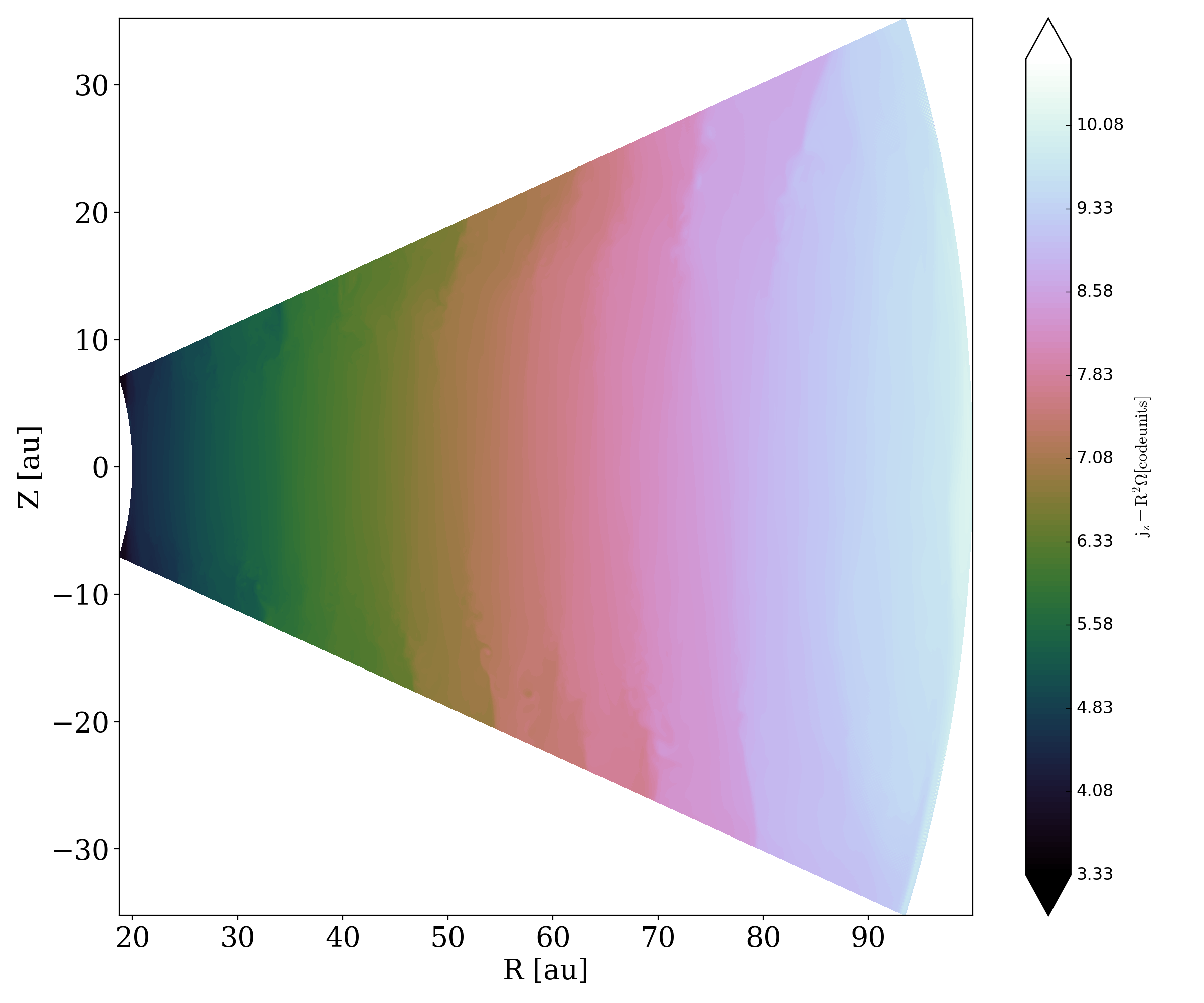}
\caption{Polar slice showing the specific angular momentum in the 3D disk simulation by \cite{Flock2020} after 350 orbits of evolution.
}
\label{fig:angmom3D}
\end{figure}
\begin{figure*}
\centering
\includegraphics[width=\linewidth]{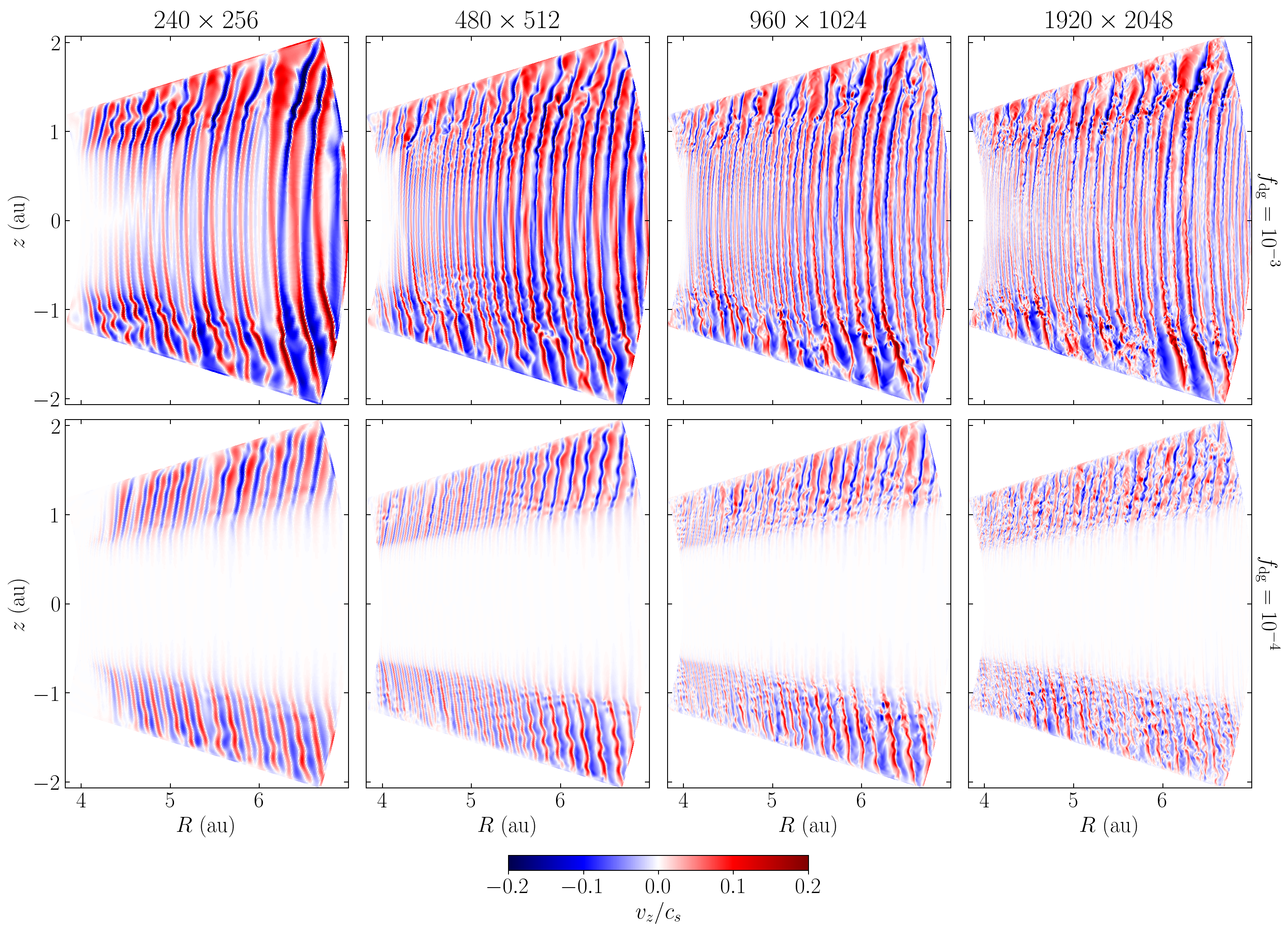}
\caption{Velocity distributions normalized by the isothermal sound speed after 300 orbits for varying resolution and dust content. In run \sftw{dg3c4\_256} (top left), it can be seen that the transition between finger and body modes is still not complete.}
\label{fig:vel_resolution}
\end{figure*}

Figure \ref{fig:angmom3D} shows a constant-$\phi$ slice of the specific angular momentum in the 3D Rad-HD disk simulation by \cite{Flock2020} after 350 inner orbits of evolution. Approximately uniform-$j_z$ bands are formed predominantly at the disk upper layers, where, as in the present work, the perturbations produced by the VSI are maximal. These bands are approximately axisymmetric, and consequently they are also visible in azimuthal averages of the same quantity. Figure \ref{fig:vel_resolution} shows the velocity distributions in our simulations after 300 orbits for all of our employed resolutions. 

\section{Radiative cooling timescale}\label{A:CoolingTimeScale}

We estimate the thermal relaxation timescale due to radiative processes by computing the damping time of small temperature perturbations in a hydrostatic disk. In this analysis we neglect mass advection and assume a constant $\rho$ and $\mathbf{v}=v_\phi\hat{\bm{\phi}}$ distribution taken from the initial condition of each Rad-HD simulation. We can therefore write the temperature in terms of the pressure using the ideal gas law, which simplifies the form of the resulting equations and the following analysis. We write the pressure and the radiation fields as $(p,E_r,\mathbf{F}_r)=(p_0+\delta p,E_{r0}+\delta E_r,\mathbf{F}_{r0}+\delta\mathbf{F}_r)$, where $(p_0,E_{r0},\mathbf{F}_{r0})$ are the hydrostatic unperturbed values and $(\delta p,\delta E_r,\delta\mathbf{F}_r)$ are time-dependent perturbations. Inserting this expression into Equation \eqref{Eq:RadHD} and keeping only linear terms in the perturbations, we obtain the following system of equations:
\begin{equation}\label{Eq:CoolingPertEq1}
    \begin{split}
        \partial_t \delta p &= - \kappa^d_P\rho_d\, c\, (\Gamma-1)(\eta \, \delta p - \delta E_r) \\
        \partial_t \delta E_r &= \kappa^d_P\rho_d\, \hat{c}\, (\eta \, \delta p - \delta E_r) - \hat{c}\nabla\cdot \delta \mathbf{F}_r \\
        \partial_t \delta \mathbf{F}_r  &= - \hat{c}\,  \nabla \cdot \delta \mathbb{P}_r - \rho_d \chi^d_R \hat{c} \, \delta \mathbf{F}_r\,,
    \end{split}
\end{equation}
where the quantity $\eta=4\, a_R T^4/p$ is evaluated at the unperturbed state, while source terms are written neglecting terms of order $\mathbf{v}/c$ and opacity variations with temperature.
The last of these equations considers a general total opacity coefficient $\chi^d_R$, while in this work we neglect scattering and assume $\chi^d_R=\kappa^d_R$. We then simplify this expression by considering one-dimensional radiation transport along a single direction $\mathbf{n}_0=\mathbf{F}_{r0}/\vert\vert\mathbf{F}_{r0}\vert\vert$. In this way the radiation pressure tensor takes the simple form $P_r^{xx}=\xi E_r$, where $x$ is the only spatial coordinate. This simplification does not alter the resulting cooling rates, as later justified in Section \ref{SS:Eigenv1}.

To evaluate the dependence of cooling rates on the perturbation lengthscale $\lambda$, we consider perturbations of the form $(\delta \hat{p}(k,t), \delta \hat{E_r}(k,t), \delta \hat{F^x_r}(k,t))\, e^{ikx}$, where $k\in\mathbb{R}$ is the wavenumber, which verifies $|k|=2\pi/\lambda$. Assuming that the background fields are uniform in space, or equivalently, that their variation lengthscales are much larger than $\lambda$, Eq. \eqref{Eq:CoolingPertEq1} becomes the following homogeneous system of ordinary differential equations:
\begin{equation}\label{Eq:CoolingPertEq2}
     \frac{\partial}{\partial t}\begin{pmatrix}
        \delta p \\ \delta E_r \\ \delta F^x_r
        \end{pmatrix} =
        A \begin{pmatrix}
        \delta p \\ \delta E_r \\ \delta F^x_r
        \end{pmatrix}\,,
\end{equation}
where the matrix $A$ is defined as 
\begin{equation}\label{Eq:CoolingPertEqAMatrix}
A = \begin{pmatrix}
-\eta c\, (\Gamma-1)/\lambda_P     &  c\, (\Gamma-1)/\lambda_P & 0 \\
 \eta \hat{c}/\lambda_P &  -\hat{c}/\lambda_P  &
- i k \hat{c}  \\
0 & - i\hat{c} k  (\xi-\xi' f) & -\hat{c}(1/\lambda_R+ i \xi' k)
\end{pmatrix},
\end{equation}
while $\lambda_P=(\rho_d\kappa^d_P)^{-1}$ and $\lambda_R=(\rho_d\chi^d_R)^{-1}$ are respectively the Planck and Rosseland mean free paths and $\xi'=\mathrm{d}\xi/\mathrm{d}f=2f/\sqrt{4-3f^2}$, which is always positive for $f>0$ and vanishes for $f=0$.

Given an initial temperature perturbation, which in this analysis takes the form $(\delta p_0, 0,0)$, the time-dependent solution of Eq. \eqref{Eq:CoolingPertEq2} can be obtained via matrix exponentiation as $e^{At}(\delta p_0, 0, 0)^\intercal$. The matrix $A$ has three different eigenvalues, $\{s_\mathrm{c},s_+,s_-\}$, and therefore all perturbations evolve as a linear combination of $\{e^{s_\mathrm{c} t},e^{s_+ t},e^{s_- t}\}$. We derive expressions for these eigenvalues in Sections \ref{SS:Eigenv1}--\ref{SS:Eigenv2} and show in Section \ref{SS:ExpSol} that $\delta p \sim e^{s_\mathrm{c} t}$, where $|s_\mathrm{c}| \ll |s_\pm|$. The cooling time is therefore given by $t_\mathrm{cool}^{-1} = |s_\mathrm{c}|$, while the other two eigenvalues correspond to the faster adjustment of the radiation fields. These conclusions remain unaltered when $\hat{c}<c$ for high enough $\hat{c}$. We study the applicability of the RSLA in Section \ref{SS:CoolingRSLA}, where we use this analysis to obtain constraints to the values of $\hat{c}$ that can be employed without introducing unphysical results.

\subsection{Eigenvalues and cooling rate}\label{SS:Eigenv1}

We start by computing the eigenvalues of $A$, whose characteristic polynomial is
\begin{equation}\label{Eq:CharPol}
\begin{split}
    P(s) &=\det (A-s I_3) \\
    &= -\left( s - s_\mathrm{thin}  \right)
    \left[
    \left( s + \frac{\hat{c}}{\lambda_P} \right)
    \left( s + \frac{\hat{c}}{\lambda_R} + i\hat{c}\xi' k \right)\right.\\
    &+ \hat{c}^2 k^2 \left(\xi-\xi' f\right)\bigg]
    - s_\mathrm{thin} \frac{\hat{c}}{\lambda_P}
    \left( s + \frac{\hat{c}}{\lambda_R} + i\hat{c}\xi' k \right)
    \,,
\end{split}
\end{equation}
where $I_3$ is the rank-3 identity matrix and
\begin{equation}\label{Eq:Sthin}
    s_\mathrm{thin}=-\frac{\eta c (\Gamma-1)}{\lambda_P}\,.
\end{equation}
In principle we can obtain all three roots of $P$ by applying a cubic formula, but instead we can get accurate analytic expressions for these roots by considering the limits of small and large $|s|$.

We first consider the limit 
\begin{equation}\label{Eq:DiffCond}
    |s|\ll \frac{\hat{c}}{\lambda_R}\,,
\end{equation}
which also guarantees $|s|\ll \hat{c}/\lambda_P$, since our employed opacities verify $\kappa^d_P>\kappa^d_R$ for all temperatures \citep{MelonFuksman2022} (this is also true in general for opacity laws of the form $\kappa^d_\nu\propto \nu^n$ as long as $n> 0.81$). In that case, the characteristic equation $P(s)=0$ loses its dependence on $\hat{c}$ and becomes
\begin{equation}
\begin{split}
    (s-s_\mathrm{thin})
    \left[
    \left( 1+ \xi' \lambda_R k \right)
    +\lambda_P\lambda_R k^2(\xi-\xi'f)
    \right] \\ = 
    - s_\mathrm{thin}\left( 1+ i \xi' \lambda_R k \right)\,,
\end{split}
\end{equation}
whose solution is
\begin{equation}\label{Eq:CoolingRate0}
    s = s_\mathrm{thin} \frac{
    \left(\xi-\xi' f\right) \lambda_R\lambda_P k^2}{
    1+ i \xi' \lambda_R k + \left(\xi-\xi' f\right)\lambda_R\lambda_P k^2 } \,.
\end{equation}
This expression can be simplified by noting that the term $i \xi' \lambda_R k$ is much smaller than $1$ for large $\lambda$ ($\lambda_R k\lesssim 1$) and much smaller than the $\mathcal{O}(\lambda_R\lambda_P k^2)$ term for small $\lambda$ ($\lambda_R k\gg 1$), and thus it can be neglected.
On the other hand, the factor $\xi-\xi'f$ only affects the value of $s_\mathrm{c}$ if $\lambda_R\lambda_Pk^2$ is not much larger than $1$, which can only occur if $f\ll 1$, in which case $\xi'f\approx f^2\ll 1$ and $\xi\approx 1/3$. Therefore we can safely replace $\xi-\xi' f$ by $1/3$, which leads to the simpler expression
\begin{equation}\label{Eq:sc}
   s_c = s_\mathrm{thin}\, \frac{
     \lambda_R\lambda_P k^2}{
    3 + \lambda_R\lambda_P k^2 } \,.
\end{equation}
Using the $k$ distributions in our simulations (Section \ref{S:StabilityAnalysis}), we estimated the relative error of this approximation to be of at most $0.01\%$.

If this analysis is repeated in 3D, the same expressions are obtained
with the difference that $\lambda_P \lambda_R k^2$ is replaced by a term of the same order of magnitude that tends to $\lambda_P \lambda_R k^2/3$ wherever $\lambda_P \lambda_R k^2\lesssim 1$. Thus, the particular form of the $\mathcal{O}(\lambda_P \lambda_R k^2)$ terms for anisotropic transport does not affect the resulting cooling rate, which always tends to $s_\mathrm{thin}$ for $\lambda_P \lambda_R k^2\gg 1$ and to Equation \eqref{Eq:sc} otherwise. Consequently, Equation \eqref{Eq:sc} still holds when considering multidimensional radiation transport. Analogously, we would have obtained the same result using any other closure than M1 or even solving the full frequency-integrated radiative transfer equation, since $P^{ij}_r$ always tends to $E_r \delta^{ij}/3$ for $\lambda_P \lambda_R k^2\lesssim 1$. Therefore, Equation \eqref{Eq:sc} is valid in general for gray radiative transfer.

As we show in Section \ref{SS:ExpSol}, this eigenvalue alone determines the radiative cooling rate, which can be computed as $t_\mathrm{cool}=|s_\mathrm{c}|^{-1}$. Specifically,
\begin{equation}\label{Eq:tcool}
    t_\mathrm{cool}=
    \frac{\rho \epsilon}{4\, a_R T^4}\,
    \frac{3 + \lambda_R\lambda_P k^2 }{\lambda_R\lambda_P k^2}\,
    \frac{\lambda_P}{c}
    \,,
\end{equation}
which is the constant-opacity limit of Equation \eqref{Eq:tcool_ms}. The wavelength-dependent behavior of $t_\mathrm{cool}$ can be understood by noting that the value of $k$ regulates how much radiation energy is transported away through advection before it can reheat the dust-gas mixture via absorption. This can be seen by comparing the relative size of the different terms in Eq. \eqref{Eq:CoolingPertEq2}. For large $k$, $E_r$ quickly evolves into a state such that $\delta E_r
\approx  \eta\delta p /(1 + \mathcal{O}(\lambda_P\lambda_R k^2))\ll \eta \delta p$, which can be seen by zeroing the time derivatives of the radiation fields. This means that $\delta E_r$ can be dropped from the cooling source term of $\delta p$, which is proportional to $(\eta \delta p - \delta E_r)$. In this way, radiation transport limits the maximum value of $\delta E_r$ enough to cause temperature perturbations to be damped solely due to optically thin cooling, which is scale-independent. This is not verified for smaller $k$ values if $\lambda_P\lambda_R k^2/3$ is not much larger than $1$, in which case $(\eta \delta p - \delta E_r)\approx \eta \delta p \frac{\lambda_R\lambda_P k^2}{3 + \lambda_R\lambda_P k^2 }$. For small enough $k$, diffusion overcomes optically thin cooling and becomes the sole relaxation mechanism.

The above derivation relies on the constancy of the opacity coefficients. If we consider instead that opacities depend on the local temperature, as we do in this work, the resulting eigenvalue corresponding to thermal relaxation is
\begin{equation}
\begin{split}
    s_c& = - \frac{c(\Gamma-1)}{\lambda_P(1+\lambda_P \lambda_R k^2/3)}\\
    &\times
    \left[
    \frac{\lambda_P\lambda_R k^2}{3}\left.\bigg(
    \eta + b_P\frac{a_R T^4 - E_r}{p}
    \right.\bigg)
    - i b_R \frac{\lambda_P \mathbf{k}\cdot\mathbf{F}_r}{p}
    \right]\,,
\end{split}
\end{equation}
where $b_P = \frac{d \log \kappa^d_P}{d \log T}$ and $b_R = \frac{d \log \chi^d_R}{d \log T}$. This expression only depends on $b_P$ far from the midplane, where $E_r\neq a_R T^4$. On the other hand, the rightmost term proportional to $b_R$ only appears in the diffusion regime, in which case it introduces an oscillatory component. Since in that case $\mathbf{F_r}=-\frac{\lambda_R}{3}\nabla E_r$, it is straightforward to prove that the relative variation of $s_c$ due to that term is at most $4 b_R \lambda / L$, where $L$ is the variation lengthscale of $E_r$. Given the initial assumption that $\lambda \ll L$, that term can be neglected in all regimes and the resulting cooling time is given by Equation \eqref{Eq:tcool_ms}.

\begin{figure*}[t]
\centering
\includegraphics[width=\linewidth]{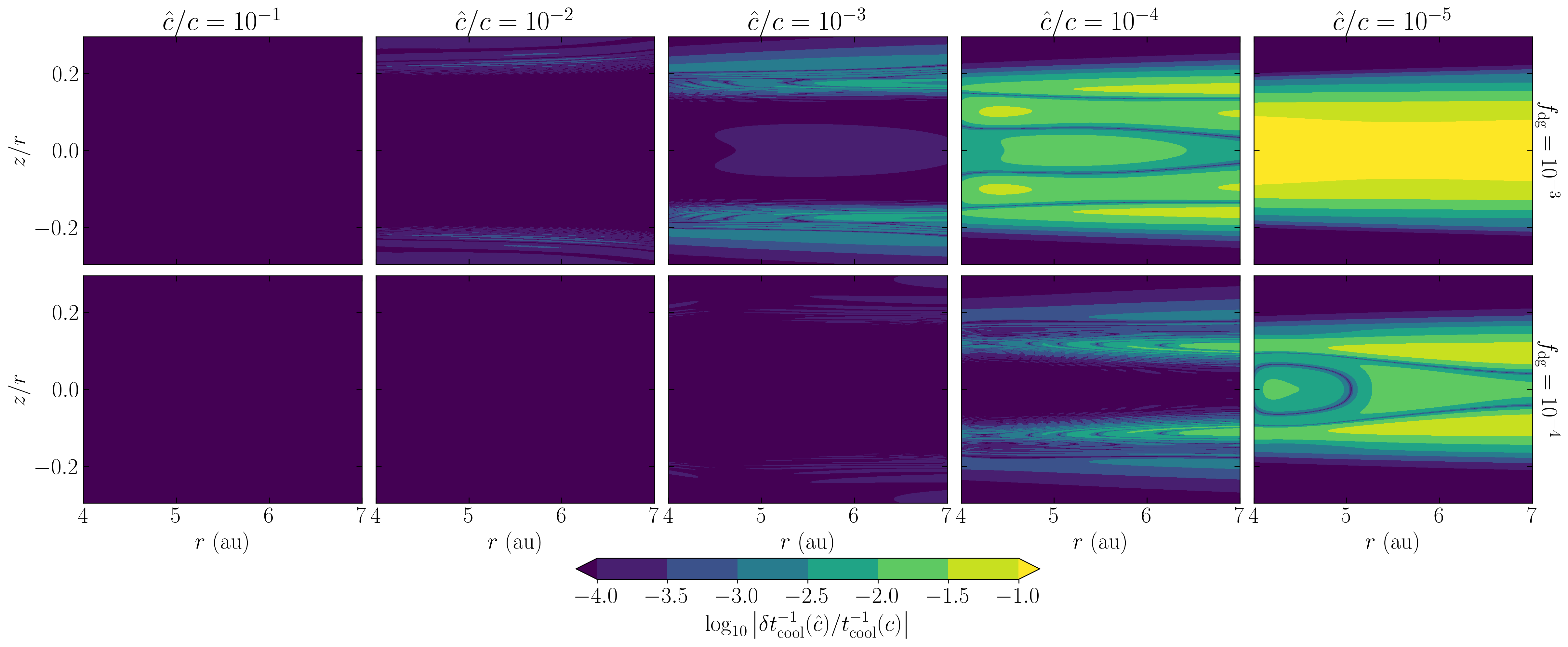}
\caption{Relative variation of the cooling rate with respect to its value when $c=\hat{c}$ as a function of $\hat{c}/c$ for $f_\mathrm{dg}=10^{-3}$ (top row) and $10^{-4}$ (bottom row).}
\label{fig:tcool_c_fdg_relvar}
\end{figure*}

\subsection{Consistency check and constraints on $\hat{c}$}\label{SS:Eigenv1Consistency}

We now test the consistency of the small $|s_c|$ condition assumed in Eq. \eqref{Eq:DiffCond}. This limit can be better understood by considering the components of $\delta E_r$ and $\delta F_r^x$ that evolve proportionally to $e^{s_c t + i k x}$, which we denote as $\delta \overline{E}_r$ and $\delta \overline{F}_r^x$. Those components must also satisfy Eq. \eqref{Eq:CoolingPertEq2}, which leads to
\begin{equation}
    \begin{split}
        \left(s_c/\hat{c} + 1/\lambda_P\right) \delta \overline{E}_r 
        &= (\eta/\lambda_P)\delta p - i k \delta \overline{E}_r \\
        \left(s_c/\hat{c} + 1/\lambda_R\right) \delta \overline{F}^x_r 
        &= - i k (\xi-\xi' f) \delta \overline{E}_r - i \xi' k \delta \overline{F}^x_r 
    \end{split}
\end{equation}
meaning that if $s_c$ satisfies Eq. \eqref{Eq:DiffCond}, the time derivatives of the radiation fields can be dropped. This means that, in that case, $\delta \overline{E}_r$ and $\delta \overline{F}^x_r$ evolve quasi-statically, as their readjustment time is much shorter than $|s_c|^{-1}$, which explains why $\hat{c}$ disappears from the equations. In particular, if $f\ll 1$, this leads to the usual expression for the radiative flux in the diffusion limit, $\mathbf{F}_r=-\frac{1}{3 \rho \chi_R}\nabla E_r$, whose perturbation in this analysis becomes $\delta \overline{F}^x_r=-i (\lambda_R k/3) \delta \overline{E}_r$. 

Using the obtained expression for $s_c$ (Equation \eqref{Eq:sc}), we can see that Eq. \eqref{Eq:DiffCond} is satisfied for all $k$ as long as $|s_\mathrm{thin}|\ll \hat{c}/\lambda_R$, which leads to the equivalent condition
\begin{equation}\label{Eq:CondRSLA}
    \eta (\Gamma-1) \frac{\lambda_R}{\lambda_P} \ll \frac{\hat{c}}{c}\,.
\end{equation}
For $\hat{c}=c$, this equation is satisfied close to the midplane, where $\eta(\Gamma-1)\lambda_R/\lambda_P\sim 10^{-5}-10^{-4}$, but not at the disk atmosphere. However, our expression for $s_c$ is still valid in that region, as we explain next. Far enough from the midplane, the timescale $|s_\mathrm{thin}|^{-1}$ becomes shorter than the light-crossing time of a distance $\lambda_R$, typically of a few hundred $\mathrm{au}$ in those regions. In such locations, $\lambda_R\lambda_P k^2\rightarrow\infty$ (otherwise $\lambda\sim \lambda_{R/P}$, which would render the uniform background assumption inapplicable), in which case $s=s_\mathrm{thin}$ provided a large-$k$ condition is satisfied independently from Eq. \eqref{Eq:CondRSLA}. To see this, we rearrange the characteristic equation as
\begin{equation}
    s = s_\mathrm{thin}(1-\varphi(s))\,,
\end{equation}
with
\begin{equation}
    \varphi(s) = \left( 1+ \frac{s\lambda_P}{\hat{c}} 
    +\frac{(\xi-\xi' f)\lambda_R\lambda_P k^2}{1+s\lambda_R/\hat{c}+i \xi' k \lambda_R}\right)^{-1}\,,
\end{equation}
from which we can see that $s=s_\mathrm{thin}$ is obtained as long as $\left|\varphi(s_\mathrm{thin})\right|\ll 1$. We can then combine both conditions, which leads to the conclusion that Eq. \eqref{Eq:sc} is valid provided
\begin{equation}\label{Eq:Condsc}
\min\left(
\frac{\eta(\Gamma-1)(\lambda_R/\lambda_P)}{c/\hat{c}},\,
|\varphi(s_\mathrm{thin})|
\right)\ll 1\,,
\end{equation}
that is, as long as either the cooling timescale is much shorter than the light-crossing time of a mean free path or the gas-dust mixture is largely optically thin along a perturbation wavelength. For the wavelengths in this work (see Section \ref{S:StabilityAnalysis}), this relation holds in the entire domain for all considered $\hat{c}$ values in Section \ref{SS:CoolingRSLA}, except for $\hat{c}/c=10^{-5}$ in the case $f_\mathrm{dg}=10^{-3}$, in which the left-hand side of Eq. \eqref{Eq:Condsc} is of order $1$ close to the midplane. We further study the effect of reducing $\hat{c}$ on the cooling timescale in Section \ref{SS:CoolingRSLA}.


\subsection{Remaining eigenvalues}\label{SS:Eigenv2}

Having verified the validity of Eq. \eqref{Eq:sc}, it only remains to show that $\delta p\sim \delta p_0 e^{s_\mathrm{c}t}$, for which we need to compute the order of magnitude of the contribution to $\delta p$ of the remaining eigenvalues. These can be obtained with good accuracy by assuming $|s|\gg |s_c|$, which transforms the characteristic equation into
\begin{equation}
\left( \frac{s \lambda_P}{\hat{c}} + 1\right)
\left( \frac{s \lambda_R}{\hat{c}} + 1 + i \xi' \lambda_R k\right)
+(\xi-\xi' f)\lambda_R\lambda_P k^2=0\,.
\end{equation}
Since we only want to obtain the order of magnitude of the two solutions to this equation, $s_\pm$, it is enough to compute their values in the optically thick and thin limits. If $\lambda_R k \ll 1$, which also guarantees $\lambda_P k \ll 1$, the two solutions of these equations are
\begin{equation}
s_+ \approx -\frac{\hat{c}}{\lambda_P},\,\,\,\,\,s_- \approx -\frac{\hat{c}}{\lambda_R}\,,
\end{equation}
while for $\lambda_P k \gg 1$, and therefore also $\lambda_R k \gg 1$, we have
\begin{equation}
s_\pm \approx 
-\frac{\hat{c}}{2}\left(\frac{1}{\lambda_R}+\frac{1}{\lambda_P}+i\xi'k\right)
\pm i \hat{c} k \sqrt{\xi - \xi' f -(\xi')^2/4}\,,
\end{equation}
where the argument of the square root is always positive if $f<1$. Therefore, $|s_\pm|$ is at least of order $\sim \max(\lambda_R^{-1} \hat{c},k \hat{c})$. We can then check the consistency of the assumption $|s_\pm|\gg |s_\mathrm{c}|$, for which it is enough to verify that $|s_\pm|\gg |s_\mathrm{thin}|$, since $|s_\mathrm{thin}|\geq|s_\mathrm{c}|$. For $\lambda_R k \ll 1$, this is equivalent to Eq. \eqref{Eq:CondRSLA}, which as already mentioned is verified in that regime for large enough $\hat{c}$. The same is true for $\lambda_R k \gg 1$, in which case the resulting condition is $\hat{c}/c\gg \eta(\Gamma-1)/(\lambda_P k)$, which in that limit holds everywhere for $\hat{c}=c$.

\subsection{Exponential solution}\label{SS:ExpSol}

With the obtained eigenvalues, we can compute an expression for $\delta p(t)$ and compare the relative size of the terms proportional to $\{e^{s_\mathrm{c} t},e^{s_+ t},e^{s_- t}\}$. The leading orders of $\delta p = (e^{At}(\delta p_0,0,0)^\intercal)_1$ can be evaluated in a rather straightforward way by applying the Cayley-Hamilton theorem, which allows us to compute $e^{At}$ as a linear combination of $\{I,\,A,\,A^2\}$ with time-dependent coefficients, where $I$ is the $3\times 3$ identity matrix.
This results in expressions of the form
\begin{equation}\label{Eq:dpexpansion_thick}
    \delta p = \delta p_0 \left[ e^{s_\mathrm{c}t} +
    \sum_{i} \mathcal{O}_i\left( \eta \frac{c}{\hat{c}} \right) e^{s_it}\right]
\end{equation}
if $\lambda_R k\lesssim 1$, and
\begin{equation}\label{Eq:dpexpansion_thin}
    \delta p = \delta p_0 \left[  e^{s_\mathrm{c}t} +
    \sum_{i} \mathcal{O}_i\left(\frac{\eta}{(\lambda_P k)^2}\frac{c}{\hat{c}}\right) e^{s_it}\right]\,,
\end{equation}
if $\lambda_P k\gg 1$\,, where $\mathcal{O}_i(x)$ denotes a term of order $x$. For $\hat{c}=c$, the leading term in these equations is proportional to $e^{s_\mathrm{c}t}$, whereas the other terms are much smaller, both close to the midplane, where $\eta\ll 1$, as well as away from the midplane, where $\lambda_P k$ is always much larger than $1$. This proves that $t_\mathrm{cool}^{-1}=|s_c|$. 
On the other hand, the relative size of these terms changes for $\hat{c}<c$, which together with the changes in the eigenvalues can affect the overall relaxation rate for sufficiently small $\hat{c}$. We study this effect in the next section.

\subsection{Dependence on the reduced speed of light}\label{SS:CoolingRSLA}
 
In the previous sections we showed that $t_\mathrm{cool}$ can be different from its physical value when $\hat{c}/c<1$, either due to changes in the value of $s_\mathrm{c}$ or in the relative size of the terms in Equations \eqref{Eq:dpexpansion_thick} and \eqref{Eq:dpexpansion_thin}. We now evaluate the overall change in the cooling time when $\hat{c}/c<1$ by numerically computing $\delta p(t)$ in our domain, defining $t_\mathrm{cool}(\hat{c})$ as the $e$-folding time of $\delta p$, i.e., $\delta p(t_\mathrm{cool}(\hat{c}))/\delta p(0)=e^{-1}$. This definition coincides with the real cooling time when $c=\hat{c}$. We achieve this by numerically computing $e^{At}$ for different times in the interval $[0,2\, t_\mathrm{cool}(c)]$ using the SciPy \citep{SciPy} \sftw{expm} matrix exponentiation routine, which follows the method in \cite{matexp}. The $A$ matrix is in turn computed at each location using the $k$ distributions obtained in Appendix \ref{A:WavelengthEstimation} for $N_\theta = 512$. 

\begin{figure}[t!]
\centering
\includegraphics[width=\linewidth]{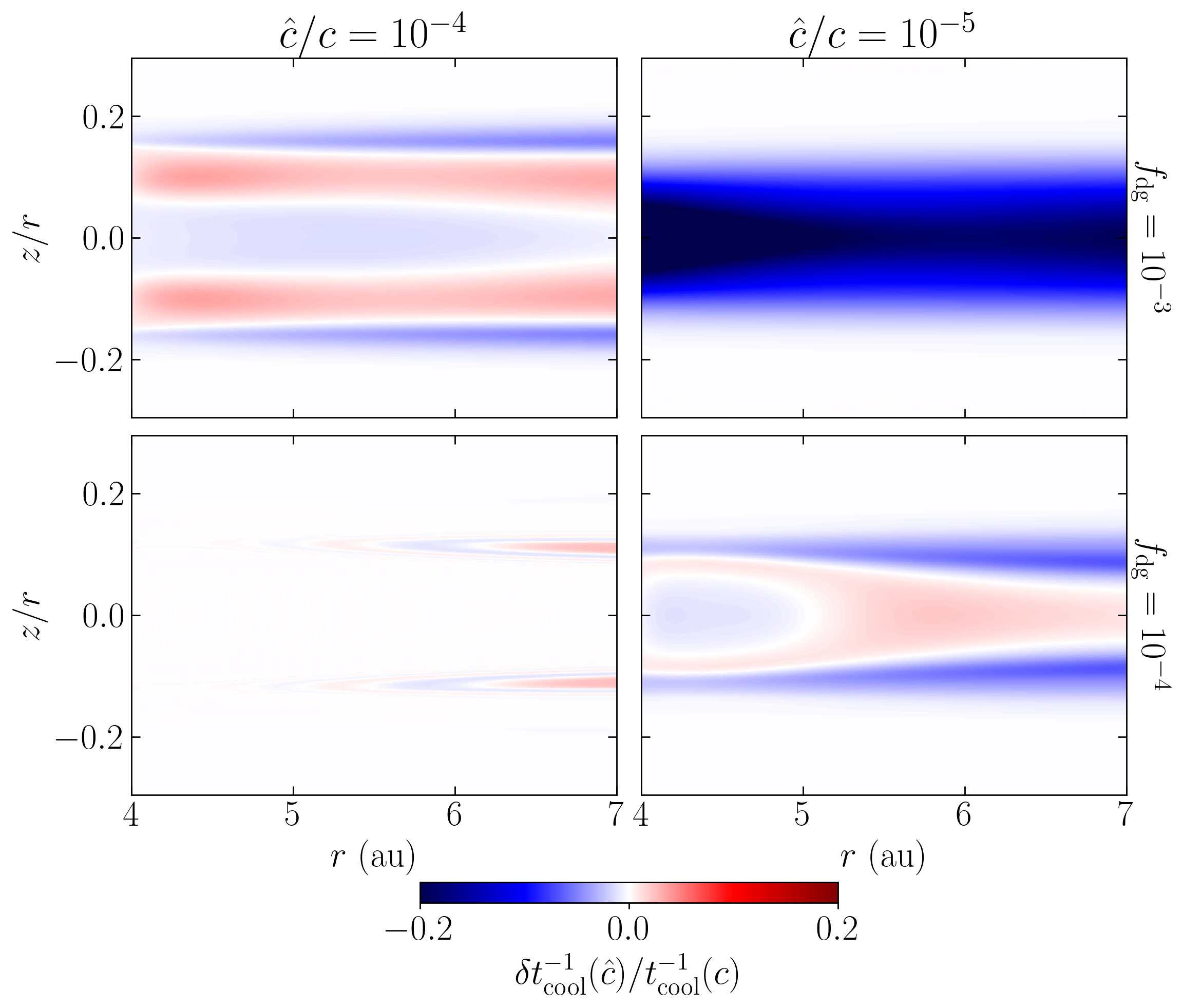}
\caption{Signed relative variation of the cooling rate with respect to its value with $c=\hat{c}$ for $\hat{c}/c=10^{-4},10^{-5}$ and both considered $f_\mathrm{dg}$ values.}
\label{fig:tcool_c_fdg_lowestc}
\end{figure}

We repeated this computation for different $\hat{c}/c$ values in the interval $[10^{-5},10^{0}]$. The obtained distribution of the relative error $|\delta t^{-1}_\mathrm{cool}(\hat{c})|/ t^{-1}_\mathrm{cool}(c)$, with $\delta t^{-1}_\mathrm{cool}(\hat{c})=t^{-1}_\mathrm{cool}(\hat{c})-t^{-1}_\mathrm{cool}(c)$, is shown in Fig. \ref{fig:tcool_c_fdg_relvar} for all considered $\hat{c}/c$ and $f_\mathrm{dg}$ values. For the two highest $\hat{c}/c$ values, we also show in Fig. \ref{fig:tcool_c_fdg_lowestc} the signed relative variation $\delta t^{-1}_\mathrm{cool}(\hat{c})/ t^{-1}_\mathrm{cool}(c)$, which shows that the cooling timescale does not always decrease with $\hat{c}$, since the $e$-folding time depends on the contribution of the different terms in Eqs. \eqref{Eq:dpexpansion_thick} and \eqref{Eq:dpexpansion_thin}. The computed relative errors grow for decreasing $\hat{c}$ and stay on the order of $1\%$ and below for all parameters except for $(\hat{c}/c,f_\mathrm{dg})=(10^{-5},10^{-3})$, in which case values of $10\%$ are reached in a significant portion of the domain covering the disk midplane. For those parameters, the obtained cooling times are always lower than those with $\hat{c}=c$. We can then expect significant deviations in the resulting VSI evolution in that case. 

An analytical constraint for the minimum value of $\hat{c}/c$ that does not alter $t_\mathrm{cool}$ can be obtained by noting that, in the midplane region where the largest errors occur for $\hat{c}/c=10^{-5}$, the most stringent condition for the value of $\hat{c}/c$ among those derived in Sections \ref{SS:Eigenv1Consistency}-\ref{SS:ExpSol} is given by Eq. \eqref{Eq:Condsc}, which guarantees that $s_\mathrm{c}$ is not affected by the reduction of $\hat{c}$. This condition is verified except for $(\hat{c}/c,f_\mathrm{dg})=(10^{-5},10^{-3})$, in which case the left-hand side of the inequality is of order $1$ close to the midplane, which is consistent with the error distributions in Fig. \ref{fig:tcool_c_fdg_relvar}.

We can now use these results to estimate the maximum possible reduction of $\hat{c}$ that does not affect the dynamics of the VSI. In \cite{Skinner2013}, it was argued that the RSLA is applicable as long as all relevant timescale hierarchies are unaffected by such reduction. However, it is not enough for this analysis to compare the order of magnitude of $t_\mathrm{cool}$ and $\Omega^{-1}$ (or any other dynamical time) to guarantee the independence of the growth of VSI modes on $\hat{c}$, since if $t_\mathrm{cool}>t_\mathrm{crit}$, the linear growth rate and the strength of the saturated instability depend on the precise value of $t_\mathrm{cool}$ \citep[][]{LinYoudin2015,Manger2018,Klahr2023a}. It is therefore necessary that $t_\mathrm{cool}$ is unchanged by the reduction of $\hat{c}$, which is approximately the case if $\hat{c}/c\geq 10^{-4}$, as already shown. It remains to be evaluated if $\hat{c}$ is much larger than the maximum gas velocity, which in our runs is on average at most $7\%$ of the isothermal sound speed. Since $\hat{c}=10^{-4}c$ is always between $30$ and $60$ times larger than that speed, we can expect this value to be high enough not to introduce unphysical results. However, this can only be assessed through testing, as we do in Appendix \ref{A:RSLAtests}.
\section{RSLA tests}\label{A:RSLAtests}

\begin{figure}[t]
\centering
\includegraphics[width=\linewidth]{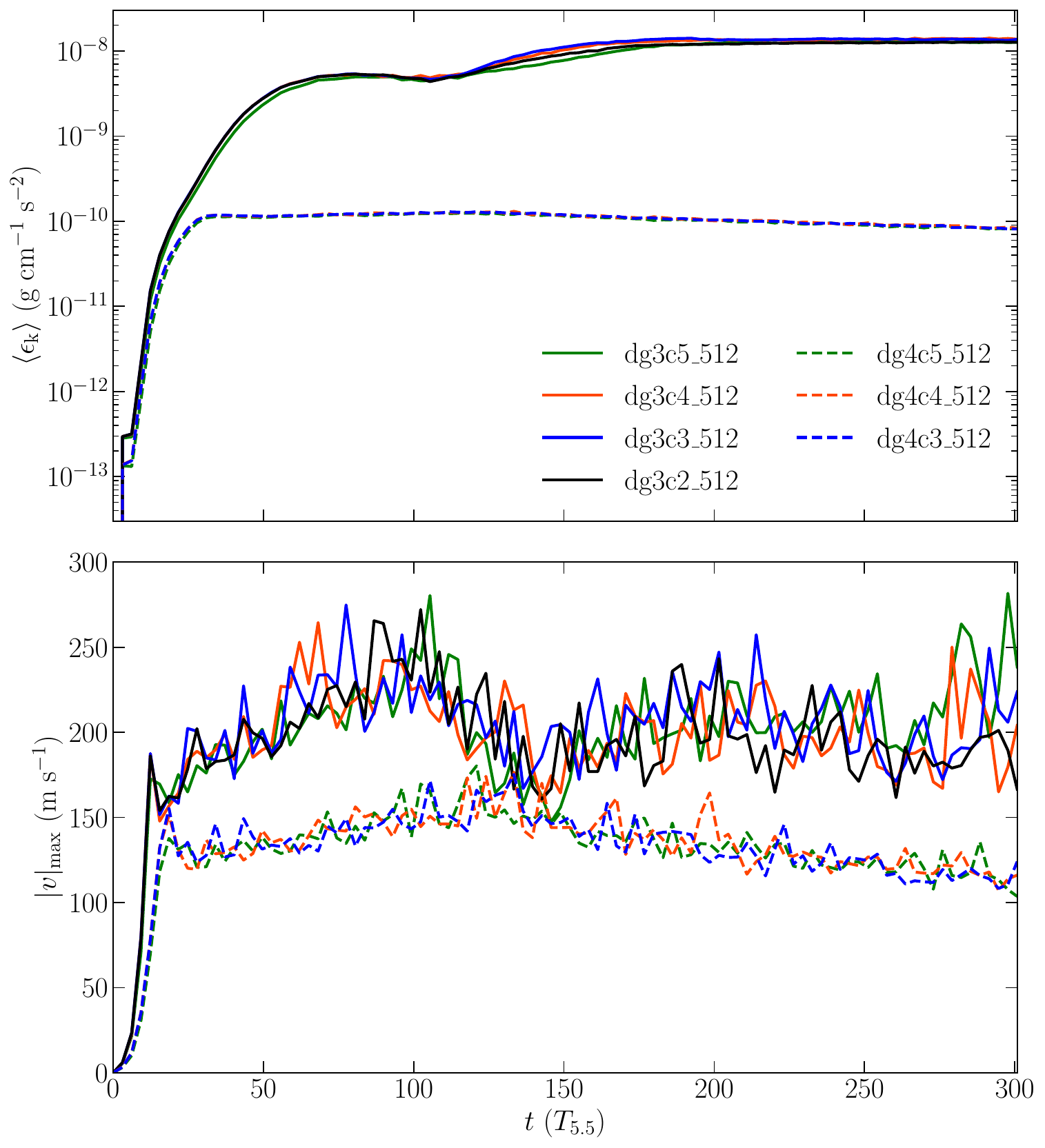}
\caption{Average kinetic energy (top) and maximum velocity (bottom) for different $\hat{c}/c$ values at a resolution of $480\times 512$. Solid and dashed lines correspond to $f_\mathrm{dg}=10^{-3}$ and $10^{-4}$, respectively.}
\label{fig:ekin_c}
\end{figure}

\begin{figure}[t]
\centering
\includegraphics[width=\linewidth]{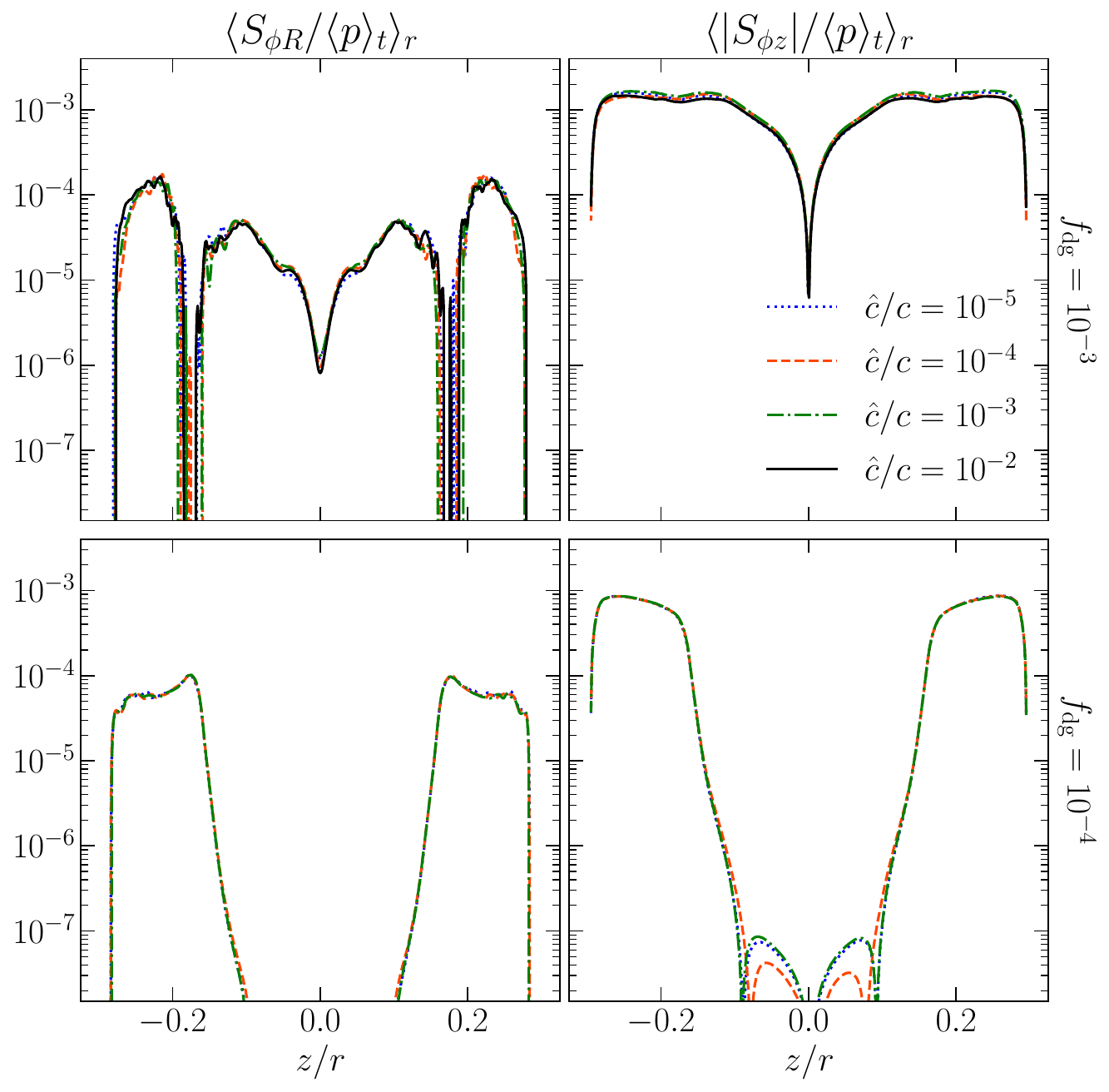}
\caption{Same as Fig. \ref{fig:Sphi_prs_1Davgs_res}  but comparing the Reynolds stress components for different $\hat{c}/c$ values at a resolution of $480\times 512$.}
\label{fig:Sphi_c}
\end{figure}

To validate the nominal choice of $\hat{c}/c=10^{-4}$ used in all simulations shown so far in this work, we tested the results of Appendix \ref{A:CoolingTimeScale} by recomputing some quantities related to the evolution of the VSI for $N_\theta=512$ using different $\hat{c}/c$ values spanning from $10^{-5}$ to $10^{-3}$, with an additional run with $\hat{c}/c=10^{-2}$ for $f_\mathrm{dg}=10^{-3}$ (see Table \ref{tt:RadHD}). We studied differences in the time evolution of the VSI by computing the kinetic energy and the maximum velocity as a function of time as in Section \ref{SS:VelEkin} (Fig. \ref{fig:ekin_c}), and we evaluated changes in the angular momentum transport in the saturated stage by computing the components of the Reynolds stress tensor as in Section \ref{SS:ReynoldsStress} (Fig. \ref{fig:Sphi_c}). For $f_\mathrm{dg}=10^{-4}$, all curves match almost perfectly except for the random oscillations in $|v|_\mathrm{max}$, as expected from the analysis in Appendix \ref{A:CoolingTimeScale}. For $f_\mathrm{dg}=10^{-3}$, we do notice that, both during the growth of the finger modes and the body modes, the runs with $\hat{c}/c=10^{-5}$ have a slightly smaller kinetic energy than the other runs, whereas the saturated steady state seems identical in all cases. The curves for all other $\hat{c}/c$ values overlap except during the growth of the body modes, where, however, no identifiable trend as a function of $\hat{c}/c$ is followed. A lower kinetic energy with $\hat{c}/c=10^{-5}$ is also observed when comparing runs \sftw{dg3c4\_1024} and \sftw{dg3c5\_1024}, with much smaller relative differences between the curves (not shown here). It must be noted that the value of $\hat{c}/c$ not only may affect the cooling time, but it also determines the maximum time step used for the time evolution of radiation fields \citep{MelonFuksman2021}, and thus some variations are also expected due to the accumulation of small numerical differences and possibly even different amounts of numerical diffusion. On the other hand, all curves overlap almost perfectly in the different saturated stages of the VSI. From these results, we conclude that our chosen value of $\hat{c}/c=10^{-4}$ is high enough not to introduce unphysical effects in any stage of the VSI. On top of this, it even seems that $\hat{c}/c=10^{-5}$ is enough to describe the steady state of the saturated VSI, and only introduces small variations in the growth rates. However, this result only applies to our employed setup, and similar testing is needed in general.

\begin{figure*}[t]
\centering
\includegraphics[width=\linewidth]{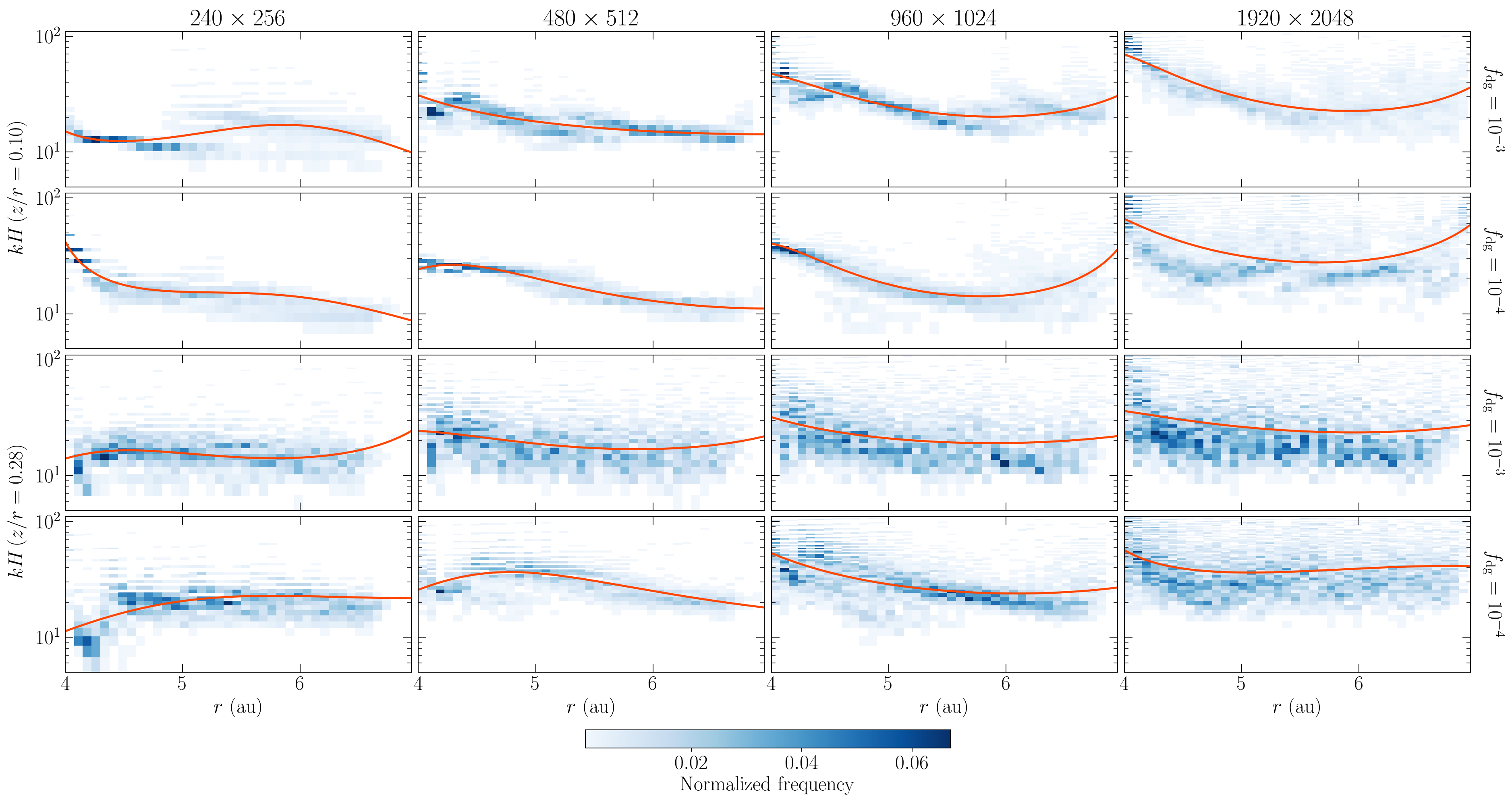}
\caption{Wavenumber histograms (normalized using the local scale height) as a function of radius for all resolutions and $f_\mathrm{dg}$ values, measured at $z/r=0.1$ and $0.28$ (top and bottom panels, respectively).
}
\label{fig:kH_r}
\end{figure*}

\section{Wavelength estimation}\label{A:WavelengthEstimation}

In this appendix we describe our procedure for the estimation of the wavelengths used in the computation of radiative cooling timescales. We achieve this by computing the distance between two consecutive sign changes of $v_\theta$ at fixed $\theta$. Given that wavelengths are typically larger above the irradiation surface than below (see, e.g., Fig. \eqref{fig:vel_2048}), we carry out this computation for two different heights on each side of the surface, namely $\theta-\pi/2=0.1$ and $0.28$. This is possible also in the VSI-suppressed region for $f_\mathrm{dg}=10^{-4}$, given that in that case we still obtain slow vertical modes below the irradiation surface (see Fig. \ref{fig:vel_2048}). To improve the statistics of our estimation, we accumulate the $\lambda$ values computed in this way every $\sim 0.8$ orbits between $t=150$ and $300$ $T_{5.5}$. The obtained wavelengths are therefore the predominant ones after saturation, and not necessarily the fastest growing ones in the linear stage. 2D histograms of $kH=2\pi H/\lambda$ and radial distance to the star are shown in Fig. \ref{fig:kH_r}, where we used a locally defined scale height $H=c_s \Omega_K^{-1}$ with $c_s^2=p/\rho$. 

At the highest resolution, maximum wavenumbers up to $k H\sim 100$ occur in every case close to the inner radial boundary, where the resolution is maximal. At that resolution, we obtain typical values of $kH$ between $\sim20$ and $\sim80$ below the irradiation surface, whereas above it values with $kH\sim10$ also occur. The dispersion of $kH$ values grows in all cases with resolution and with distance from the midplane, since it is mainly caused by KH eddies. In \cite{LinYoudin2015} it was proposed that the maximum wavenumbers, if not limited by numerical diffusion, are limited by the effective turbulent viscosity following the approximate relation $(kH)^2\lesssim q H/\alpha$, where $\alpha$ parameterizes the produced viscosity as $\nu=\alpha c_s H$. Even though the Reynolds stresses produced by the VSI are largely anisotropic (Section \ref{SS:ReynoldsStress}) we can test this prescription using $\alpha\approx \alpha_r$, since the wave vector is mainly radial. For our simulation parameters and the obtained $\langle\alpha_r\rangle_p$ values, this gives the condition $kH\lesssim 70$, which is similar to the obtained wavelengths. However, this analysis is not very precise, as larger values of up to $\sim 280$ can be obtained if $\langle \alpha_R\rangle_p$ is used instead. 
 
\begin{figure*}[t!]
\centering
\includegraphics[width=\linewidth]{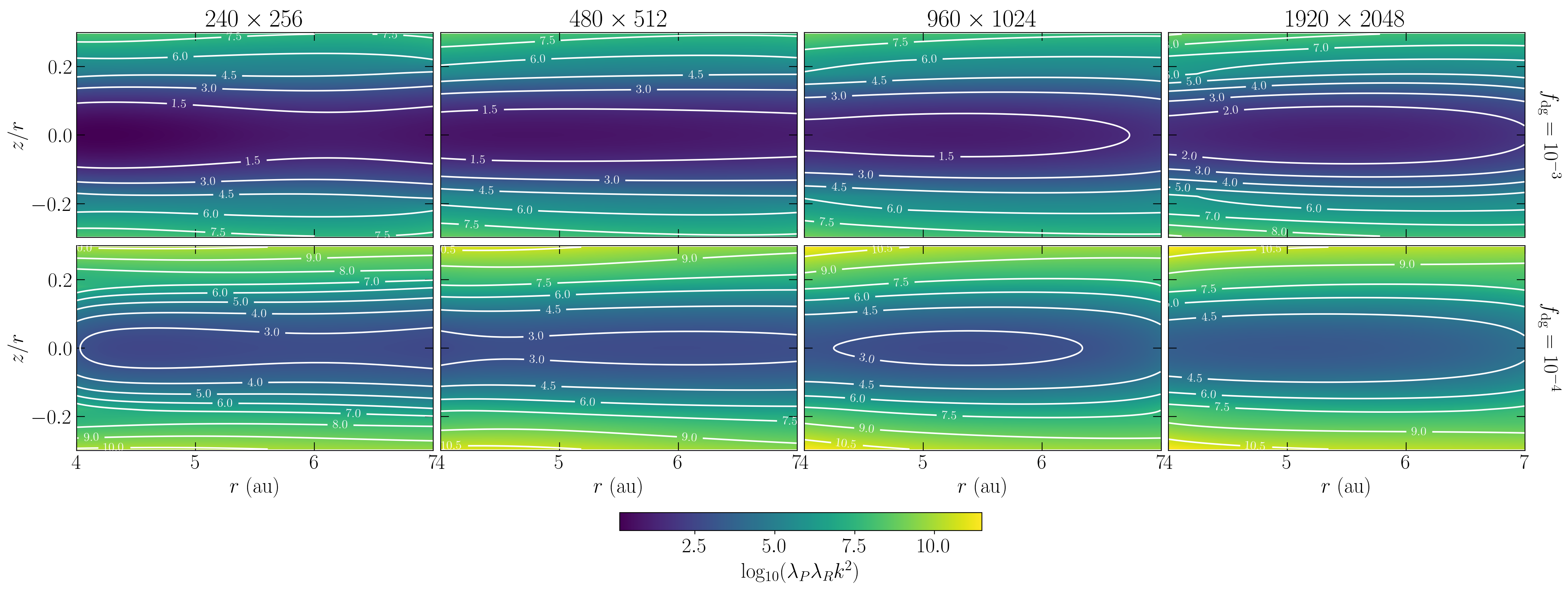}
\caption{Estimated distributions of the parameter $\lambda_P\lambda_R k^2$ for all resolutions and $f_\mathrm{dg}$ values, indicating the optical depth of the VSI modes ($\lambda_P\lambda_R k^2\gg 1$ indicates optically thin cooling).}
\label{fig:llk_rth}
\end{figure*}
 
As a side note, Fig. \ref{fig:kH_r} shows discontinuous wavenumber jumps in some cases, for which different explanations can be found in the literature. On the one hand, \cite{Stoll2014} proposed that these can result from the lengthscale cut-off imposed on the VSI modes by the employment of a finite-sized grid with finite resolution. This would be in principle supported by the variation in the discontinuous jump location as resolution is changed (Fig. \ref{fig:kH_r}). On the other hand, the simulations in \cite{Svanberg2022} show wavelength jumps at fixed radial locations, which were proposed by the authors to correspond to turning points where a linear analysis predicts the radial wavenumber to go to zero. However, we have not seen signs of such turning points in $r-t$ maps of the angular momentum at different heights at any of our resolutions (not shown here), although this could be due to the relatively smaller radial extent of our domain.
 
 
 Since we are only interested in obtaining a rough estimate of $k$ to be used in Eq. \eqref{Eq:tcool}, we fit the measured wavelength distributions as a function of $r$ using a third-order polynomial. The resulting $k(r)$ functions, shown in Fig. \ref{fig:kH_r} as $kH(r)$, are then taken as representative values of $k$ in the entire regions above and below $z/r \sim 0.2$, where the wavelength transition approximately occurs. A 2D distribution of $k$ is then obtained by connecting the fitted values in both regions via a linear taper function in the region limited by $z/r=0.2\pm 0.02$. As shown in Fig. \ref{fig:llk_rth}, the resulting $\lambda_R\lambda_P k^2$ values are always much larger than $1$ for all of our employed resolutions, meaning that temperature perturbations cool down at the optically thin rate even in vertically optically thick regions of the disk (see Fig. \ref{fig:hydrostaticTshear}). This also justifies the rather approximate way in which we computed $k$, which introduces negligible errors in the estimated cooling times, given that the minimum estimated $\lambda_R\lambda_P k^2$ values are $\mathcal{O}(10)$, and therefore $\frac{3+\lambda_R\lambda_P k^2}{\lambda_R\lambda_P k^2}\approx 1$
\end{appendix}

\end{document}